\documentclass[useAMS,usenatbib]{mn2e}

\usepackage{graphicx}
\usepackage{natbib}
\usepackage{subfigure}
   \usepackage{amsmath}
   \usepackage{amsfonts}   
   \usepackage{amssymb}    
\usepackage{longtable,lscape}
\usepackage{Times}
\usepackage[T1]{fontenc}
\usepackage{aecompl}
\usepackage{pslatex}

\def\aap{A\&A}
\def\apjl{ApJL}
\def\apjs{ApJS}

\title[Star-forming galaxies at $z \sim 2$]{The UV to FIR spectral energy distribution of star-forming galaxies in the redshift desert\thanks{{\it Herschel} is an ESA space observatory with science instruments provided by European-led Principal Investigator consortia and with important participation from NASA.}}
\author[I. Oteo et al.]
{\parbox{\textwidth}{I. Oteo,$^{1,2}$\thanks{E-mail: \texttt{ioteo@iac.es}},
\'A. Bongiovanni$^{1,2,3}$,
G. Magdis$^{4}$, 
A.M. P\'erez-Garc\'ia$^{1,2,3}$, 
J. Cepa$^{1,2}$, 
H. Dom\'inguez S\'anchez$^{1,2}$
A. Ederoclite$^{5}$, 
M. S\'anchez-Portal$^{3,6}$, and
I. Pintos-Castro$^{1,2,7}$
}\vspace{0.4cm}\\
$^{1}$Instituto de Astrof{\'i}sica de Canarias (IAC), E-38200 La Laguna, Tenerife, Spain\\
$^{2}$Departamento de Astrof{\'i}sica, Universidad de La Laguna (ULL), E-38205 La Laguna, Tenerife, Spain\\
$^{3}$Asociaci\' on ASPID. Apartado de Correos 412, La Laguna, Tenerife, Spain\\
$^{4}$Department of Physics, University of Oxford, Keble Road, Oxford OX1 3RH\\
$^{5}$Centro de Estudios de F\'isica del Cosmos de Arag\' on, Plaza San Juan 1, Planta 2, Teruel, 44001, Spain\\
$^{6}$Herschel Science Centre (ESAC). Villafranca del Castillo, Spain\\
$^{7}$Centro de Astrobiolog\'{i}a, INTA-CSIC, P.O. Box - Apdo. de correos 78, Villanueva de la Ca\~nada Madrid 28691, Spain\\
}

\begin{document}

\date{Accepted ??. Received ??; in original form ??}

\pagerange{\pageref{firstpage}--\pageref{lastpage}} \pubyear{2002}

\maketitle

\label{firstpage}

\begin{abstract}

We analyse the rest-frame UV-to-near-IR spectral energy distribution (SED) of Lyman break galaxies (LBGs), star-forming (SF) $BzK$ ($sBzK$), and UV-selected galaxies at $1.5 \lesssim z \lesssim 2.5$ in the COSMOS, GOODS-N, and GOODS-S fields. Additionally, we complement the multi-wavelength coverage of the galaxies located in the GOODS fields with deep FIR data taken from the GOODS-\emph{Herschel} project. According to their best-fitted SED-derived properties we find that, because of their selection criterion involving UV measurements, LBGs tend to be UV-brighter, bluer, have a less prominent Balmer break (are younger), and have higher dust-corrected total SFR than $sBzK$ galaxies. In this way, $sBzK$ galaxies represent the general population of SF galaxies at $z \sim 2$ better than LBGs. In a colour--mass diagram, LBGs at $z \sim 2$ are mostly located over the blue cloud, although galaxies with older age, higher dust attenuation, and redder UV continuum slope deviate to the green valley and red sequence. Furthermore, for a given stellar mass, LBGs tend to have bluer optical colours than $sBzK$ and UV-selected galaxies. We find clean PACS ($100\,\mu$m or $160\,\mu$m) individual detections for a subsample of 48 LBGs, 89 $sBzK$, and 91 UV-selected galaxies, that measure their dust emission directly. Their ${\rm SFR_{total} = SFR_{UV} + SFR_{IR}}$ cannot be recovered with the dust-correction factors derived with their continuum slope and the IRX-$\beta$ relations for local starbursts, similar to what happens at higher redshifts. This has implications, for example, in the definition of the main sequence (MS) at $z \sim 2$, which is sensitive to the dust-correction factors adopted. In an SFR--mass diagram, PACS-detected galaxies are located above the \cite{Daddi2007} MS and thus their star formation is probably driven by starburst. This is in agreement with the shape of their IR SEDs. PACS-detected galaxies with redder UV continuum slope and higher stellar mass are more attenuated. We find that for a given UV continuum slope the dustiest galaxies at higher redshifts are more attenuated and that for a given stellar mass the dustiest galaxies at higher redshifts have stronger FIR emission. This suggests an evolution of their dust properties. However, we do not find significant evolution in the relation between dust attenuation and stellar mass with redshift, at least at $z \leq 2.5$. There is a subpopulation of 17, 26, and 27 LBGs, $sBzK$, and UV-selected galaxies, respectively, that are detected in any of the SPIRE ($250\,\mu$m, $350\,\mu$m and $500\,\mu$m) bands. We speculate that this sample of SPIRE-detected LBGs is the bridging population between sub-mm galaxies and LBGs.

\end{abstract}



\section{Introduction}\label{intro}

Several methods have been traditionally employed to look for star-forming (SF) galaxies in the high-redshift universe. One of the most used and successful is the Lyman break or dropout technique, that segregates the so-called Lyman break galaxies (LBGs). It is based on sampling the Lyman break feature of galaxies with a combination of two broad band filters, blue-ward and red-ward ones, each located on both sides of the Lyman break. Many samples of LBGs have been found and studied at different redshifts, mostly at $z \gtrsim 3$ where the Lyman break is located in the optical \citep{Madau1996,Steidel1996,Steidel1999,Steidel2003,Stanway2003,Giavalisco2004LBGs,Bunker2004,Verma2007,Iwata2007,Coe2012}. One of the advantages of the Lyman break compared to other selection techniques that segregate SF galaxies (see later in the text) is that it can be used for carrying out evolutionary studies by sampling the Lyman break at different redshifts with different broad band filter combinations.

Although the Lyman break technique has been historically used for finding galaxies in the high-redshift universe, it can also be employed to look for LBGs at intermediate redshifts, $0.8 \lesssim z \lesssim 2.5$. In this case, UV measurements coming from space-based telescopes are required for sampling the Lyman break. The combination of FUV and NUV GALEX channels or other filter sets in space-based UV  telescopes allows us to identify LBGs at $0.8 \lesssim z \lesssim 1.2$ \citep{Burgarella2006,Burgarella2007,Nilsson2011_LBG,Burgarella2011,Basu2011,Oteo2013a,Oteo2013_ALHAMBRA_PACS,Chen2013} and the combination of NUV and optical U-band filters segregates LBGs at $1.5 \lesssim z \lesssim 2.5$, the so-called \emph{redshift desert} \citep{Hathi2010,Ly2009,Ly2011,Haberzettl2012,Hathi2013}. The number of LBGs reported and studied at $.08 \lesssim z \lesssim 2.5$ is much lower than that at higher redshifts, despite the redshift range being very important since it is then that the peak of the cosmic star formation of the universe is thought to have taken place.

Apart from the Lyman break technique, some other methods have been employed to segregate and analyse galaxies at $1.5 \lesssim z \lesssim 2.5$. \cite{Adelberger2004} defined several selection criteria by employing different combinations of optical colours to produce samples in different redshift ranges: $GRi$ for redshifts $0.85 \lesssim z \lesssim 1.15$, $GRz$ for $1.0 \lesssim z \lesssim 1.5$, and $U_nGR$ for $1.4 \lesssim z \lesssim 2.1$ and $1.9 \lesssim z \lesssim 2.7$. The galaxies selected in this way have been traditionally called BM/BX galaxies. Another ground-based optical colour selection technique is the $BzK$ method, which is aimed at finding galaxies in the redshift range $1.4 \lesssim z \lesssim 2.5$ and classifying them as SF or passively evolving systems \citep{Daddi2004}. \cite{Franx2003} developed a technique designed to look for high-redshift galaxies with intense Balmer breaks at $z \geq 2$. These are called distant red galaxies (DRGs). The presence of a Balmer break can indicate the presence of evolved stellar populations.

Analysing and understanding the differences and similarities between the different kinds of galaxies selected with different selection criteria is essential for understanding galaxy formation and evolution. \cite{Grazian2007} have reported that the selection of galaxies with BX/BM/LBG criteria is sensitive to moderately obscured SF galaxies, but misses dusty starburst (SB) objects. They also found that the $BzK$ criterion is highly efficient at $1.4 \lesssim z \lesssim 2.5$, but when galaxies become faint in the $K$ band and red in the $z - K$ colour, it is difficult to distinguish between SF and evolved galaxies. \cite{Ly2011} have reported that the average galaxy stellar mass, reddening and star formation rates decrease systematically from the $sBzK$ population to the LBGs, and to the BX/BMs. \cite{Haberzettl2012} have found that NUV data provide greater efficiency for selecting SF galaxies. They also report that, although the BM/MX and $BzK$ techniques are very efficient for detecting galaxies within $1 \lesssim z \lesssim 3$, the galaxies found are biased against those SF galaxies which are more massive and contain a noticeable number of red stellar populations. They argue that an NUV-based LBG selection criterion is therefore more suitable for comparing with populations found at $z \gtrsim 3$.

The analysis of the physical properties of LBGs and other SF galaxies at different redshifts has traditionally been carried out by fitting their UV-to-mid-IR photometry to stellar population templates, the so-call SED-fitting technique. This allows us to estimate the age, dust attenuation, stellar mass and SFR of the galaxies studied \citep{Nilsson2011_LBG,Basu2011,Haberzettl2012} to within certain limitations. For example, a typical source of uncertainty comes from the degeneracy between age, star formation history (SFH) and dust attenuation. The inclusion of the emission lines in the SED-fitting procedure also introduces changes in the values of the SED-derived parameters \citep{Zackrisson2008,Schaerer2009,Schaerer2010,Schaerer2011,Schaerer2013,deBarros2013}.

The presence of dust in galaxies, which absorbs the rest-frame UV light of massive stars and reradiates it at FIR wavelengths, prevents the determination of the total SFR in galaxies based solely on the rest-frame UV luminosity. Dust corrections are required but might be uncertain \citep{Wuyts2011_SED,Oteo2013_z3,Oteo2013_ALHAMBRA_PACS}. The best procedure to obtain reliable determinations of dust attenuation and the total SFR of galaxies is by combining UV and FIR measurements \citep{Buat2005,Magdis2010,Buat2010,Wuyts2011_SED,Oteo2011,Oteo2012a,Oteo2012b}. Unfortunately, few LBGs have been individually detected in the FIR so far, mainly at $z \sim 2$ \citep{Chapman2000,Chapman2009,Siana2009,Rigopoulou2010,Magdis2010LBGs}. \cite{Vijh2003} demonstrate that with careful analysis of the UV slope, coupled with appropriate dust attenuation models, they can identify some of the most heavily attenuated specimens in LBG samples at $z \sim 3$. \cite{Finkelstein2009a} find that LBGs at $z \sim 4$ are on average 60\% more likely to be detected than Ly$\alpha$ emitters, implying that LBGs are more dusty, and thus indicating an evolutionary difference between the two populations. \cite{Magdis2010LBGs} studied for the first time, using Photodetector Array Camera and Spectrometer \citep[PACS,][]{Poglitsch2010} data, the FIR SED of infrared-luminous LBGs at $z \sim 3$. Although none of their galaxies is individually detected with \emph{Herschel} \citep{Pilbratt2010}, a stacking analysis suggests a median IR luminosity of $L_{\rm IR} = 1.6 \times 10^{12} L_\odot$. Also with a stacking analysis, \cite{Lee2012} studied the FIR emission of LBGs at $3.3 \lesssim z \lesssim 4.3$ and found that their IR-to-UV luminosity ratio ($L_{\rm IR} / L_{\rm UV}$) is low  compared to that observed for $z \sim 2$ LBGs. At $z \sim 3$ and using stacked Spectral and Photometric Imaging REceiver \citep[SPIRE,][]{Griffin2010} detections, \cite{Davies2013} find that a significant fraction of LBGs at that redshift is obscured and that the dust attenuation of the galaxies is not as high as that predicted by their UV slope. \cite{Burgarella2011} report the first SPIRE-250$\mu$m and SPIRE-350$\mu$m FIR individual detections of LBGs by employing data from the HerMES project \citep{Oliver2010}. They found SPIRE detections for 12 LBGs at $0.7 \lesssim z \lesssim 1.6$ and only one at $z \sim 2$. All these galaxies are high mass, luminous IR galaxies and have redder NUV-U and U-R colours than other SPIRE-undetected galaxies. \cite{Oteo2013_ALHAMBRA_PACS} report the PACS-100$\mu$m/160$\mu$m detections of a sample of 42 GALEX-selected LBGs at $z \sim 1$ located in the COSMOS field. They found that PACS-detected LBGs are dustier, more massive and have a redder UV continuum than PACS-undetected LBGs. Their total IR luminosities place them in the LIRG regime. None of them has a ULIRG nature, even though ULIRGs could have been detected by PACS. This is in agreement with the total IR luminosities of the SPIRE-detected LBGs of \cite{Burgarella2011}. \cite{Oteo2013_z3} found a population of LBGs at $z \sim 3$ that are individually detected in the FIR. Their total IR luminosity place them in the ULIRG or hyper-luminous galaxies regime. They also found evidence that the IR emission of LBGs might have changed with redshift in the sense that the dustiest LBGs at higher redshifts have stronger FIR emission than the dustiest LBGs at lower redshifts. The low number of individually FIR-detected LBGs reported so far, mostly at $z \geq 1.2$, has prevented a detailed study of their FIR emission, which is essential, for example,  to obtaining accurately their dust attenuation and total SFR. Therefore, it is clear that a deeper analysis of the FIR emission of LBGs at $z \sim 2$ is needed. Additionally, FIR emission might give important clues about the differences between LBGs, $sBzK$ and UV-selected galaxies.

In this paper we try to understand better the differences and similarities between the various kinds of SF galaxies in the redshift desert by carrying out a multiwavelength analysis from the rest-frame UV to the FIR. One of the main goals of the work isto  analyse the physical properties of SF galaxies named differently according to their selection criteria in order to see whether they correspond to  similar/dissimilar populations and examine the degree of overlapping between the different samples. To do that, we adopt a reference sample of LBGs at $z \sim 2$ and compare their SED-derived physical properties and their PACS-100$\mu$m/160$\mu$m and SPIRE-250$\mu$m/350$\mu$m/500$\mu$m FIR emission with those for a sample of SF $BzK$ galaxies and a general population of UV-selected SF galaxies located within the same redshift range. We do not compare LBGs with DRGs or BM/BX galaxies since their redshift distributions are significantly different although with similar median values.

The paper is organized as follows: In Sections \ref{data} and \ref{selec} we describe the photometric data and selection criteria for selecting the different galaxies adopted in this work, respectively. In Section \ref{SED} we explain the SED-fitting procedure and analyse the SED-derived physical properties of the galaxies studied for the purposes of comparison. We report the PACS and SPIRE FIR detections and study the FIR SED of our galaxies in Section \ref{FIR_SED}, using this sample to set their differential properties and net contribution to the overall star formation rate density at $z \sim 2$. Finally, the main conclusions of the study are presented in Section \ref{conclu}.

Throughout this paper we assume a flat universe with $(\Omega_m, \Omega_\Lambda, h_0)=(0.3, 0.7, 0.7)$, and all magnitudes are listed in the AB system \citep{Oke1983}.

\section{Data sets}\label{data}

As a first step in our selection of galaxies at $1.5 \lesssim z \lesssim 2.5$, we generate a multiwavelength photometric catalogue by combining all the available information in the COSMOS, GOODS-N and GOODS-S fields from GALEX-UV to IRAC-mid-IR. This includes GALEX observations taken within the framework of the Deep Imaging Survey (DIS), the \cite{Santini2009} multiwavelength catalogue from the U-band to MIPS-24$\mu$m, the photometric catalogues of \cite{Capak2004} and \cite{Capak2007}, the mid-IR observations of the S-COSMOS survey \citep{Sanders2007}, and the deep K and IRAC data of \cite{Wang2010}.

\begin{figure}
\centering
\includegraphics[width=0.45\textwidth]{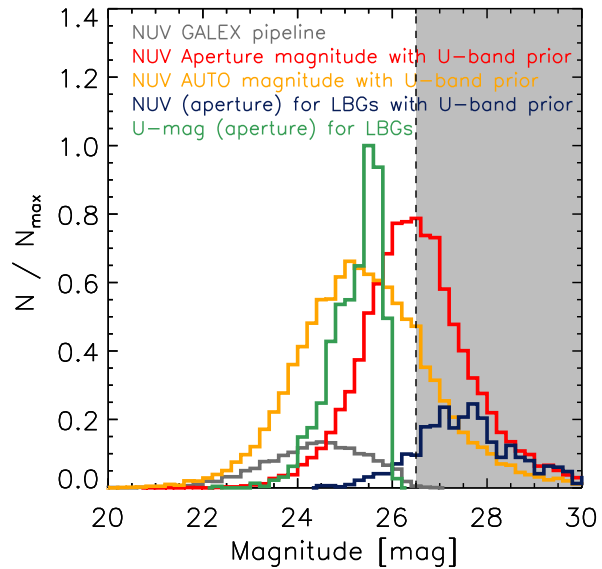}
\caption{Distribution of the NUV magnitude obtained with the GALEX pipeline (grey histogram), 3$''$ aperture NUV magnitude extracted by using $U$-band priors (red histogram), and NUV AUTO magnitude extracted with $U$-band priors (orange histogram). Only sources with extracted NUV flux are represented. For comparison, we also show the aperture NUV (blue) and $U$-band (green) aperture magnitudes of our LBGs at $z \sim 2$ with measurements in the NUV channel. The grey shaded zone represents faint magnitudes for which a NUV flux cannot be considered as genuine detections (see also \citet{Salim2009}). This Figure shows the UV extraction results in the GOODS-N field.
              }
\label{extraction}
\end{figure}

On the UV side, we employ GALEX data strictly to do a preliminary selection of the LBGs candidates, as explained in Section \ref{lbgs}. When working with GALEX data it should be taken into account that the large PSF of the GALEX images decreases the detection limiting magnitude observations. Actually, the GALEX pipeline produces catalogues by using SExtractor \citep{Bertin1996} aperture photometry. This procedure might be reliable for resolved and shallow images but when dealing with deep surveys, such photometry may suffer from blending and source confusion. Since we use GALEX data only to sample the Lyman break at $z \sim 2$ but not for SED-fitting (see Sections \ref{lbgs} and \ref{SED}) the deblending and confusion are not crucial. Despite this, for extracting and avoiding those problems we perform aperture photometry on GALEX images by employing $U$-band coordinates as priors. In this process we use SExtractor in its dual mode, with detections in the U-band and measurements in the GALEX NUV images.\footnote{GALEX images are taken from the MAST archive and correspond to the data release 7} As in \cite{Haberzettl2012} we do not deconvolve the NUV images. After extracting the NUV fluxes with the $U$-band priors, we find that detections have $NUV < 26.5$ at $3\sigma$. This is the criterion we adopt for considering an NUV detection in our catalogues and it is similar to the one employed in previous studies analyzing high-redshift UV-detected galaxies \citep{Salim2009,Haberzettl2012}. Figure \ref{extraction} represents the distributions of the NUV magnitudes obtained with the GALEX pipeline and with our method with $U$-band priors. It can be seen that when using $U$-band priors we can increase the actual depth of the NUV observations, limited to about 25.5 in blind UV extraction (see also \cite{Burgarella2006,Burgarella2007,Burgarella2011}). The publicly available GALEX data in the COSMOS field had been already treated with the EM-algorithm, aimed at resolving the blended objects in the far and the near UV using the information (position and shape) available from existing, well resolved catalogues in the visible range \citep{Gillaume2006}. With a list of optical prior positions, the algorithm measures their UV fluxes on the GALEX  images by adjusting a GALEX PSF model. The algorithm was run on the four NUV and the four FUV GALEX images covering the COSMOS field obtained as a product of the GALEX pipeline processing. The prior optical photometric information corresponds to a u*-band mosaic (and its SExtractor-derived catalogue) based on CFHT-u* observations. Comparing our method and the PSF-fitting method in the COSMOS field we obtain similar results (most LBGs undetected in the NUV filter or well below the limit $NUV \sim 26.5$) and, therefore, for our purposes of measuring the Lyman break, we consider the U-band-driven method to be valid. This is further supported by the fact that, for the sources with an available spectrum, their spectroscopic redshifts indicate that they are mostly within $1.5 < z < 2.5$ (see Section \ref{zphot}). This is the expected redshift range of the NUV/U-selected LBGs and, therefore, the selection of the sources has been carried out in the right way.


On the FIR side, we use PACS and SPIRE FIR data taken within the framework of the GOODS-\emph{Herschel} Open Time Key Program \citep{Elbaz2011}. In this project, GOODS-S was observed with PACS-100$\mu$m/PACS-160$\mu$m and GOODS-N with PACS-100$\mu$m, PACS-160$\mu$m and also with SPIRE-250$\mu$m, SPIRE-350$\mu$m, and SPIRE-500$\mu$m. The catalogues used here are 80\% complete at the flux levels of (1.7, 5.5, 10, 13, 22) mJy in (PACS-100$\mu$m, PACS-160$\mu$, SPIRE-250$\mu$m, SPIRE-350$\mu$m, SPIRE-500$\mu$m), respectively, in GOODS-N and (1.3, 3.9) mJy in (PACS-100$\mu$m, PACS-160$\mu$m), respectively, in GOODS-S. At the resolution of \emph{Spitzer} and \emph{Herschel}, most of the FIR-detected sources in the GOODS-\emph{Herschel} data are point-like sources. Therefore, a PSF-fitting technique was applied to perform photometry in previously known prior positions. In a first step, sources in MIPS-24$\mu$m were extracted using prior IRAC positions taken from the GOODS \emph{Spitzer} Legacy programme \citep{Dickinson2003}. Then, the MIPS-24$\mu$m-detected sources with signal-to-noise ratio $S/N>3$ were employed as a prior information to extract sources in the \emph{Herschel} images. More details on source extraction can be consulted in the GOODS-\emph{Herschel} release documentation.\footnote{http://hedam.oamp.fr/GOODS-Herschel/} It should be remarked that the width of the IRAC PSF prevents us from distinguishing sources separated by less than 4$''$. Therefore, confusion problems might arise when performing the match between optical-based and FIR-based catalogues (see Section \ref{FIR_SED}).

It should be noted that in the final multiwavelength catalogue we include sources both detected and undetected in the UV with GALEX and in the FIR with \emph{Herschel}.

\begin{table*}

\caption{\label{SED_bands}Optical and near-, mid-IR filters used in the rest-frame UV-to-near-IR SED fits carried out in this study for each of the cosmological fields analysed.}

\centering

\begin{tabular}{ll}

\hline\hline

Field & Filters \\

\hline

COSMOS & u* (Mega-Prime), $B_{J}$, $V_{J}$, $g^{+}$, $r^{+}$, $i^{+}$, $z^{+}$ (Suprime-Cam), $K$ (FLAMINGOS), and IRAC bands\\

GOODS-S & $U_{35}$, $U_{38}$ (WFI) $U$ (VIMOS), $B$, $V$, $i$, $z$ (ACS), $J$, $H$, $K$ (ISAAC), and IRAC bands \\

GOODS-N & $U$ (MOSAIC), $B$, $V$, $I$, $z'$ (Suprime-Cam), $HK'$ (QUIRC), $K$ (WIRCam), and IRAC bands \\

\hline

\end{tabular}

\end{table*}

\section{Selection of the sources}\label{selec}

\subsection{Lyman break galaxies}\label{lbgs}

\begin{figure*}
\centering
\includegraphics[width=0.45\textwidth]{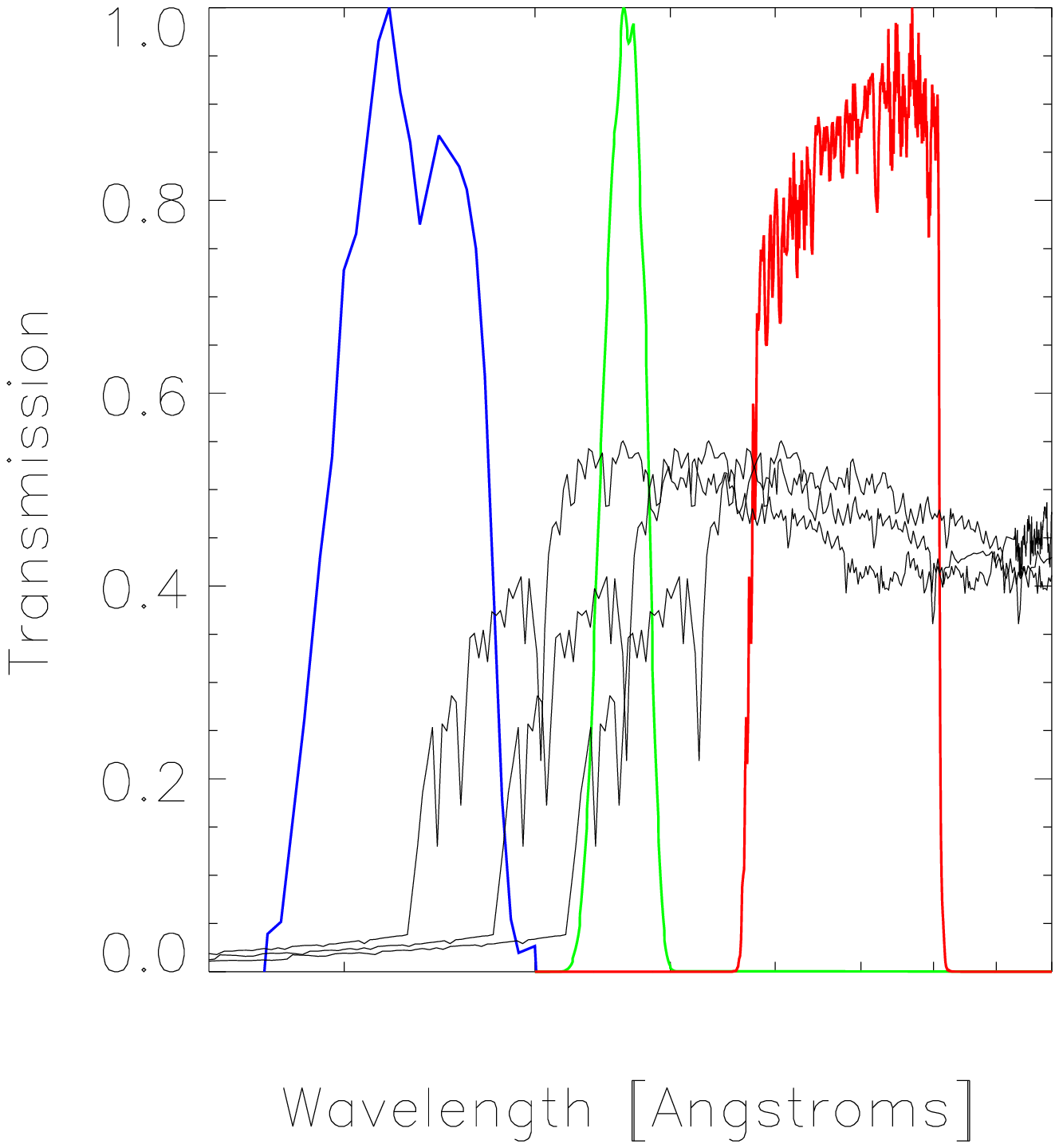}
\includegraphics[width=0.45\textwidth]{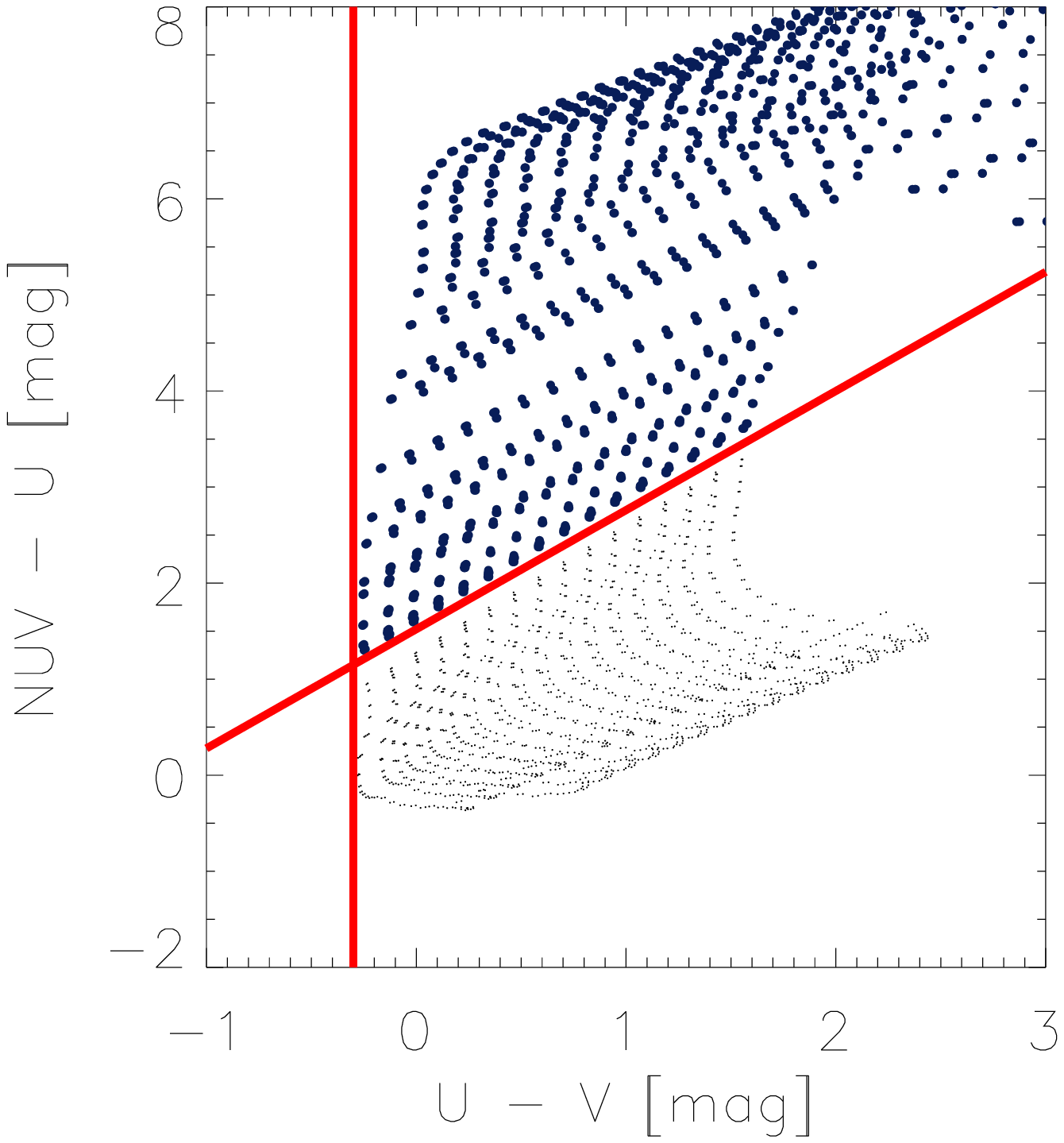}
\caption{Elaboration of the analytical selection criterion of our LBGs. \emph{Left plot:} Transmission curves of the NUV, U and V filters employed to look for LBGs at $1.5 \lesssim z \lesssim 2.5$ in the GOODS-S field. We also plot the same BC03 template redshifted at $z=1.5$, 2.0, and 2.5 with the aim of showing how the previous filters sample the Lyman break and the redward zones of the SED of a galaxy as a function of redshift. \emph{Right plot:} Colour--colour diagram employed to look for our LBGs at $1.5 \lesssim z \lesssim 2.5$. Large blue points are SF galaxies at $z \geq 1.5$, small black points are SF galaxies at $z < 1.5$, and the red lines delimit the selection window employed to formulate the analytical selection criterion to look for our LBGs.
              }
\label{selection}
\end{figure*}

With the aim of looking for LBGs at $1.5 \lesssim z \lesssim 2.5$ we use the classical drop-out technique with the broad band filters GALEX-NUV, $U$ and $V$. The NUV and U filters are the blueward and redward filters, respectively, that sample the Lyman break in that redshift range. The combination of $U$ and $V$ filters samples the UV continuum redward of the Lyman break and is used to rule out interlopers (see left plot of Figure \ref{selection}). In order to formulate an analytical selection criterion for our LBGs we convolve the transmission curves of the previous broad band filters with a large set of \citet[][hereafter BC03]{Bruzual2003} templates associated with different values of age and dust attenuation (included via the \cite{Calzetti2000} law) and shifted in wavelength according to different values of redshift spanning $0 \leq z \leq 4$. For each value of redshift, we represent the location of each template in a colour--colour space (see right plot of Figure \ref{selection}). Since the drop-out technique with the set of filters NUV, $U$, and $V$ is expected to isolate galaxies at $1.5 \lesssim z \lesssim 2.5$ we represent in the right plot of Figure \ref{selection} with large blue dots the galaxies located at $z \geq 1.5$, while galaxies at lower redshifts are represented by small black dots. The region defined by the locus of the blue dots represents the selection window for our LBGs. Analytically,

\begin{equation}\label{selection_LBGs}
{\rm NUV} - U \geq 1.24 \times (U - V) + 1.52
\end{equation}

In addition to this, we also impose non-detection in the FUV channel. In the sample of LBGs we also include those galaxies which, although being undetected in the GALEX-NUV channel, are bright enough in the U-band to ensure a true Lyman break between the two bands. Actually, due to the depth of the observations employed in the present work, the application of Equation \ref{selection_LBGs} entails that most of the selected LBGs (above 95\%) are undetected in the NUV channel or have NUV magnitude fainter than the 3$\sigma$ limit. The non-detection of LBGs in the blueward filter is usual in LBG searches \citep[see for example][]{Haberzettl2012}. As expected, the LBGs detected in the NUV band correspond to the U-band brightest galaxies. A large $NUV - U$ colour could be due to underestimation of the NUV flux because of the large PSF of the UV observations. However, we have visually inspected the optical and NUV images of the NUV-detected LBGs and have checked that they are all well isolated galaxies in terms of the NUV PSF and, therefore, their UV fluxes are well extracted. In this way, the $NUV - U$ colour of the NUV-detected LBGs is compatible with the presence of a Lyman break.

In addition to the previous analytical selection criterion, we restrict the photometric redshift (see Section \ref{zphot}) of the galaxies to be within $1.5 \leq z \leq 2.5$ in order to have as clean a sample of interlopers as possible and avoid the inclusion of lower redshift galaxies and stars \citep{Ly2009,Haberzettl2012}. Actually, the z-phot criterion reduces the colour-selected sample by about 50\% in each field. These interlopers are mostly galaxies with photometric redshifts $1.0 \lesssim z \lesssim 1.5$, with no difference in the NUV detection fraction compared to the selected LBGs. These galaxies are not included in the sample since, despite meeting the drop-out criterion, their Lyman break is still in the NUV band and cannot therefore be strictly considered as LBGs at $z \sim 2$. Furthermore, we avoid the contamination from AGN by removing all the sources that have X-ray detections in the catalogues of \cite{Alexander2003,Elvis2009,Xue2011} in GOODS-N, COSMOS and GOODS-S fields, respectively. With all these considerations, we end up with a sample of 3207, 681 and 1300 LBGs in COSMOS, GOODS-S and GOODS-N, respectively. Figure \ref{extraction} shows the distribution of the NUV (blue histogram) and $U$-band (green histogram) magnitudes of our LBGs at $z \sim 2$ in the GOODS-N field. On the UV side, only galaxies with an NUV measurement are considered. Most LBGs are not detected in the NUV channel because of their strong Lyman break. Those for which we recover an NUV flux, the corresponding magnitudes are below the limit $NUV \sim 26.5$ and, consequently, are also undetected in practice. The separation between the green and blue histograms is a measurement of the strength of the Lyman break.

\subsection{BzK galaxies}\label{bzk}

\cite{Daddi2004} proposed a method, the $BzK$ technique, to look for SF and passively evolving galaxies in the redshift desert. This is a two-colour selection based on $B$-, $z$-, and $K$-band photometry that also allows us to distinguish between SF and passively evolving systems. Imposing $BzK = (z-K)_{\rm AB} - (B-z)_{\rm AB} > -0.2$ one can actively  segregate SF galaxies at $z > 1.4$ (the so-called $sBzK$ galaxies) independently of their dust attenuation since the reddening vector in the $BzK$ diagram is parallel to the $BzK$ limiting line. Imposing $BzK < -0.2$ and $(z-K)_{\rm AB} > 2.5$, one can select passively evolving galaxies (the so-called $pBzK$ galaxies) at $z \geq 1.4$ \citep{Daddi2004}. To place our LBGs in the wider context of galaxies at $1.5 \lesssim z \lesssim 2.5$ and study the bias produced by the drop-out selection criterion for selecting SF galaxies, we also segregate galaxies by using the $BzK$ technique. Again, in this process, we discard those galaxies with X-ray detections to avoid AGN contamination. As we did with LBGs, we also limit the photometric redshift of the $BzK$ galaxies to be within $1.5 \lesssim z \lesssim 2.5$. This z-phot criterion reduces the sample by about 50\%

Regarding $sBzK$ galaxies, we end up with a sample of 9539, 2472 and 2192 in COSMOS, GOODS-S and GOODS-N, respectively. The $sBzK$ galaxies are used in next sections as a comparison sample with our LBGs, apart from the fact that such sources constitute a genuine population of star-forming galaxies. At this point, and for this reason, it is convenient to note that LBGs and $sBzK$ galaxies do not have to be completely disjoint samples. Actually, as will be shown in Section 4.3, there is an overlap between them. On the other hand, selection of $pBzK$ galaxies was focused only on the COSMOS field, isolating a sample of 127 sources. We do not recover pBzK galaxies in the GOODS fields since they are typically faint in the U-band, despite most of the galaxies in the GOODS photometric catalogues being detected at that wavelength. It should be noted that this limitation has no implications for the conclusions of the paper, since $pBzK$ galaxies are employed only to constrain the location of the red sequence of galaxies at $z \sim 2$ in a colour--mass diagram (see Section \ref{CMD}), but are not directly compared to our selected LBGs and $sBzK$ galaxies.

\subsection{General population of UV-selected galaxies}








In order to have a more global vision of star formation at $1.5 \lesssim z \lesssim 2.5$ we also build a sample of UV-selected galaxies in that redshift range with no limitation on their rest-frame UV or optical colours. In the redshift range where our LBGs and $sBzK$ galaxies are located we can trust the photometric redshifts (see Section \ref{SED}) to build a larger sample of SF galaxies. The selection of galaxies based solely on photometric redshift avoids the bias introduced by the different colour selection criteria, such as those for LBGs or $BzK$ galaxies. In this way we construct a sample of UV-selected galaxies formed by sources whose SED-derived photometric redshifts (see Section \ref{SED}) span within $1.5 \leq z_{\rm phot} \leq 2.5$, are detected in the $B$-band, which samples the rest frame FUV in that redshift range and the galaxies do not have X-ray counterparts within 2$''$ of their optically based position. This sample is formed by 8100, 2767, and 2581 galaxies in COSMOS, GOODS-S and GOODS-N, respectively. This is a pure photometric redshift and UV selection; therefore, this sample contains all the LBGs and $sBzK$ galaxies (see Section \ref{comparison} for more details).

\section{UV-to-mid-IR SED fitting}\label{SED}

For each galaxy in the multiwavelength photometric catalogue previously built in Section \ref{selec} we perform an SED-fitting procedure with BC03 templates. Table \ref{SED_bands} summarizes the filter sets employed for carrying out the SED fits in each of the three cosmological fields. The objective of this process is twofold: determining the photometric redshift of the galaxies while at the same time obtaining their SED-derived physical properties, such as rest-frame UV luminosities, age, stellar mass, dust attenuation, UV continuum slope, dust-corrected total SFR, sSFR and amplitude of the Balmer break. In this process, we include all the available photometry from UV to IRAC. We do not employ redder wavelengths since they have a significant contribution of dust emissio,n which is not considered in the BC03 templates.

Two of the most important features of the SED of galaxies to be fitted are the Balmer break and the UV-continuum slope. The sampling of these two features at the same time enables us to have better determinations of the photometric redshift of the galaxies and, therefore, more accurate determinations of their physical properties. Furthermore, the amplitude of the Balmer break is an indication of the age of the galaxies, and the UV continuum slope is a measurement of reddening. Additionally, the rest-frame K-band emission is a good tracer of the stellar mass of a galaxy. At the expected redshift range of our LBGs and $BzK$ and UV-selected galaxies, $1.5 \lesssim z \lesssim 2.5$, the Balmer break is located between the optical and near-IR and the rest-frame K-band emission is redshifted to IRAC bands. Therefore, with the aim of sampling all these features and fitting the SEDs of our selected galaxies as accurately as possible, we only consider galaxies that are detected in all optical bands from B and at least the near-IR K filter, and IRAC-3.6$\mu$m and IRAC-4.5$\mu$m mid-IR channels.

We carry out the SED fits by using the Zurich Extragalactic Bayesian Redshift Analyzer \citep[ZEBRA,][]{Feldmann2006} code which, in its maximum-likelihood mode, employs a $\chi^2$ minimization algorithm over the templates to find the one that best fits the observed SED of each input object. We built a set of BC03 templates associated with different physical properties of galaxies by using the \verb+GALAXEV+ software. In this process we adopt a \cite{Salpeter1955} initial mass function (IMF), distributing stars from 0.1 to 100 M$_\odot$, and select a fixed value for metallicity of $Z=0.2Z_{\odot}$. We choose a fixed value for metallicity since this parameter does not alter significantly the shape of the BC03 templates and, consequently, tend to suffer from large uncertainties \citep{deBarros2013}. Dust attenuation is included in the templates via the \cite{Calzetti2000} law and parameterized through the colour excess in the stellar continuum, $E_s(B-V)$, for which we choose values ranging from 0 to 0.7 in steps of 0.025. We also include intergalactic medium absorption adopting the prescription of \cite{Madau1995}. Regarding SFR, we consider time-constant models. We have also performed SED fits with BC03 templates associated with exponentially declining SFH. However, the $\chi^2$ of the results are similar for different values of the SFH time scale and, consequently, it is not possible to distinguish between different kinds of SFH with a SED-fitting method \citep[see also for example][]{Oteo2013_z3}. In this way, we adopt the BC03 templates associated with a constant SFR since it has been broadly employed in the literature and, furthermore, is in agreement with the assumptions made to build the \cite{Kennicutt1998} calibrations between rest-frame and IR luminosities and SFR that will be employed in this work. In this way, the values of the SED-derived properties reported in this work should be understood as those obtained with the assumption of a constant SFR and considering that others SFHs might lead to slightly different results.

In the constant SFR scenario, different values of the SFR do not alter the shape of the BC03 templates, and the way to proceed is to obtain the UV-derived SFR from the normalization of the templates to the observed photometry: once the normalization is done we obtain the rest-frame UV luminosity associated with each normalized template, $L_{1500}$, by integrating it with a top-hat filter centred in rest-frame 1500 \AA. Throughout this study, we consider the rest-frame UV luminosities expressed as $L_{\rm UV}$ to be in $\nu L_{\nu}$ units, whereas the rest-frame UV luminosities expressed as $L_{1500}$, are in $L_{\nu}$ units. $L_{1500}$ is converted into SFR by using the \cite{Kennicutt1998} calibration:

\begin{equation}\label{SFRUV}
\textrm{SFR}_{UV,uncorrected}[M_{\odot}\textrm{yr}^{-1}] = 1.4 \times 10^{-28}L_{1500}
\end{equation}

It should be pointed out that Equation \ref{SFRUV} applies to galaxies with continuous star formation over a time scale of $10^8$ or longer. For younger populations, Equation \ref{SFRUV} gives underestimated values of the ongoing SFR \citep{Leitherer1995}. A percentage of the galaxies studied in this work have SED-derived ages below that threshold (see Sections \ref{sp} and \ref{comparison}). However, these galaxies are expected to be older, since such objects are unlikely to have formed immediately prior to the epoch of observation (see, for example, \cite{Sawicki2012} for a discussion). An age below 100 Myr is an indication that those galaxies are experiencing a recent burst of star formation that is masking any older underlying population. Furthermore, the application of the \cite{Kennicutt1998} calibration has been widely employed in previous studies analysing the properties of high redshift SF galaxies and therefore enables an easier comparison with the results in the literature. Additionally, the use of the \cite{Kennicutt1998} relation between the rest-frame UV luminosity and SFR is compatible with the use of the \cite{Kennicutt1998} calibration between the total IR luminosity and the SFR \citep[see for example][]{Magdis2010}.

The SFR derived with the Equation \ref{SFRUV} is uncorrected for the attenuation that the dust produces in the rest-frame UV continuum. Once the SED-derived $E_s(B-V)$ is known for each source, we can correct $L_{1500}$ from dust attenuation by multiplying it by the dust correction factor 10$^{0.4A_{1500}}$, where $A_{1500}$ is the dust attenuation in 1500\AA. The $A_{1500}$ values are obtained from the SED-derived value of $E_s(B-V)$ assuming the \cite{Calzetti2000} law. Using the dust-corrected $L_{1500}$ in Equation \ref{SFRUV} we obtain an estimation of the dust-corrected total SFR. The dust-corrected total SFR can be also recovered by using the dust correction factors derived from the UV continuum slope \citep{Bouwens2012,Oteo2013a} and the application of an IRX-$\beta$ relation \citep{Meurer1999,Overzier2011,Takeuchi2012}. All these dust correction methods will be used throughout the study. The stellar mass of the galaxies is derived from the BC03+GALAXEV code.

The age of a galaxy is related to the amplitude of the Balmer break in the sense that it is stronger for older galaxies. We calculate the amplitude of the Balmer break for each galaxy from its best-fitted BC03 template by integrating it with two top-hat filters centred in rest-frame 4500 \AA\ and 3500 \AA. With the fluxes obtained we calculate the rest-frame luminosities in 4500 \AA\ and 3500 \AA, $L_{R}$ and $L_{L}$, respectively. We define the amplitude of the Balmer break as the ratio between these two luminosities: $L_R/L_L$.

In this work we also aim to study the UV continuum of our selected galaxies. According to \cite{Calzetti1994}, the UV continuum can be parameterized by its spectral slope, $\beta$, defined by assuming that the UV continuum behaves as a power-law function: $f_\lambda \propto \lambda^\beta$, where $f_\lambda$ is the flux density of the UV continuum in wavelength units. The study of the UV continuum slope is relevant because it is related to different properties of galaxies such as metallicity, age and SFH, but most importantly to dust attenuation. Actually, several relations have been found between dust attenuation and the UV continuum slope \citep{Meurer1999,Boissier2007,Buat2012,Overzier2011,Takeuchi2012,Nordon2013} that have been traditionally applied to recover the total bolometric luminosity and dust correction factors of different kinds of galaxies when FIR information is unavailable \citep{Nilsson2009_letter,Bouwens2011,Overzier2011}. Furthermore, one advantage of studying the UV continuum rather than the continuum in other rest-frame wavelengths is that it is easy to sample with UV, optical and near-IR filters for a wide range of redshift, from the local universe up to $z \sim 8$. Traditionally, the UV continuum slope has been determined by using single rest-frame UV colours \citep[see e.g.,][]{Kong2004,Buat2010}. Recently, other methods have been employed. For example, \cite{Bouwens2011} measure $\beta$ by fitting all the available photometric points which sample the UV continuum at the redshift of their galaxies with a power-law function (\emph{power-law method}). \cite{Buat2012} determine the UV continuum slope of galaxies at $0.95 < z < 2.2$ with intermediate-band filters, which are more sensitive to the possible presence of a bump in the UV continuum \citep[see also][]{Burgarella2005,Buat2011_bump}. \cite{Finkelstein2012,Oteo2013a} measure $\beta$ directly from the best-fit BC03 templates of their galaxies (\emph{SED-fitting method}). \cite{Finkelstein2012} also perform simulations in order to analyse the differences in the values of the UV continuum slopes when employing these three commented methods. They find that both the SED-fitting and power-law methods result in a smaller scatter at all magnitudes than the single-colour method, and that the number of galaxies 'catastrophically' scattered (with uncertainties $\Delta \beta > \pm 1$) is much smaller when using the SED-fitting or power-law methods than when using rest-frame UV colours. Supported by these results, the authors argue that the SED-fitting method provides the best results for the UV slope, since the best-fit template associated with each galaxy reproduces its photometric SED in the rest-frame UV accurately according to the low $\chi^2_r$ values of the SED-fitting results. Therefore, we apply the SED-fitting method to derive the UV continuum slope of our galaxies. In this process, we use the rest-frame wavelength range [1300, 3000] \AA, which includes all the windows defined in \cite{Calzetti1994} in their definition of the UV continuum slope. The previous range is sampled with the broad-band filters from $U$ up to $z$ at the redshift of our galaxies. Finally, we have checked that the assumption of a constant SFR when building the BC03 templates does not bias the derived values of the UV continuum slope. The differences are lower than 0.2 in most cases. This is expected since $\beta$ is derived by using the best fitted template for each galaxy and those templates fit properly the UV continuum of the galaxies regardless the adopted temporal dependence of the SFH.

\begin{figure*}
\centering
\includegraphics[width=0.45\textwidth]{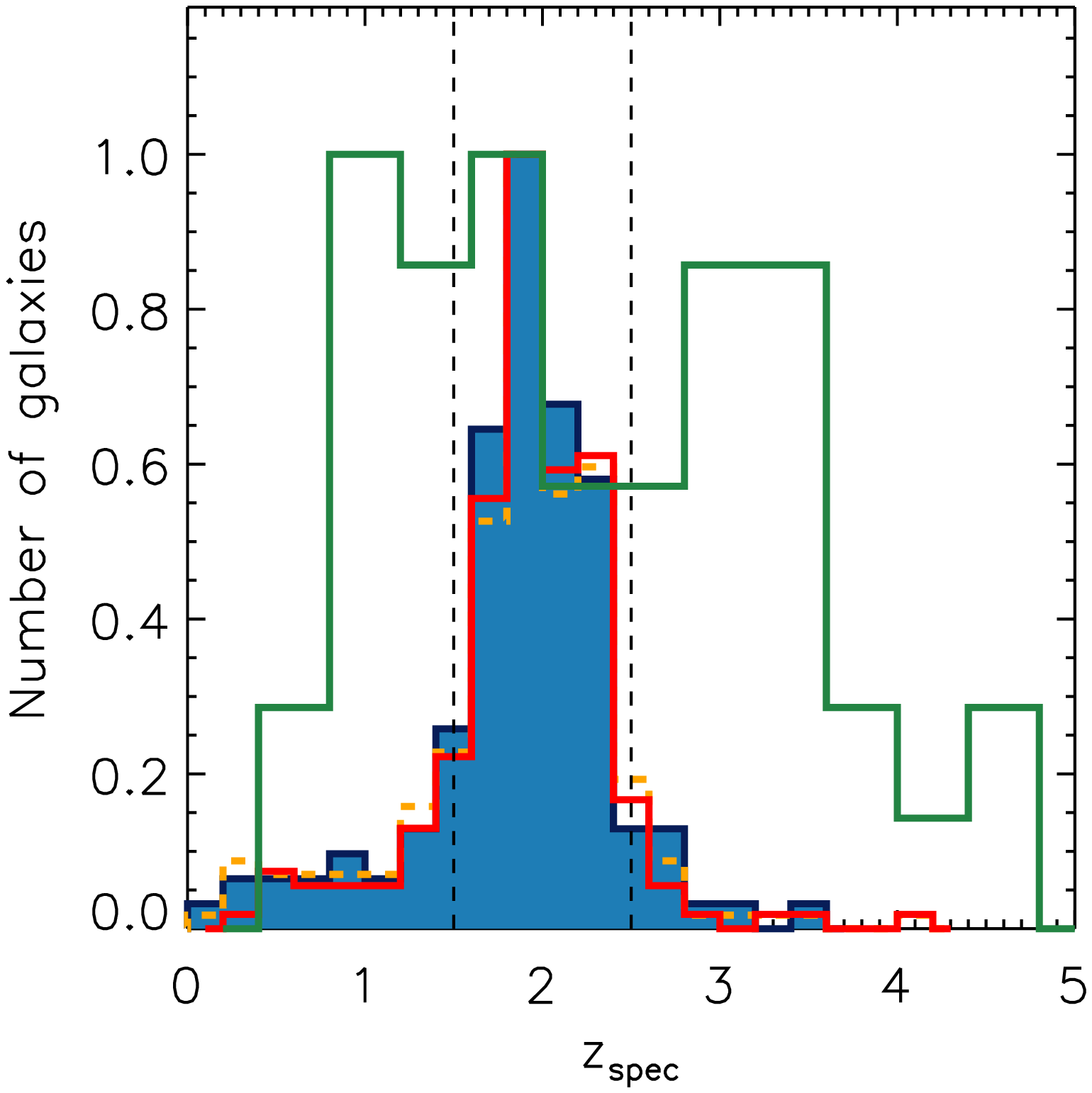}
\includegraphics[width=0.45\textwidth]{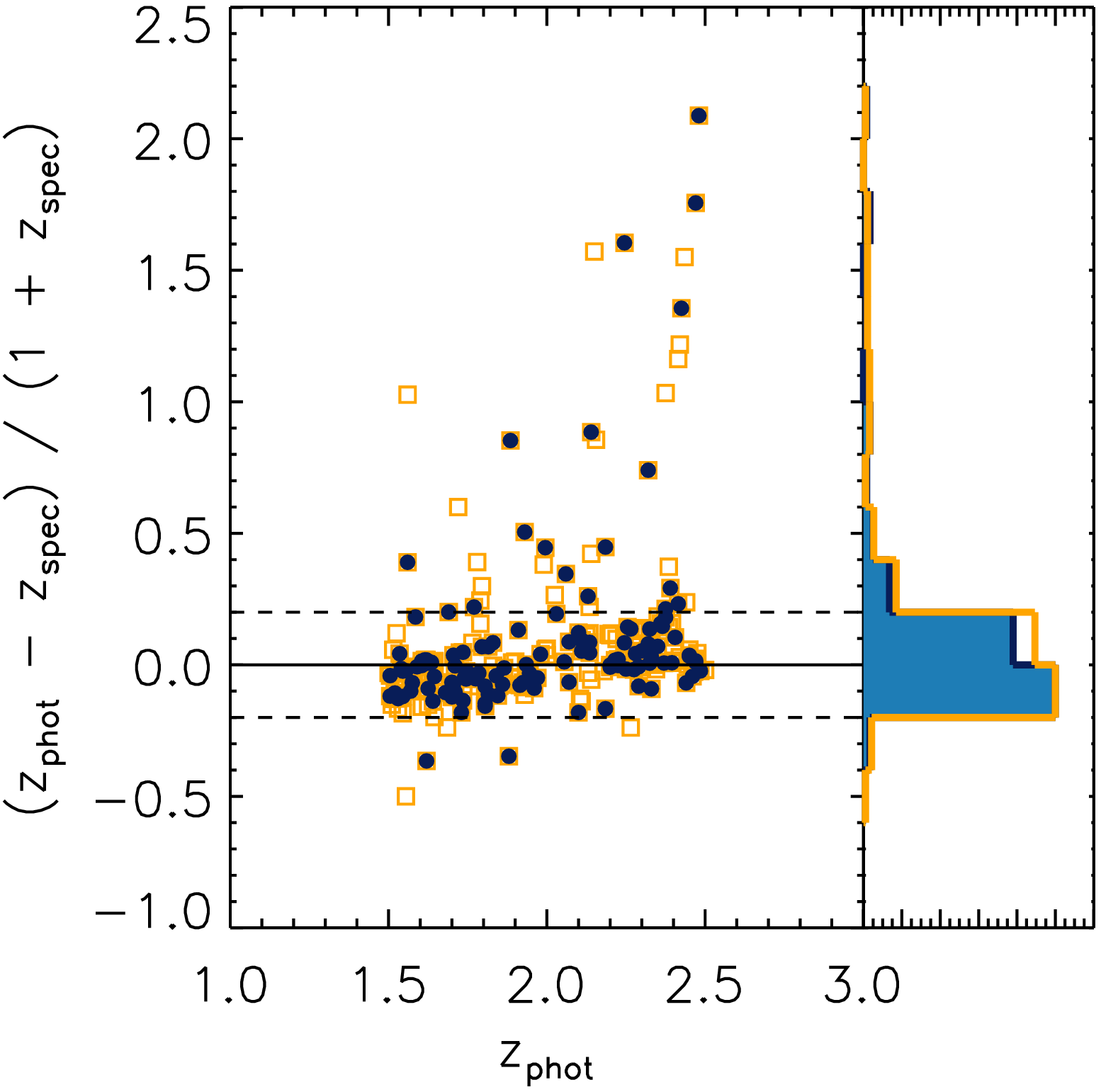}
\caption{\emph{Left}: Distribution of the spectroscopic redshifts for LBGs (blue shaded histogram), $sBzK$ galaxies (red histogram), and UV-selected galaxies (orange dashed histogram) with available spectroscopic information. Vertical dashed lines represent the redshift limits where the LBGs are expected to be located according to their selection criterion, $1.5 \lesssim z \lesssim 2.5$. For comparison, we show with a green histogram the distribution of the spectroscopic redshifts for DRGs. \emph{Right}: Accuracy of the photometric redshift determination for our LBGs (blue symbols) and UV-selected galaxies (orange symbols) at $1.5 \lesssim z \lesssim 2.5$ that have available spectroscopic redshifts. We define the accuracy of the photometric redshifts as $\sigma_{\rm zphot} = (z_{\rm phot}-z_{\rm spec})/(1+z_{\rm spec})$. Horizontal solid line indicates where photometric and spectroscopic redshifts would agree. Horizontal dashed lines represent redshift accuracy values of $\sigma_{\rm zphot} = \pm 0.2$. On the right side of the plot we include the distribution of the $\sigma_{\rm zphot}$ values for both the LBGs (blue shaded histogram) and the general population of UV-selected galaxies at $1.5 \lesssim z \lesssim 2.5$ (orange histogram). In both panels, histograms have been normalized to their maxima in order to clarify the representation.
              }
\label{zphot_zspec}
\end{figure*}

\subsection{Photometric redshifts}\label{zphot}

One of the goals of an SED-fitting process is the calculation of photometric redshifts. When spectroscopic redshifts are not available, accurate values of the photometric redshifts are required to have reliable estimates of the SED-derived properties. We will use high-quality photometric redshifts derived with broad and medium band observations for the galaxies with that information available (76\% of the galaxies in GOODS-S and 92\% in COSMOS). The photometric redshifts, taken from \citep{Ilbert2013} and \cite{Cardamone2010}, are used to redo the SED fits with our adopted photometry. For the remaining sources we use our own estimations obtained with the SED-fitting procedure explained in the previous Section.

We have checked whether the selection techniques we are employing for selecting SF galaxies in the redshift desert truly isolate sources within $1.5 \leq z \leq 2.5$. The left panel of Figure \ref{zphot_zspec} shows the distribution of spectroscopic redshift of those LBGs with available spectroscopic information (blue shaded histogram) from the catalogues of \cite{Lilly2007}, \cite{Barger2008}, \cite{Popesso2009} and \cite{Balestra2010}. It can be seen, as expected by the filters used, that the drop-out selection criterion employed in this work tends to segregate galaxies at $1.5 \lesssim z \lesssim 2.5$,  the percentage of contaminants, i.e.\ galaxies a different redshift range, being very low. We also represent the distribution of the spectroscopic redshift of the $BzK$ galaxies (orange histogram) and the general population of UV-selected galaxies (grey histogram) with available spectroscopic information. It can be seen that both selection criteria also give good results for the redshift of the sources, and the number of interlopers is very low. This indicates that the selection criteria employed in this work truly segregate galaxies at $1.5 \lesssim z \lesssim 2.5$.

We also plot in the left panel of Figure \ref{zphot_zspec} the distribution of spectroscopic redshift of the DRGs selected through $J-K > 2.3$ (Vega) with the available optical spectrum. It can be seen that the redshift distribution of these galaxies span a wider redshift range than those for LBGs, $sBzK$ and UV-selected galaxies. This also happens to BM/BX galaxies. Their redshift distributions are not similar to those for LBGs and $BzK$ and UV-selected galaxies. Due to this difference in the redshift distribution, the physical properties of DRGs and BM/BX galaxies should not be directly compared to those of LBG and $sBzK$ galaxies. Therefore, DRGs and BM/BX galaxies will not be considered in the following sections. The selected pBzK galaxies cannot be compared to our LBGs and sBzK galaxies either since, as  was mention in Section 3.2, their SFR and foreseeable evolution are completely different although they are located in a similar redshift range.

The second step in the analysis of our results should be to check whether the SED-derived photometric redshifts are accurate enough to enable a further study of the physical properties of our selected galaxies. To this purpose, we compare the SED-derived photometric redshifts with the spectroscopic ones for those galaxies with available spectroscopic information in the three fields considered (see right panel of Figure \ref{zphot_zspec}). If we define the accuracy of the photometric redshift as $\sigma_z = (z_{\rm phot}-z_{\rm spec})/(1+z_{\rm spec})$ we find that most of the LBGs have values $\sigma_z < 0.2$. The average value of the photometric accuracy is 0.09 with a standard deviation of $\Delta \sigma_z = 0.35$. Therefore, the photometric redshifts might be considered accurate enough for the purpose of this work. In a similar redshift range to the one studied in this work, \cite{Haberzettl2012} and \cite{Kurczynski2012} obtained $z_{\rm phot}$ accuracies similar to our determinations.

\subsection{Stellar populations of LBGs at $1.5 \lesssim z \lesssim 2.5$}\label{sp}

\begin{figure}
\centering
\includegraphics[width=0.22\textwidth]{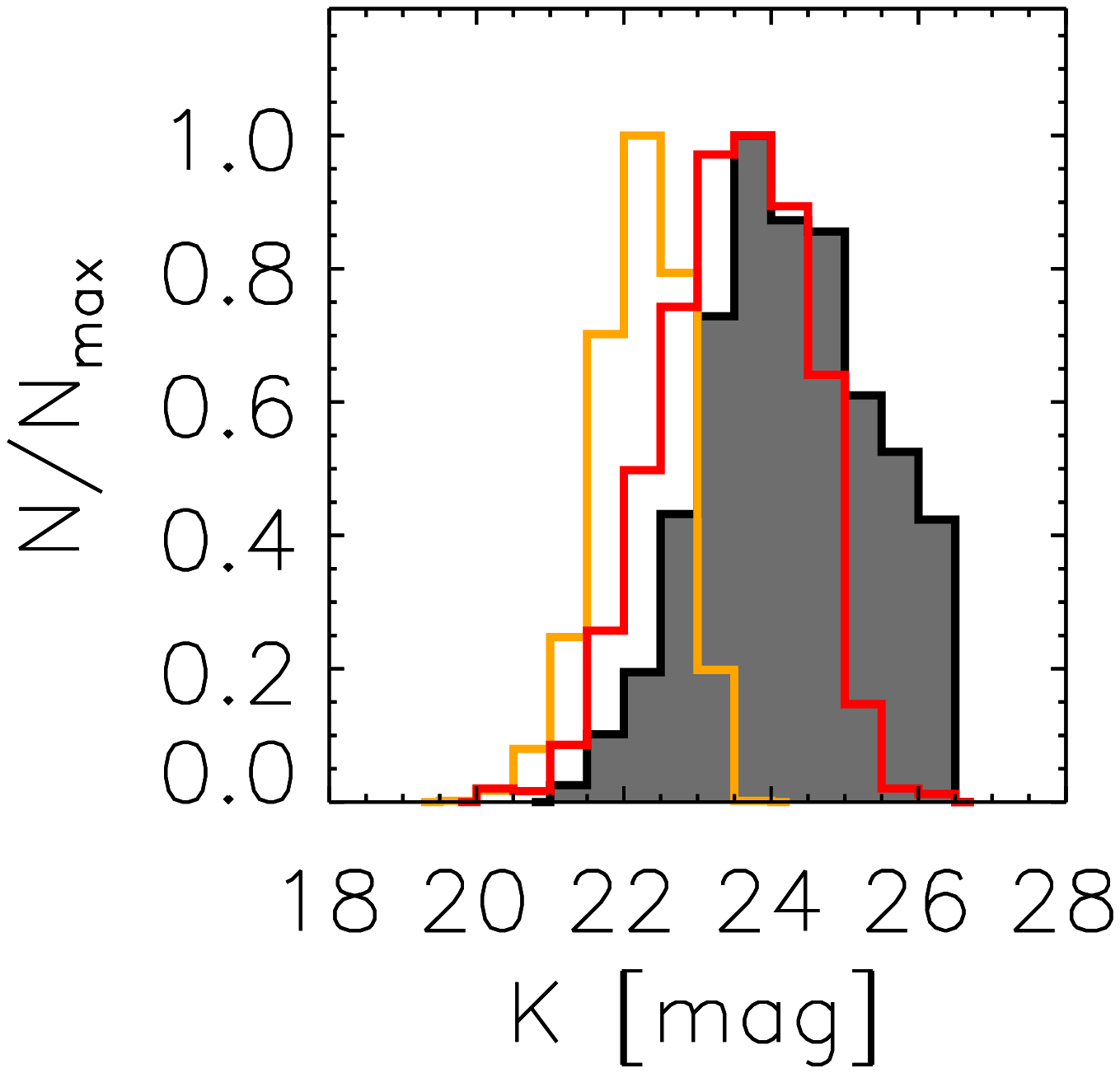}
\includegraphics[width=0.22\textwidth]{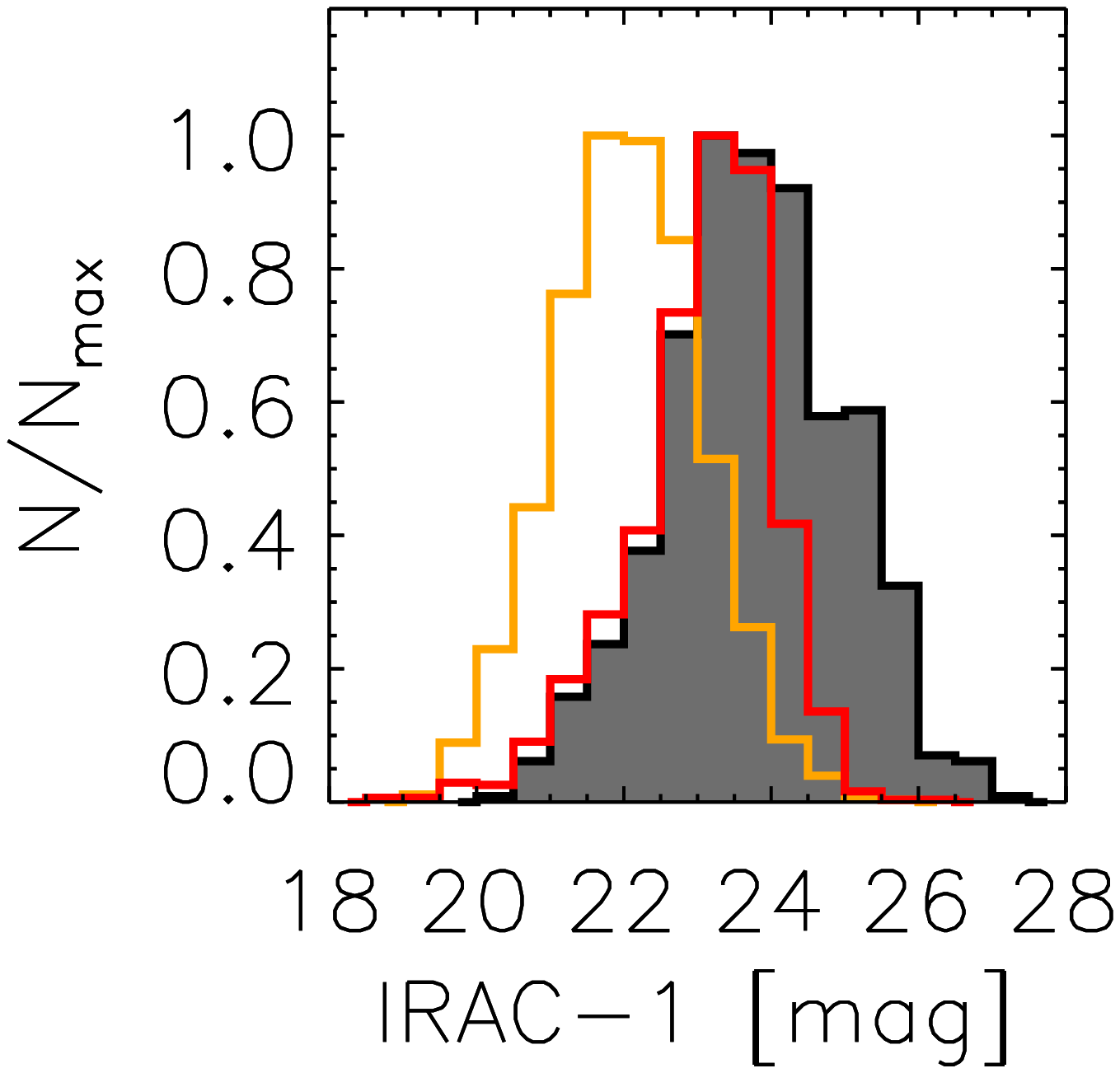}
\caption{Distribution of the $K$ (\emph{left}) and IRAC-3.6$\mu$m (\emph{right}) magnitudes for the LBGs located in the GOODS-S (grey shaded histograms), GOODS-N (red histograms) and COSMOS (orange histograms) fields. Histograms have been normalized to their maxima in order to clarify the representations.
              }
\label{kirac}
\end{figure}

As indicated in Section \ref{selec}, we have compiled photometric information in three different cosmological fields, GOODS-N, GOODS-S and COSMOS. Each field has its own photometric observations and, consequently, the depth of the observations in different wavelengths is not the same. Furthermore, the photometric coverage is not the same in all the fields. As an example for the $K$ and IRAC-3.6$\mu$m bands, we represent in Figure \ref{kirac} the observed magnitudes in those bands of the LBGs selected in the GOODS-S (grey shaded histogram), GOODS-N (red histogram) and COSMOS (orange histogram) fields. It can be seen that the near-IR and mid-IR observations are much deeper in the GOODS fields than in the COSMOS field. Furthermore, among GOODS-S and GOODS-N, the observations are deeper in the GOODS-S field and also provide a more homogeneous photometric coverage. If we wanted to gather the properties of the LBGs in the three fields at the same time, we would have to limit the observed magnitude to the shallowest survey. However, this would imply the loss of a large number of sources along with a strong bias in the near- and mid-IR as a consequence of the shallow data in the COSMOS field. Therefore, we decided to focus in this section only in the GOODS-S field, where the observations are deeper than in the other fields and the photometric coverage is perfectly suitable for the analysis of the physical properties of SF galaxies at $1.5 \lesssim z \lesssim 2.5$. The SED-derived properties of the LBGs in GOODS-N and COSMOS fields will be also presented for illustrating the differences in the SED-derived physical properties that the luminosity bias introduces.

\begin{figure*}
\centering
\includegraphics[width=0.2\textwidth]{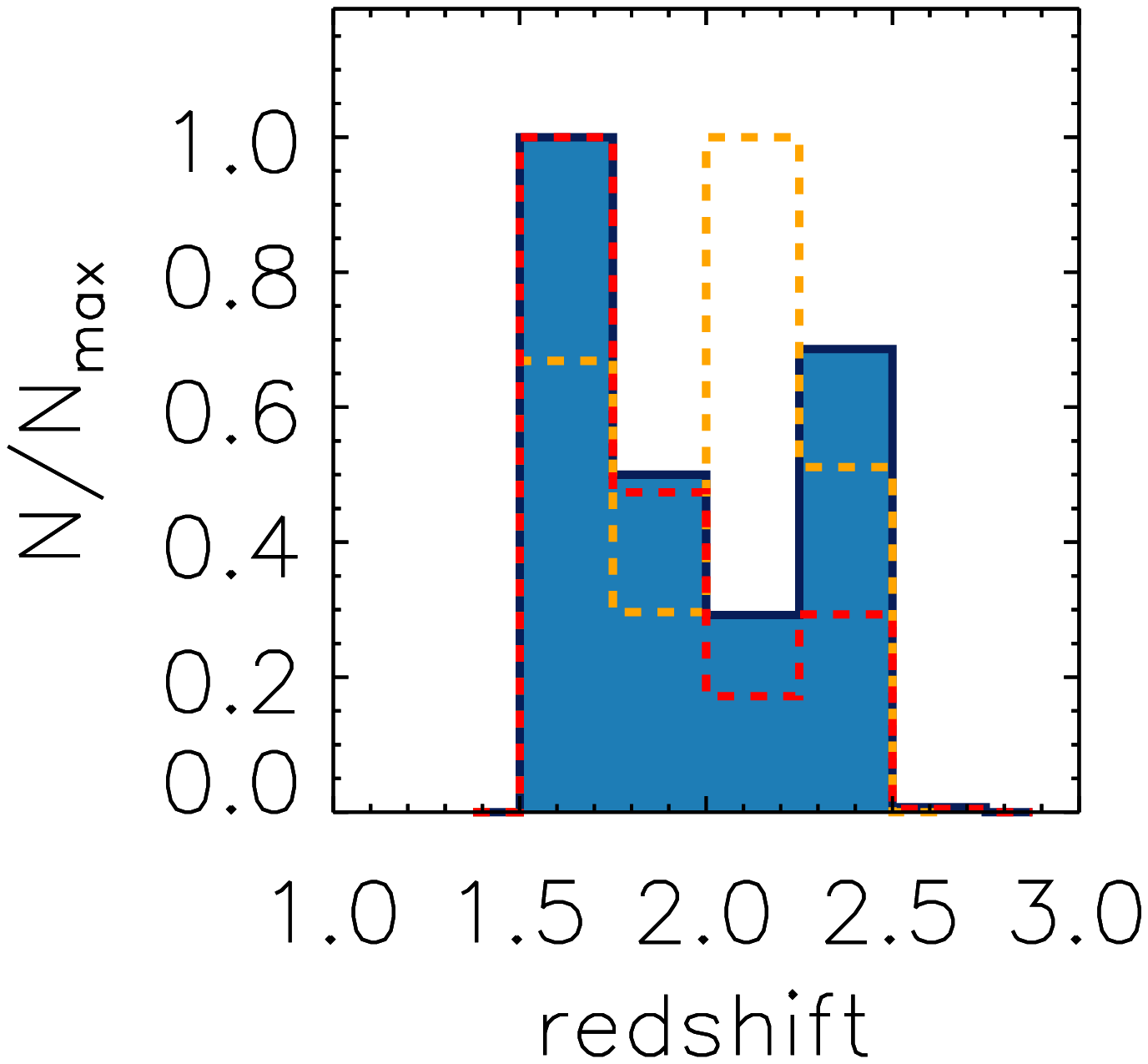}
\includegraphics[width=0.2\textwidth]{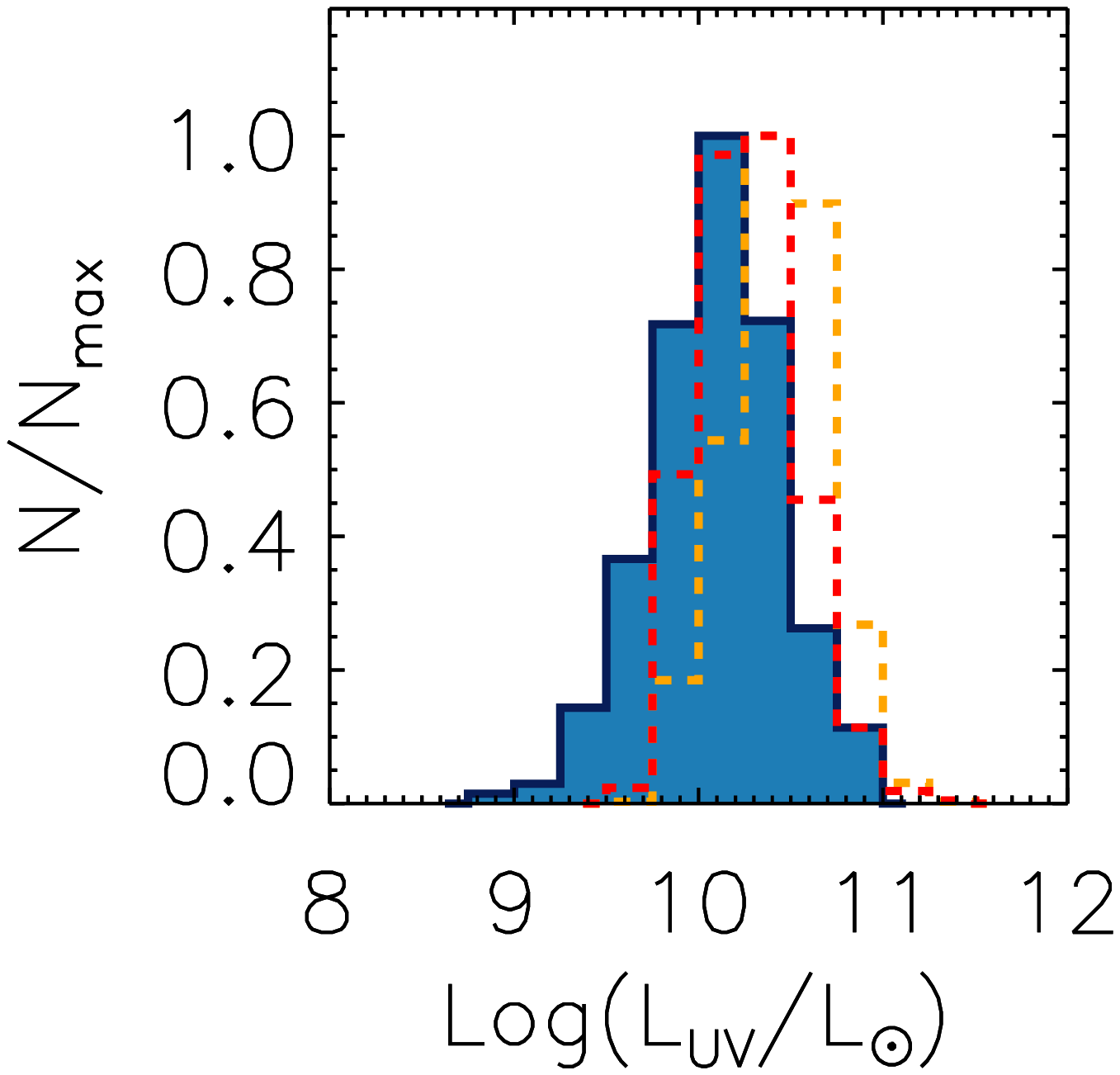}
\includegraphics[width=0.2\textwidth]{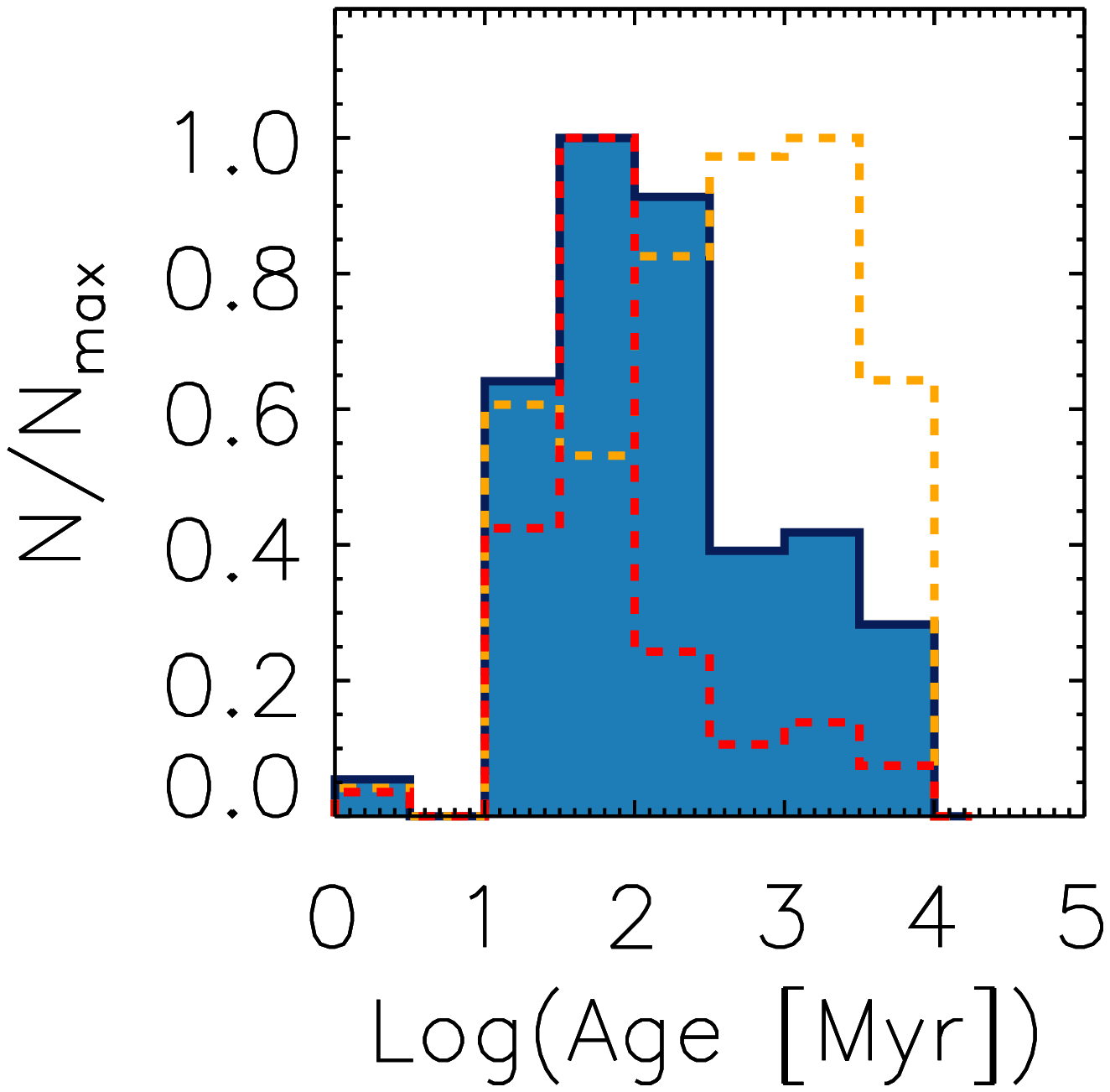} \\
\includegraphics[width=0.2\textwidth]{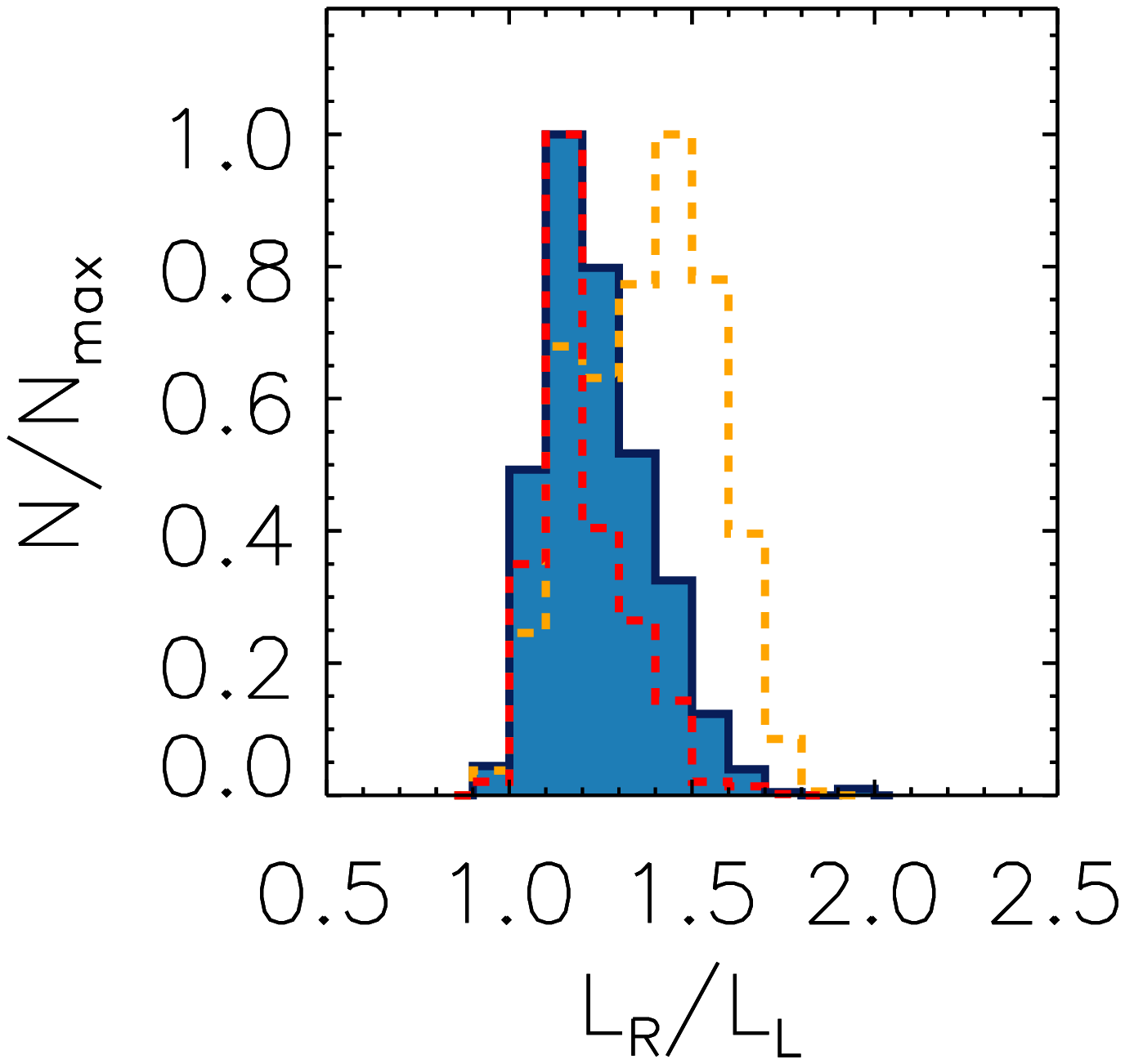} 
\includegraphics[width=0.2\textwidth]{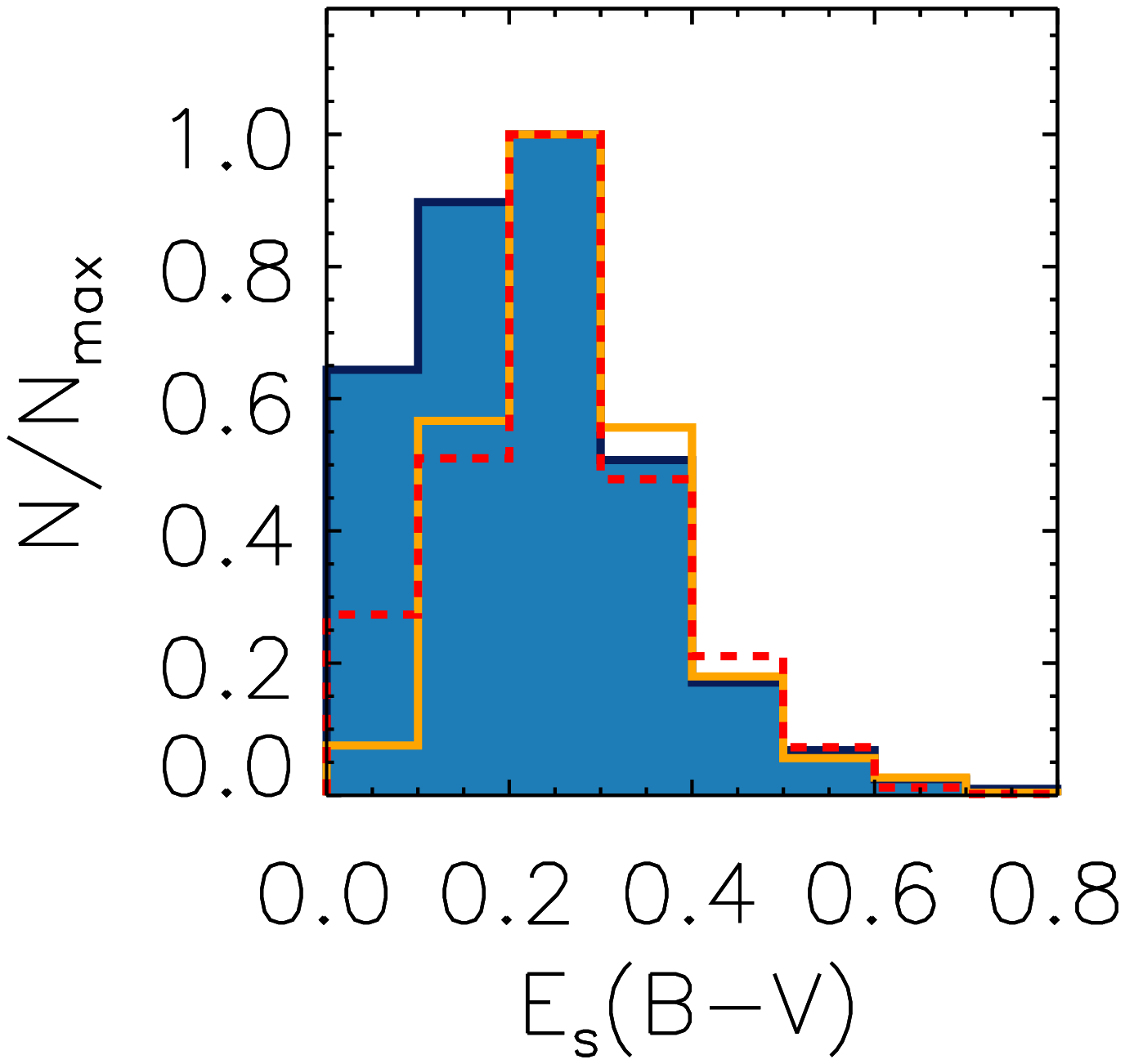}
\includegraphics[width=0.2\textwidth]{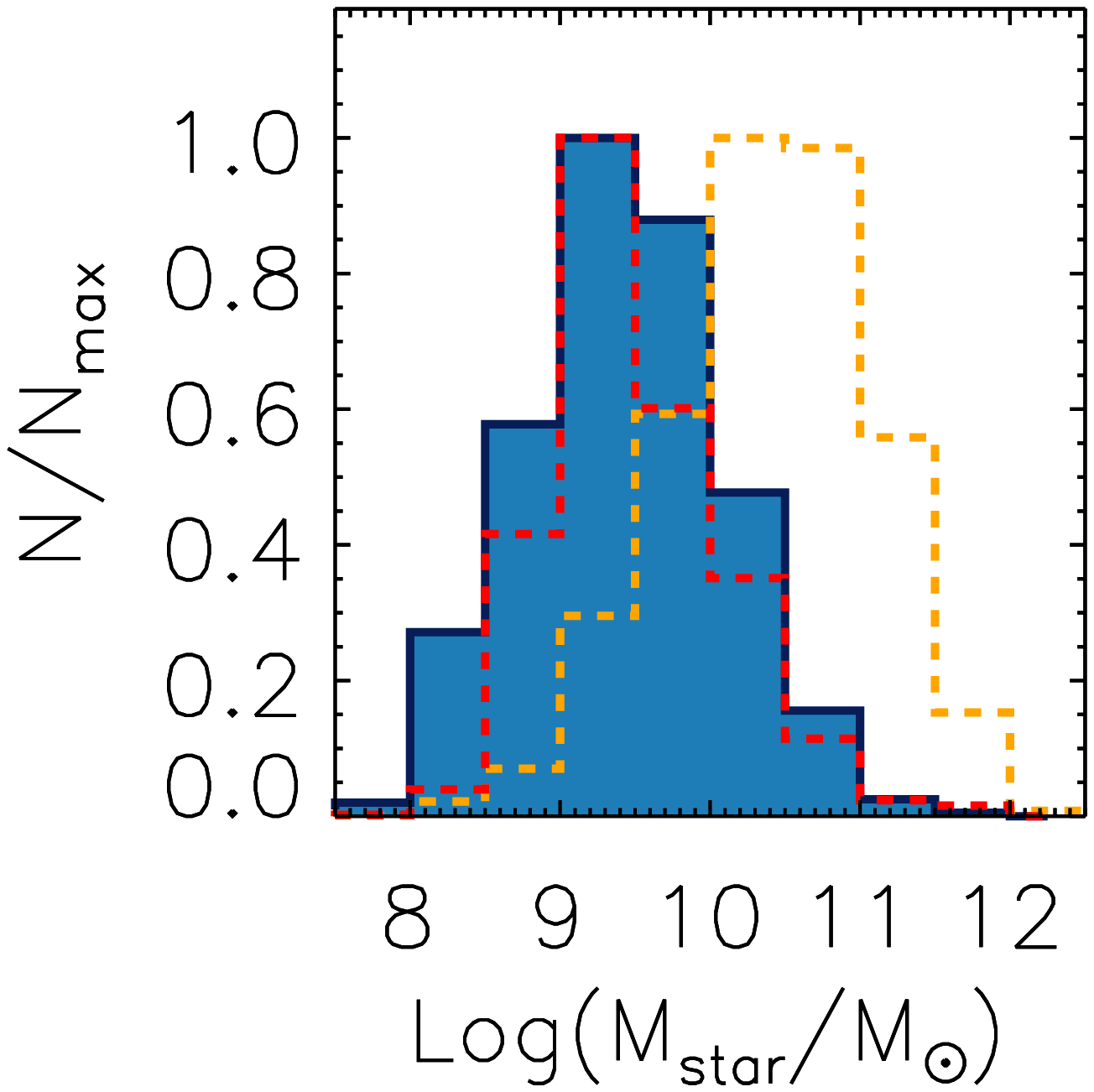} \\
\includegraphics[width=0.2\textwidth]{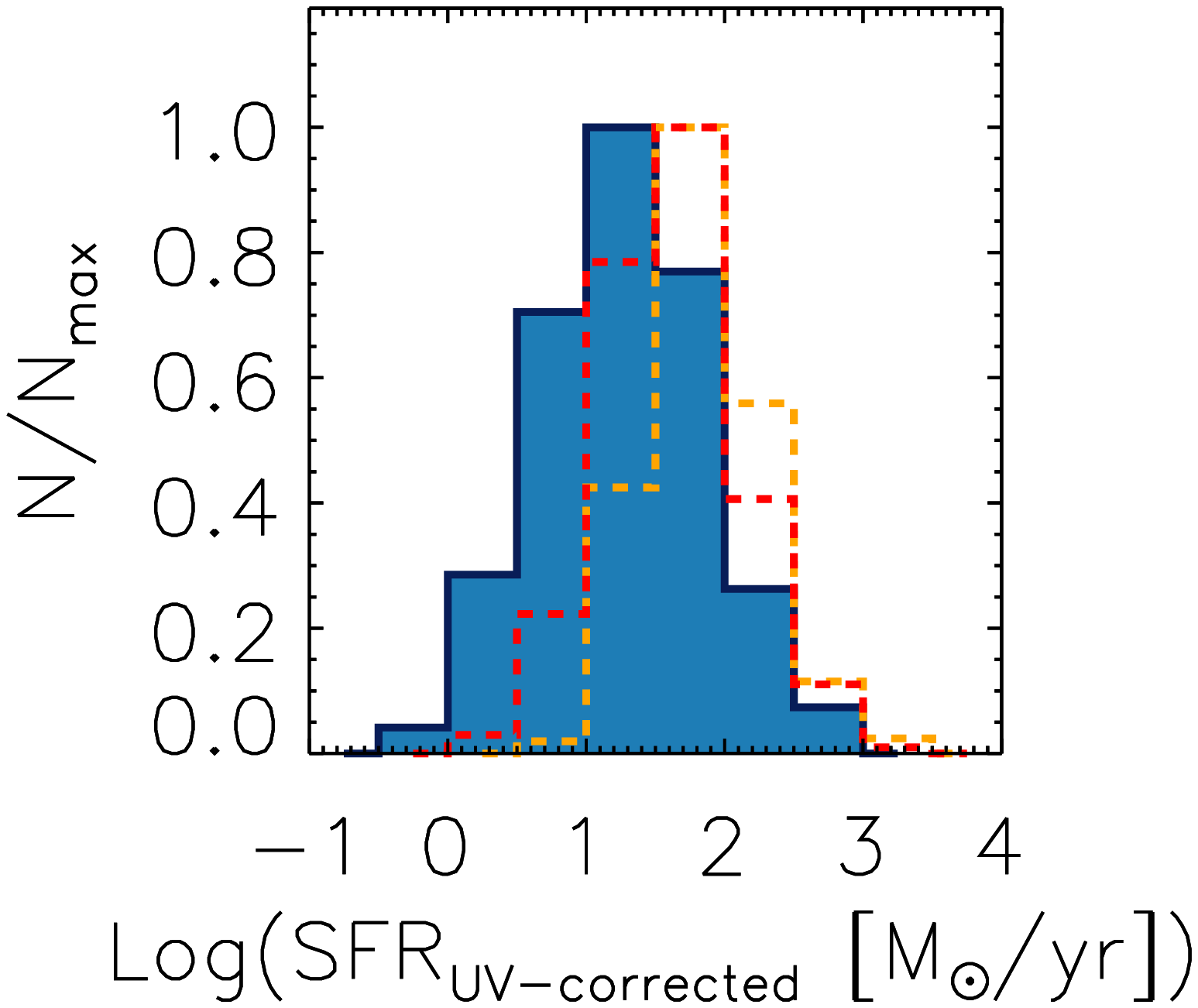}
\includegraphics[width=0.2\textwidth]{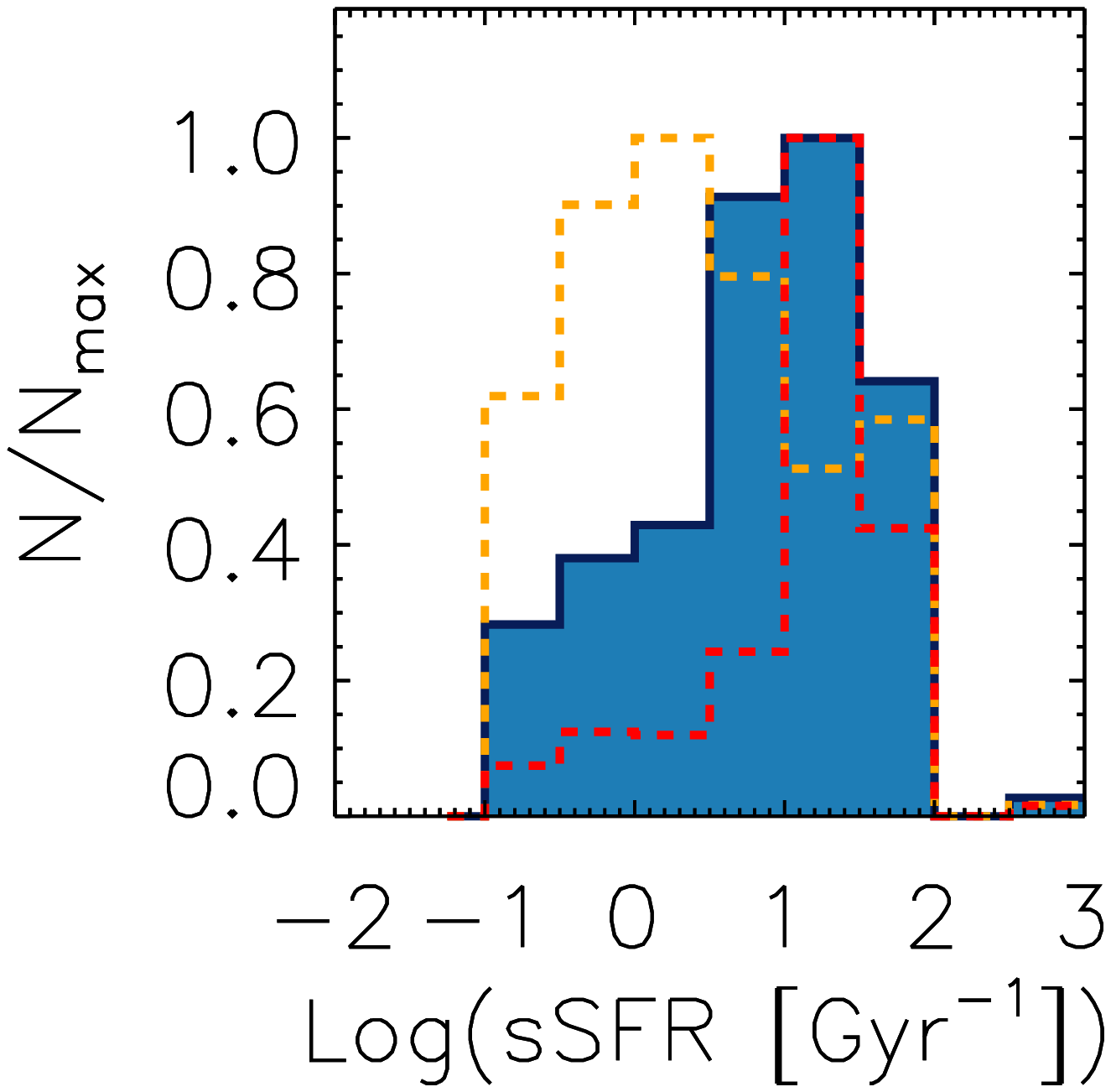}
\includegraphics[width=0.2\textwidth]{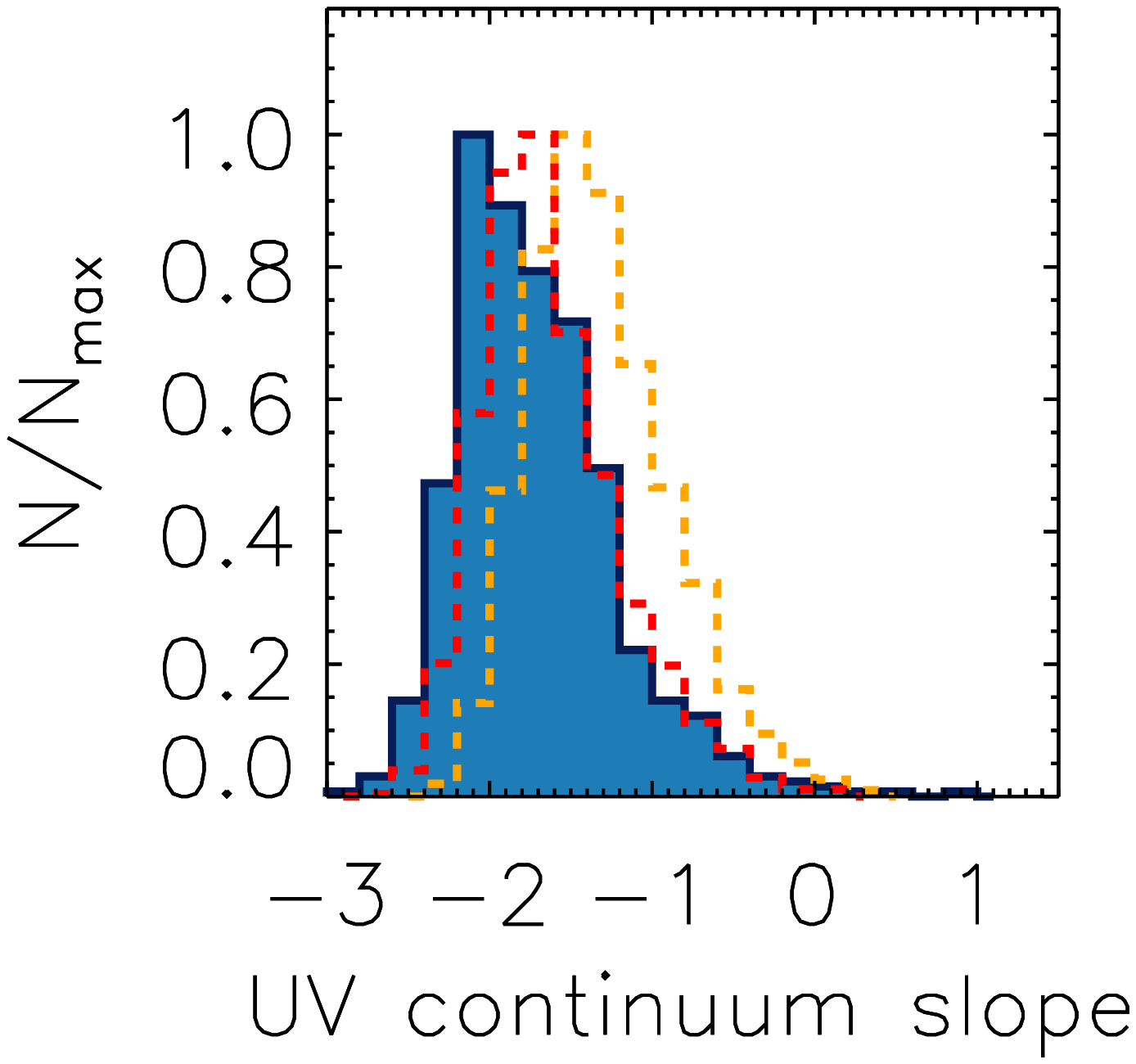}
\caption{Best-fitted photometric redshift and SED-derived rest-frame UV luminosity, age, amplitude of the Balmer break, dust attenuation as parameterized by the reddening in the stellar continuum $E_s(B-V)$, stellar mass, dust-corrected total SFR, specific SFR, and UV continuum slope for our selected LBGs at $1.5 \lesssim z \lesssim 2.5$ located in the GOODS-S region (blue shaded histograms). The dashed red and orange distributions refer to LBGs at $1.5 \lesssim z \lesssim 2.5$ located in the GOODS-N and COSMOS fields, respectively. The BC03 templates used to obtain those properties have been built by assuming a constant SFR, Salpeter IMF and fixed metallicity $Z=0.2Z_\odot$. Histograms have been normalized to their maxima with the aim of clarifying the representations.
              }
\label{properties}
\end{figure*}

Figure \ref{properties} shows the distributions of the best-fitted SED-derived photometric redshifts, rest-frame UV luminosity, age, amplitude of the Balmer break, dust attenuation, stellar mass, dust-corrected total SFR, specific SFR (sSFR) and UV continuum slope of our LBGs located in the GOODS-S field (blue shaded histograms), GOODS-N (red dashed histograms) and COSMOS field (orange dashed histograms). Focusing on the LBGs in the GOODS-S field, their SEDs indicate that they are young galaxies with a median age of 111 Myr, have intermediate dust attenuation with a median value of $E_s(B-V) = 0.2$ and their median stellar mass is $\log{\left( M_*/M_\odot \right)} = 9.43$. The median amplitude of the Balmer break is $L_R/L_L = 1.22$ and has a median UV continuum slope of -1.78. When correcting for dust attenuation, we find that the dust-corrected total SFR of our LBGs have a median value of 18.1 $M_\odot\,{\rm yr}^{-1}$, implying a median  sSFR of 9.0 ${\rm Gyr}^{-1}$. These values have been obtained with the best-fitted values for each individual galaxy and are compiled in Table \ref{SED_properties}. Comparing the distributions corresponding to LBGs located in different fields, it can be concluded that, as a consequence of the shallower near- and mid-IR data in COSMOS, the LBGs selected in that field are more massive and dustier and have more intense Balmer break, higher dust-corrected total SFR and a redder UV continuum than those in the GOODS fields. This implies that the depth of the photometric observations has a great impact on the derivation of SED-derived physical properties, and one should therefore proceed with care when comparing the properties of galaxies obtained with different photometric observations.

\subsection{Comparison between Lyman break, $sBzK$, and UV-selected galaxies}\label{comparison}

\begin{figure*}
\centering
\includegraphics[width=0.2\textwidth]{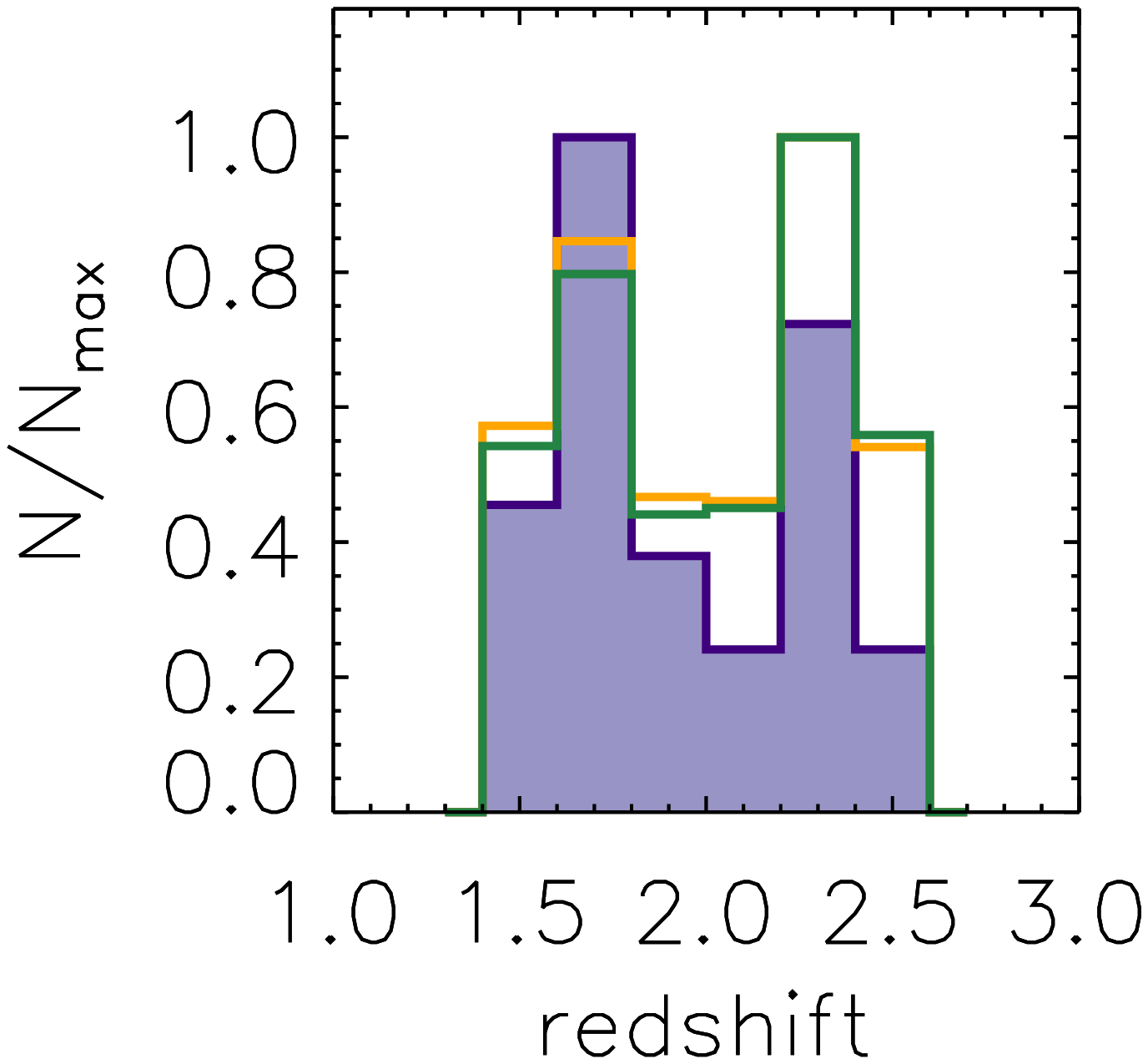}
\includegraphics[width=0.2\textwidth]{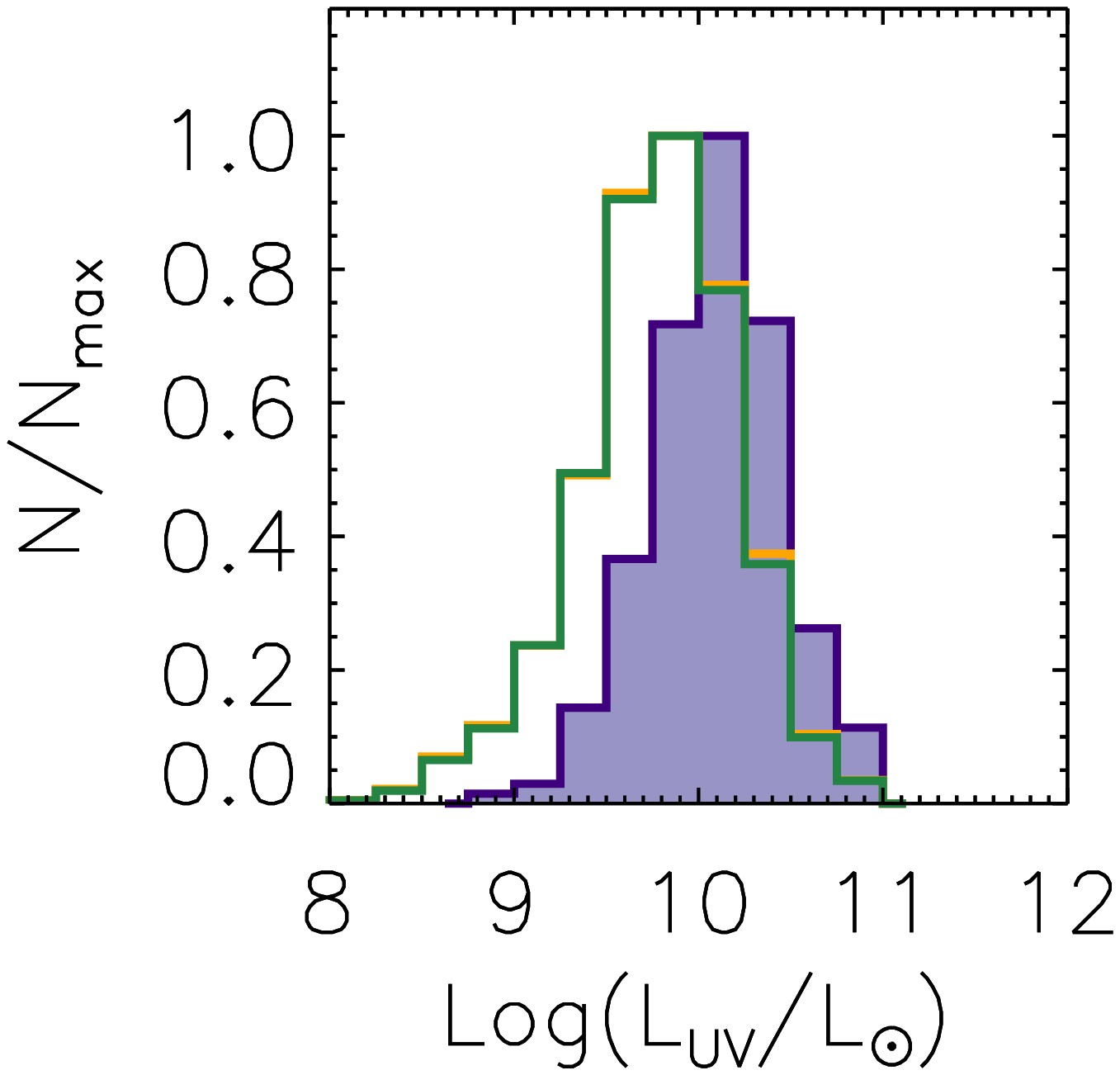}
\includegraphics[width=0.2\textwidth]{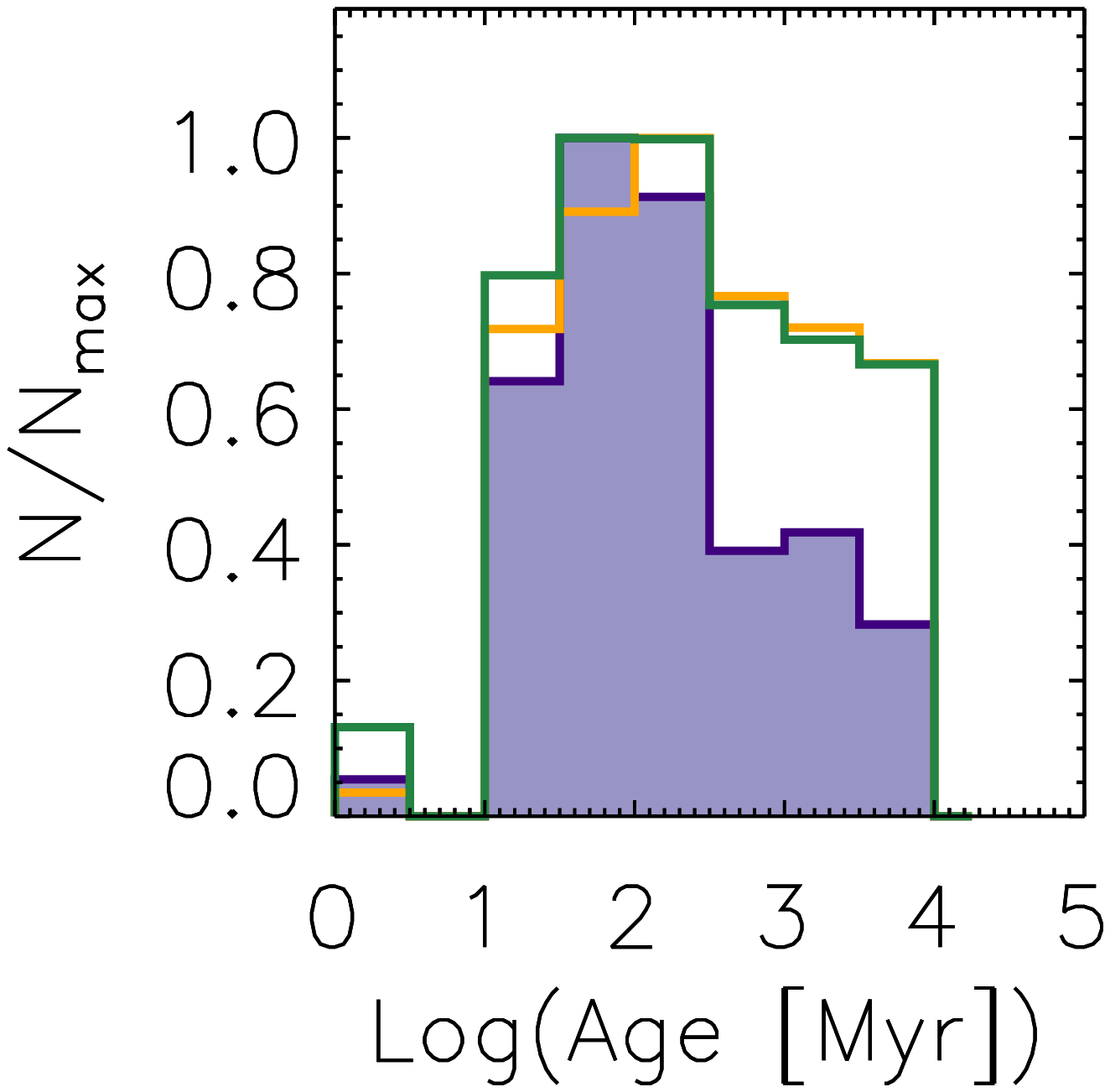} \\
\includegraphics[width=0.2\textwidth]{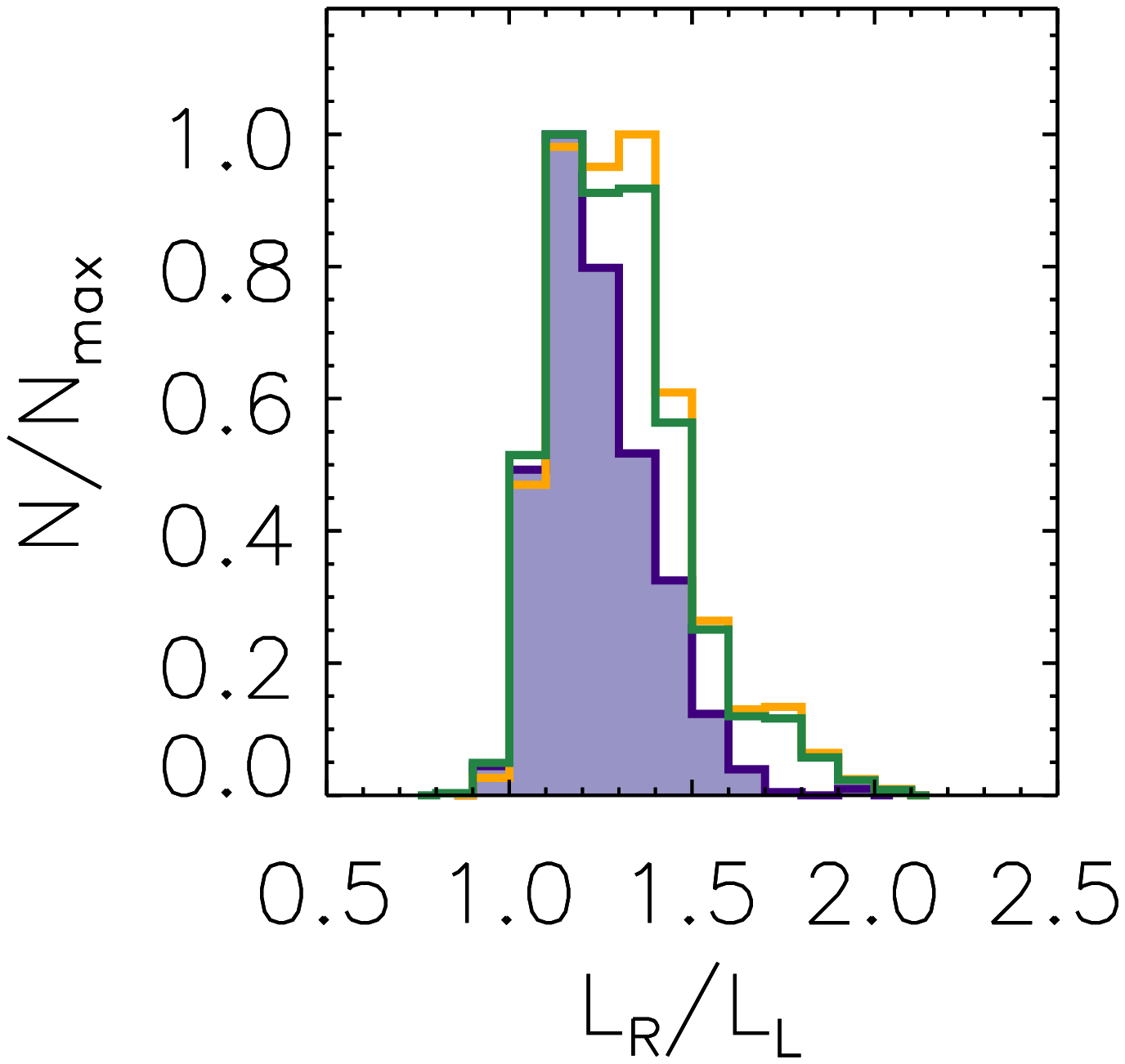}
\includegraphics[width=0.2\textwidth]{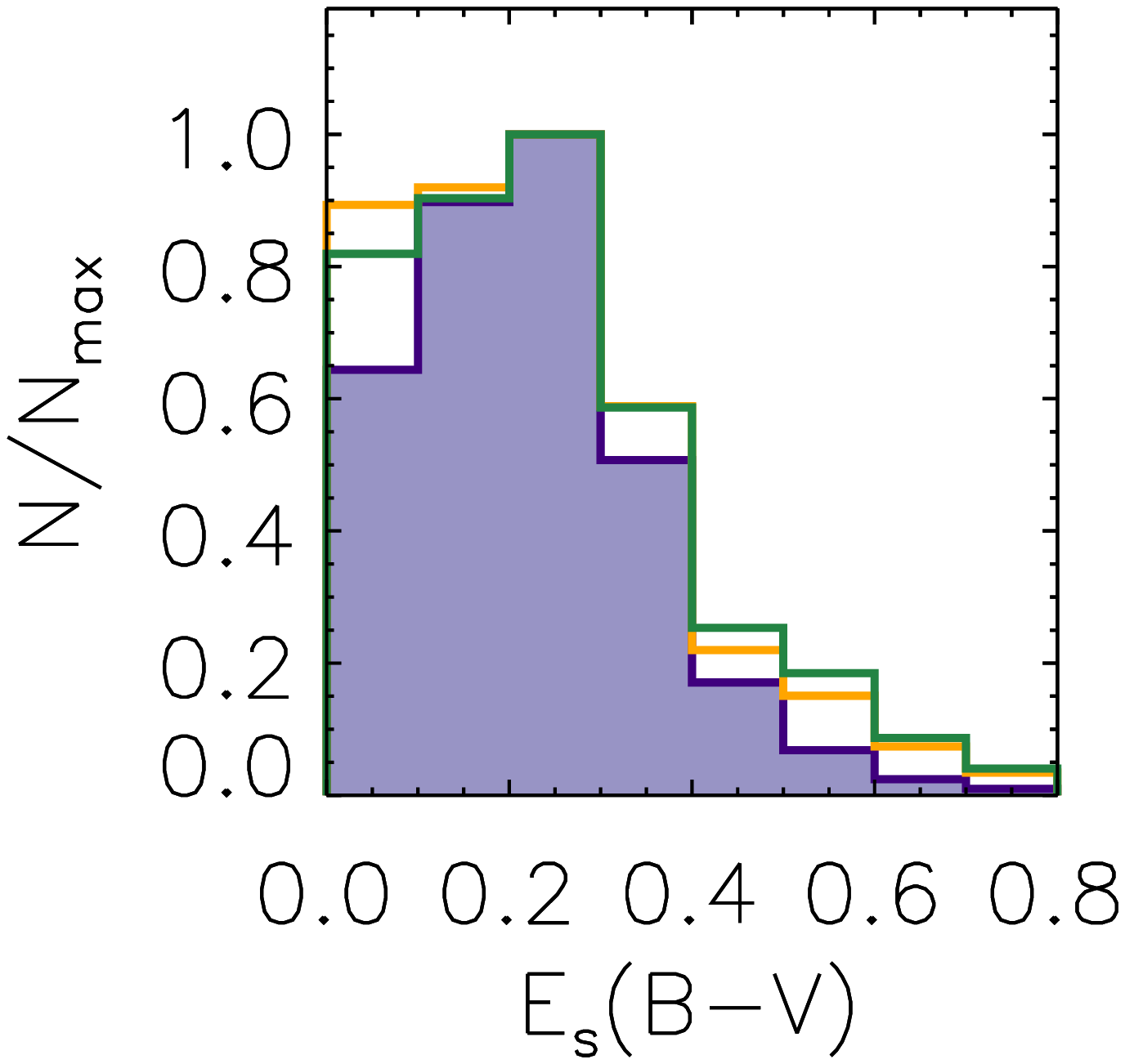}
\includegraphics[width=0.2\textwidth]{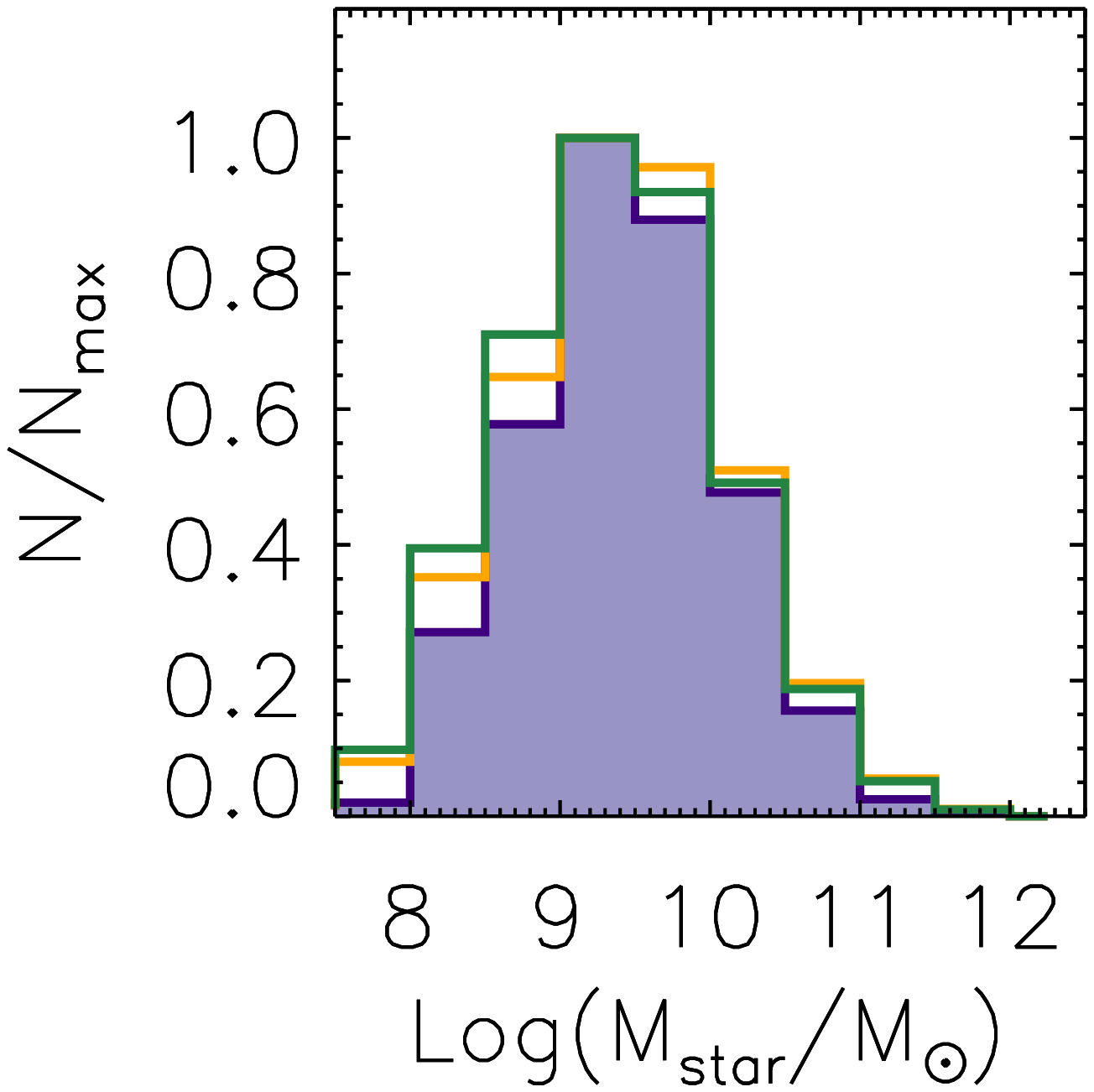}\\
\includegraphics[width=0.2\textwidth]{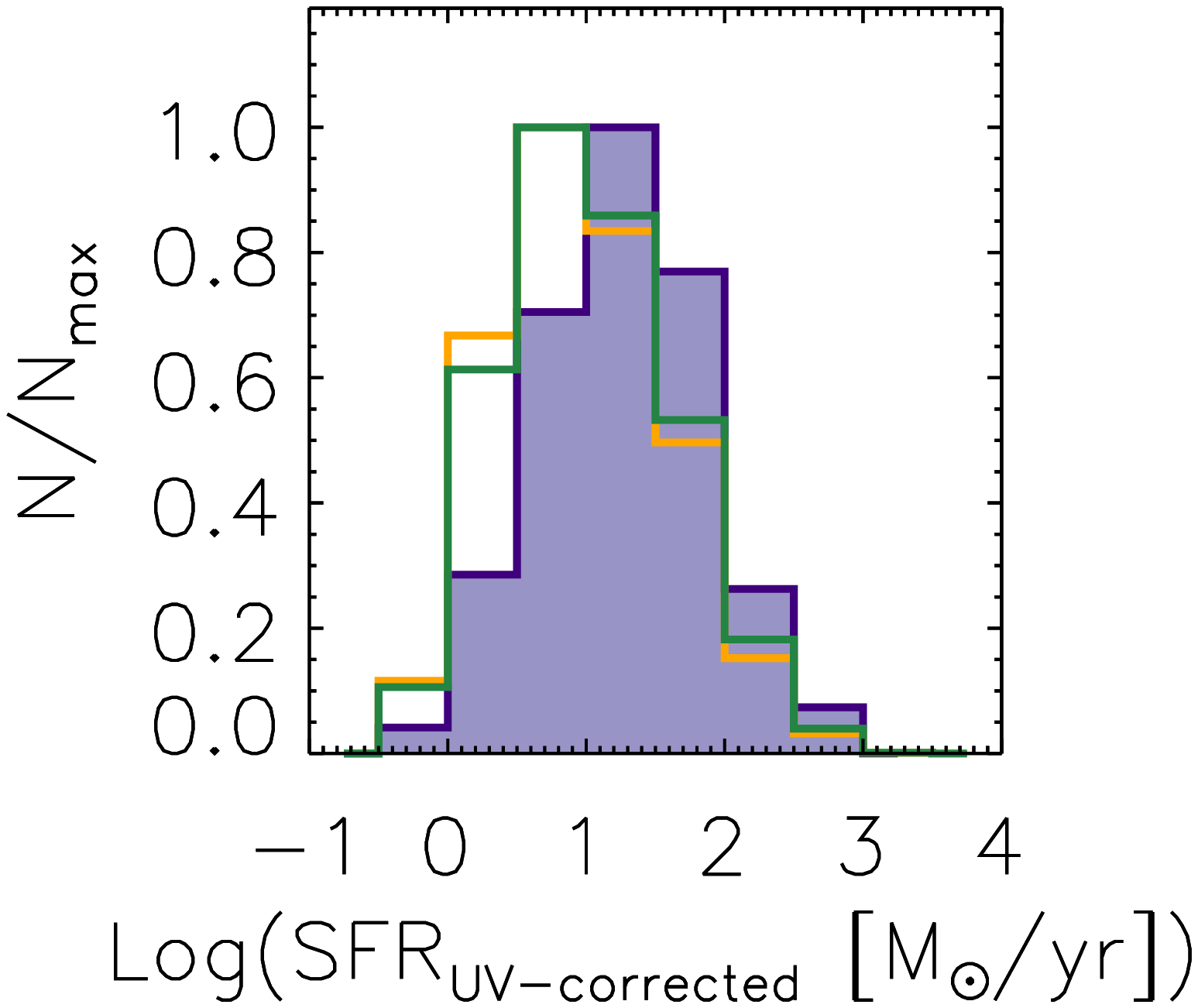}
\includegraphics[width=0.2\textwidth]{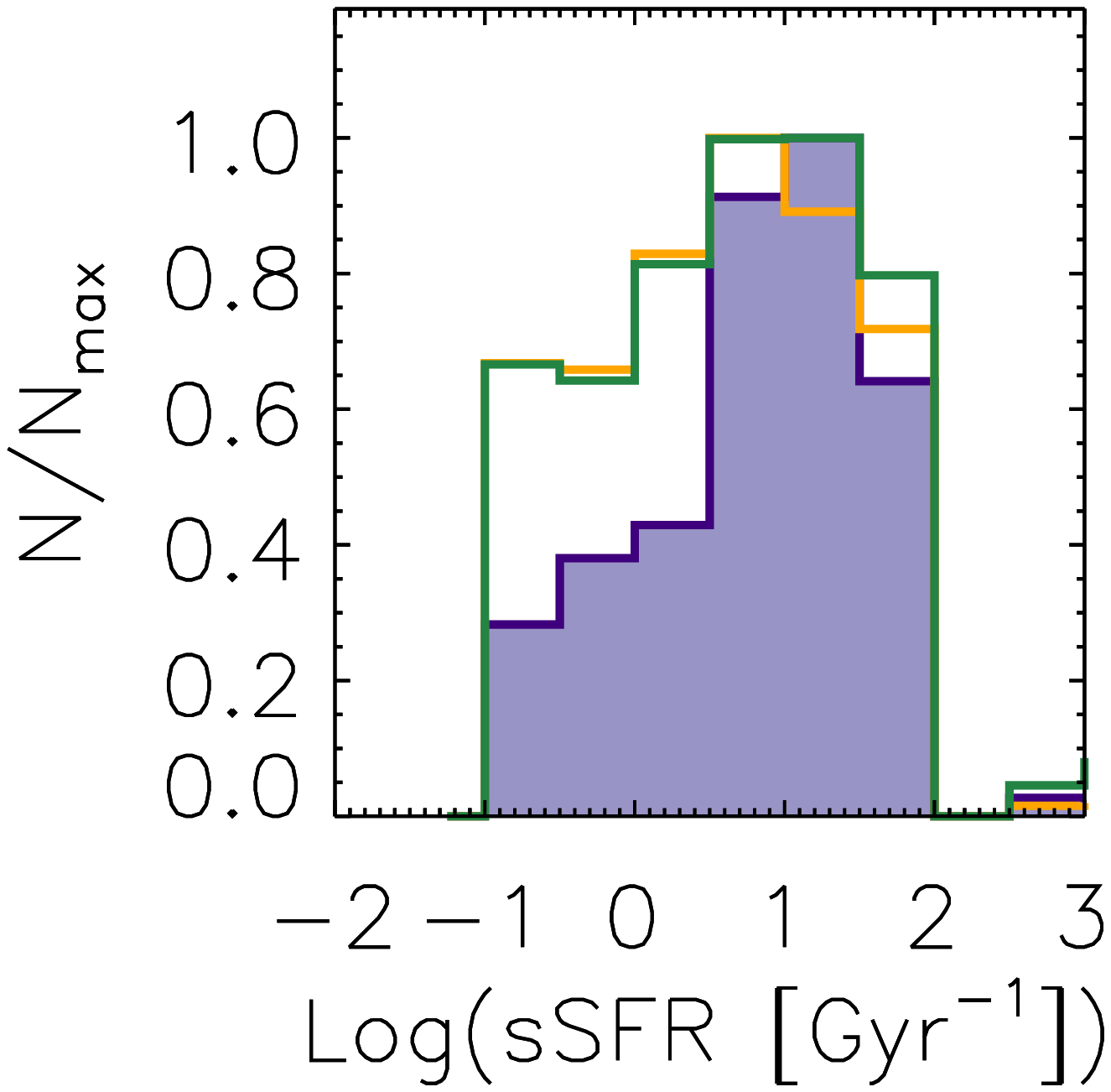}
\includegraphics[width=0.2\textwidth]{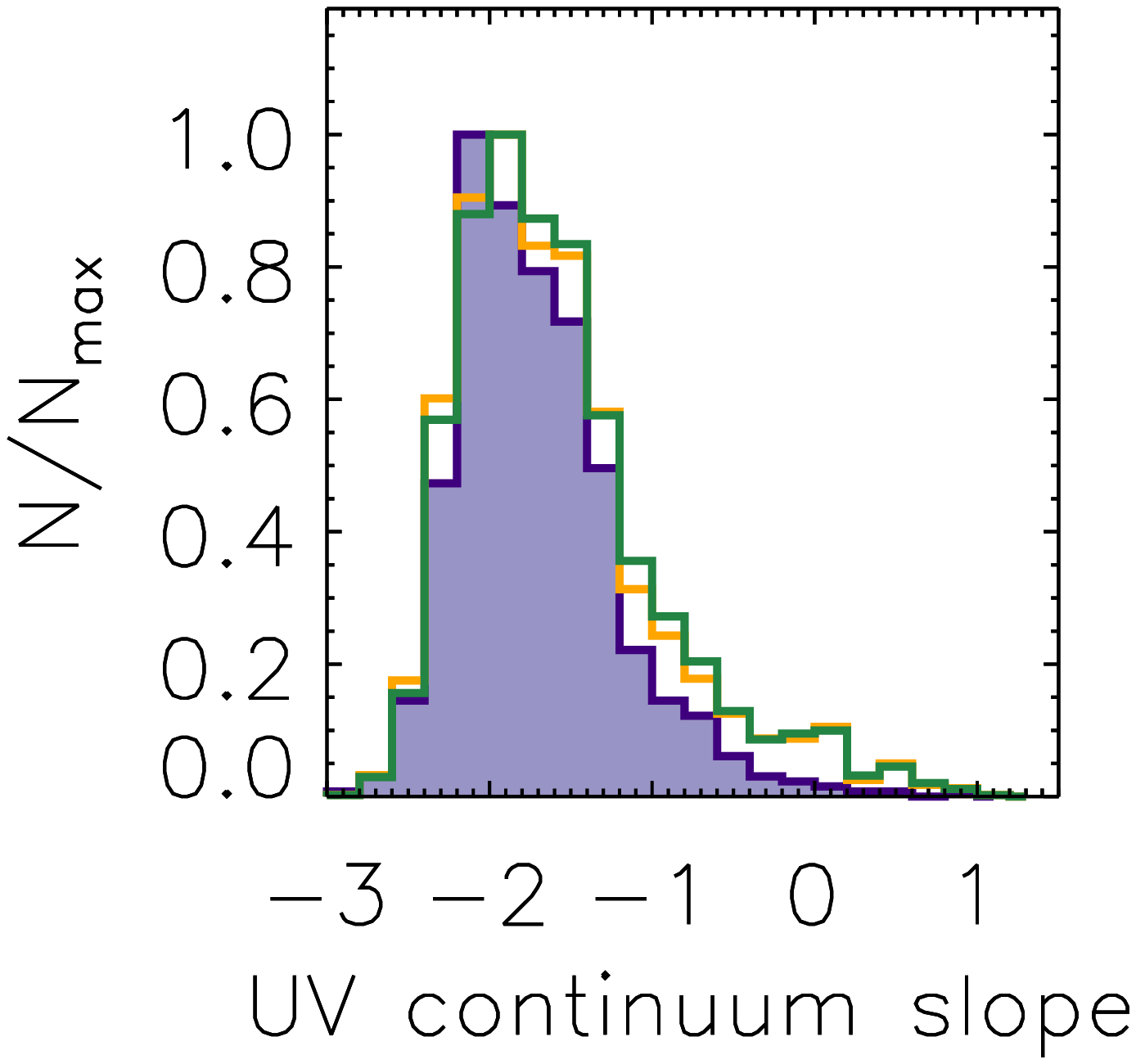}
\caption{Best-fitted photometric redshift and SED-derived rest-frame UV luminosity, age, amplitude of the Balmer Break, dust attenuation, stellar mass, dust-corrected total SFR, specific SFR and UV continuum slope for our LBGs (purple shaded histograms), sBzK galaxies (orange histograms) and UV-selected galaxies (green histograms) located in the GOODS-S field. The BC03 templates used to obtain those properties have been built by assuming a constant SFR, Salpeter IMF and fixed metallicity $Z=0.2Z_\odot$. Histograms have been normalized to their maxima with the aim of clarifying the representations.
              }
\label{BzK}
\end{figure*}

Figure \ref{BzK} shows the distributions of the best-fitted SED-derived photometric redshift, rest-frame UV luminosity, age, amplitude of the Balmer break, dust attenuation, stellar mass, dust-corrected total SFR,  sSFR and UV continuum slope for our LBGs, $sBzK$ and UV-selected galaxies at $1.5 \lesssim z \lesssim 2.5$ in the GOODS-S field. The median values of the distributions of these parameters for the three populations of SF galaxies are shown in Table \ref{SED_properties} along with the results of a Kolmogorov--Smirnov (K-S) test for each parameters and for different pairs of kinds of galaxies. As a consequence of their different selection criteria (the $sBzK$ criterion does not involve rest-frame UV color), LBGs tend to be brighter and slightly bluer in the UV than $sBzK$ galaxies. LBGs also have a less prominent Balmer break (compatible with their younger ages) and higher dust-corrected total SFR and sSFR. However, we do not find significant difference between their median stellar mass or dust attenuation. These findings are supported by the results of K-S tests. Additionally, the distributions for LBGs and $sBzK$ do not span completely disjoint sets of values for each parameter, but the values for LBGs are contained within the distributions for $sBzK$ galaxies. This indicates both that the SED-derived physical properties of $sBzK$ galaxies are more diverse than those for LBGs, and that LBGs and $sBzK$ are not completely different populations, there being an overlap between the two samples.

\begin{figure}
\centering
\includegraphics[width=0.45\textwidth]{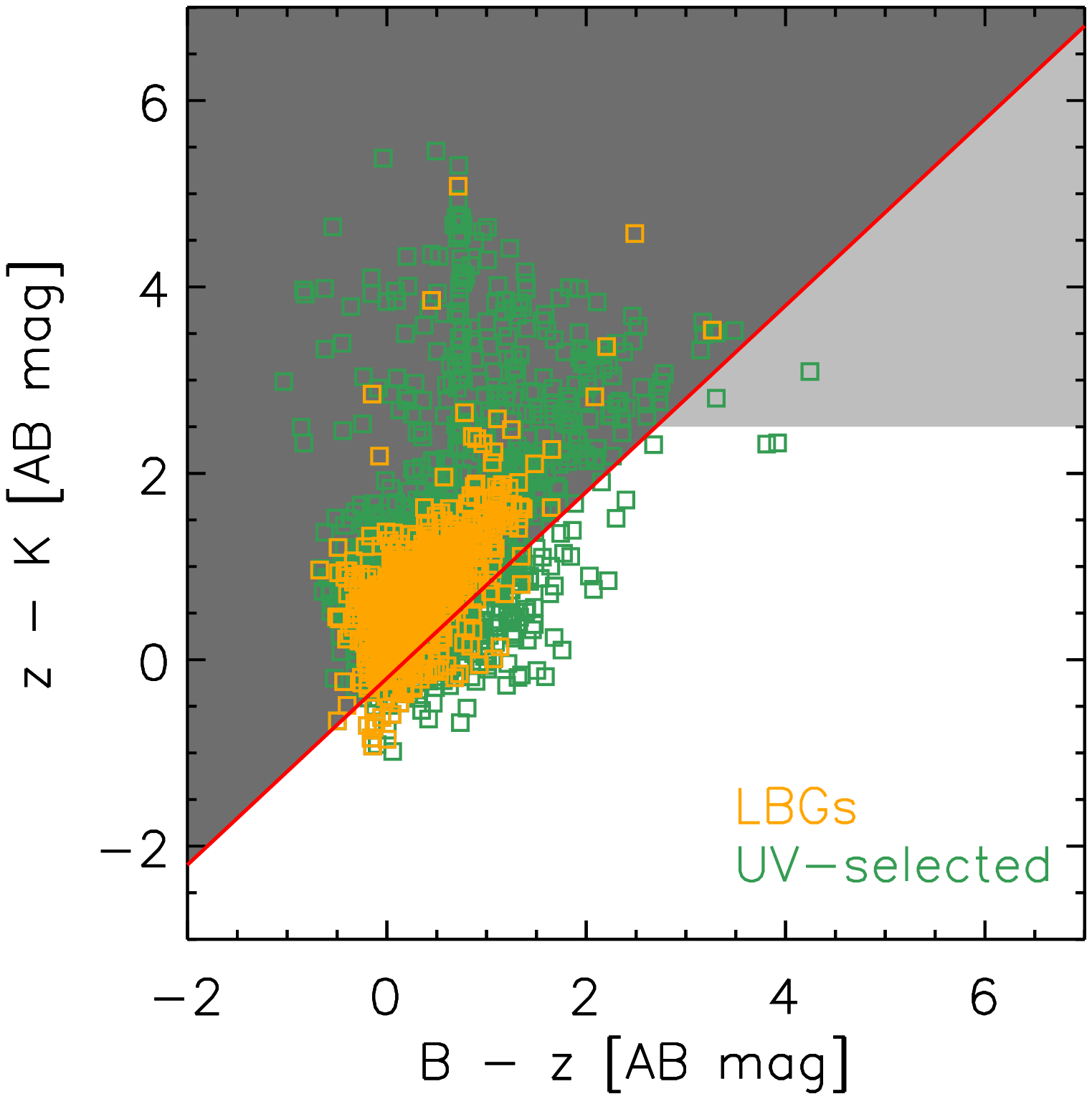}
\caption{Colour--colour diagram employed in \citet{Daddi2004} to look for $BzK$ galaxies. The dark grey shaded zone and the light grey shaded zone represent the areas where $sBzK$ and $pBzK$, respectively, are located according to the selection criteria defined in \citet{Daddi2004}. LBGs and UV-selected galaxies are indicated with orange and green open squares, respectively.
}
\label{bzk_color}
\end{figure}

As can also be seen in Figure \ref{BzK}, LBGs also tend to be UV-brighter, younger and have higher dust-corrected total SFR and sSFR than a general population of UV-selected galaxies at the same redshift. However, there is no significant difference in the SED-derived dust attenuation or stellar mass. Therefore, at $1.5 \lesssim z \lesssim 2.5$ the drop-out selection criterion involving NUV space-based data selects the UV-brightest galaxies at $z \sim 2$ which, at the same time, turn out to be younger, have slightly bluer UV continuum and higher dust-corrected total SFR and sSFR.

The SED-derived properties of $sBzK$ galaxies are very similar to those of the general population of UV-selected galaxies. The median values for $sBzK$ and UV-selected galaxies of the properties presented in Figure \ref{BzK} are represented in Table \ref{SED_properties}. It should be pointed out that we do not limit the $K$ magnitude of our $sBzK$ galaxies as done in previous studies \citep{Daddi2005,Kurczynski2012}. For example, \cite{Daddi2005} studied $sBzK$ galaxies at $z \sim 2$ that are limited to $K_{\rm Vega} < 20.5$ in the GOODS-N. The $B$-band and $z$-band photometry in their study is taken from \cite{Capak2004}, as is the present work in the same field. The only photometric difference is the $K$-band catalogue. They report that the average mass of their objects is $1\times 10^{11} M_{\odot}$ and have an average color of $B-z = 1.50$ that translates into a dust attenuation of $E_s(B-V)$ = 0.4 by using the \cite{Calzetti2000} law. These values are higher than the median for our selected $sBzK$ owing to their $K$ magnitude limitation. Actually, if we limit our $K_{s}$ magnitude to $K_{\rm Vega} < 20.5$, we obtain an average stellar mass of our $sBzK$ galaxies in the GOODS-N field of $6.3 \times 10^{10} M_{\odot}$ and an average dust attenuation of $E_s(B-V) = 0.4$, in agreement with the \cite{Daddi2005} results. The limitation in the $K_s$ magnitude results  the selected $sBzK$ galaxies being more massive, redder in the UV continuum (higher UV continuum slopes), having higher SED-derived dust attenuation and also a higher dust-corrected total SFR. However, there is no significant difference in age or specific SFR. Therefore, we include in the analysis galaxies that are typically less massive and with lower dust-corrected total SFR than those studied in \cite{Daddi2005} or \cite{Kurczynski2012} and are consequently expected to cover a wider region of, for example, the SFR--mass plane (see Section \ref{sfrmass}).





























In order to further explore the overlapping between LBGs, $sBzK$ and UV-selected galaxies, we show in Figure \ref{bzk_color} the colour--colour diagram employed in \cite{Daddi2004} to look for $BzK$ galaxies and the location in that diagram of our LBGs and UV-selected galaxies located in the GOODS-S field. We show with a dark grey shaded zone and with a light grey shaded zone the location of $sBzK$ and $pBzK$ galaxies, respectively. LBGs and UV-selected galaxies are represented with orange and green open squares, respectively. Among the 681 LBGs, 624 satisfy the $sBzK$ criterion. This represents about 90\% of the galaxies, hence confirming that most LBGs could also have  been selected as $sBzK$ galaxies, and that there is therefore an overlap between the two samples. The remaining LBGs do not meet the $sBzK$ criterion because of their bluer $z - K$ colours. On the other hand, only about 25\% of the $BzK$ galaxies satisfy the Lyman break selection criterion. This is a consequence of the UV-bright continuum of the galaxies selected through the drop-out technique. Most UV-selected galaxies (90\%) could also have  been selected by the $sBzK$ criterion, although there is a population (10\%) of UV-selected galaxies with blue $z-K$ colours that would be missed, as  happened with the LBGs. As can be seen in Figure \ref{bzk_color}, these UV-selected-not-$BzK$ galaxies have a similar $B-z$ colour to that of the whole population of UV-selected galaxies, and what therefore makes them different is mainly their bluer $z-K$ colours. These galaxies turn out to be younger, less massive and have higher dust-corrected total SFR and sSFR than both the whole population of UV-selected galaxies and the $sBzK$ galaxies. The younger age and the lower mass are compatible with the fact that at $z \sim 2$ the $z - K$ colour measures the jump in flux between the 3000 and 7000 \AA\ rest-frame, approximately. This jump is sensitive to the Balmer break and therefore to the age. This population of blue $z-K$ galaxies produces the small differences in the histograms shown in Figure \ref{BzK} between $sBzK$ and UV-selected galaxies.

\begin{figure}
\centering
\includegraphics[width=0.45\textwidth]{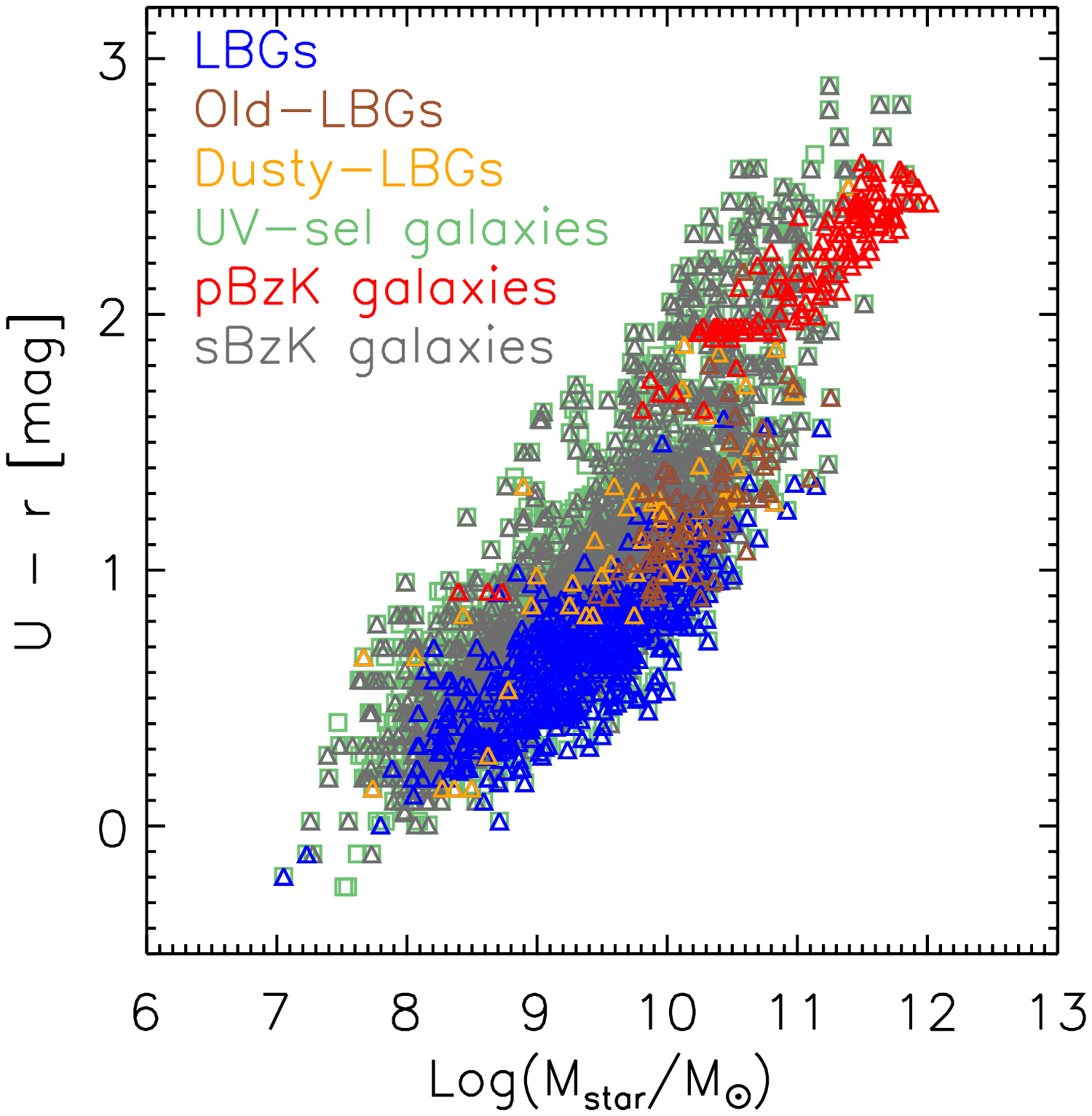}
\caption{Location of our LBGs and UV-selected galaxies at $z \sim 2$ in a colour versus stellar mass plane. Old LBGs and dusty LBGs are shown with brown and orange triangles, respectively, whereas the remaining LBGs are represented with blue triangles. Old LBGs are those whose SED-derived age is higher than 2 Gyr and dusty LBGs are those whose SED-derived dust attenuation is $E_s(B-V) > 0.4$. Our selected UV-selected galaxies are shown with open green triangles. For comparison, and with the aim of showing the location of the blue cloud and the red sequence at $z \sim 2$, we also represent the position of $sBzK$ (grey triangles) and $pBzK$ (red triangles) galaxies.
              }
\label{color_color}
\end{figure}

In conclusion, it seems that the SED-derived physical properties of $sBzK$ galaxies are more similar to those of the general population of UV-selected galaxies than to those of the LBGs; thus, SF galaxies selected through the $BzK$ criterion are more representative of the general population of SF galaxies at $z \sim 2$ than LBGs. The selection of LBGs through their rest-frame UV colours implies a strong bias towards galaxies which are UV-brighter, bluer, more massive and have higher values of the dust-corrected total SFR.

\begin{table*}
\caption{\label{SED_properties}Median values of the SED-derived properties of our selected LBGs, $sBzK$ and UV-selected galaxies located in the GOODS-S field. The results of the K-S tests for three possible combinations of galaxies are also shown.}
\centering
\begin{tabular}{lcccccc}
\hline\hline
Property & LBGs & $sBzK$ & UV-sel & K-S test LBGs/$sBzK$ & K-S test LBGs/UV-sel & K-S test $sBzK$/UV-sel\\
\hline

$\log{\left(L_{\rm UV}/L_\odot\right)}$	&	10.1	&	9.8		&	9.8		&	$<$0.001		&	$<$0.001		&	$>$0.99	\\

Age [Myr]							& 	111 		& 	251 		& 	201		&	 $<$0.001		&	$<$0.001		&	0.03	\\

$E_s(B-V)$ 						&	0.2		&	0.2		&	0.2		&	$<$0.001		&	$<$0.001		&	0.05		\\

$\log{\left(M_*/M_\odot\right)}$			&	9.4		&	9.4		&	9.4		&	0.11	&	0.004	&	0.20	\\

$SFR_{\rm total}$ [$M_\odot$/yr]	&	17.5	&	8.7		&	9.4	&	$<$0.001	&	$<$0.001	&	0.26		\\

UV slope						&	-1.79		&	-1.71	&	-1.69 & $<$0.001 & $<$0.001 & 0.68 \\

sSFR [${\rm Gyr}^{-1}$]					&	9.00		&	3.98		&	4.98 & $<$0.001 & $<$0.001 & 0.02	\\	

\hline
\end{tabular}
\end{table*}

\subsection{Colour--stellar mass diagram}\label{CMD}

An important tool for analysing the properties of SF galaxies is their location in a colour--magnitude or colour--stellar mass diagram. This kind of diagram has traditionally been used to separate local non-SF galaxies earlier than the Sa morphological type from local SF galaxies later than the Sb morphological type. In a colour space, the former tend to populate the so-called \emph{red sequence} and the latter are located in the so-called \emph{blue cloud} \citep{Hogg2002,Strateva2001}. This behaviour translates into a bimodal distribution of the colour of galaxies that allows us to study the nature of different samples of galaxies by looking at their position in colour space. It also enables us to look for galaxies with different SF natures by imposing conditions on their location in such a diagram. This bimodality in the local universe has been proven to apply at higher redshifts, at least up to $z \sim 1.6$ \citep{Bell2004,Nicol2011,Williams2009,Franzetti2007,Cirasuolo2007,Taylor2009,Weiner2005,Blanton2003}. In this study, we consider $sBzK$ and $pBzK$ galaxies as typical examples of galaxies that populate the blue cloud and the red sequence of galaxies at $1.5 \le z \leq 2.5$, respectively. This assumption is employed here with the aim of defining a blue cloud and red sequence for galaxies at the redshift range of our LBGs. It should be noted that at red $z-K$ colours the sample of $sBzK$ galaxies could be highly contaminated by passively evolving galaxies \citep{Grazian2007}. Figure \ref{color_color} shows a colour versus stellar mass diagram for galaxies at $z \sim 2$, where $sBzK$ galaxies are represented by grey open triangles and $pBzK$ galaxies are represented by red open triangles. The $U-r$ colours shown in that plot are obtained by integrating the best-fitted templates of each galaxy with the transmission curves of the $U$ and $r$ SDSS filters \citep{Abazajian2009} shifted in wavelength according to the redshift of each source. In this way, these colours are K-corrected. It can be seen, as expected from their selection criteria and their different natures, that $sBzK$ and $pBzK$ galaxies tend to populate two different regions of the diagram, although some $sBzK$ galaxies deviate towards the zone where $pBzK$ galaxies are located. Our LBGs at $1.5 \lesssim z \lesssim 2.5$ are represented in Figure \ref{color_color} with blue, orange and brown triangles. It can be seen that they mostly populate the blue cloud at their redshift, reinforcing the idea that they are active SF galaxies. Furthermore, for a given stellar mass, LBGs are among the bluest galaxies. Some LBGs tend to deviate from the blue cloud to the red sequence. We arbitrarly distinguish two classes of LBGs: those whose SED-derived dust attenuation is $E_s(B-V) \geq 0.4$ (\emph{dusty-LBGs}) and those whose SED-derived age is ${\rm Age} \geq 2000 \, {\rm Myr}$ (\emph{old-LBGs}). Dusty LBGs and old LBGs are represented by orange and brown triangles, respectively. It can be seen that old and dusty LBGs that tend to shift from the blue cloud to the red sequence. This behaviour of dusty LBGs deviating towards the red sequence is similar to that found in GALEX-selected LBGs at $z \sim 1$ \citep{Oteo2013a}. Furthermore, Oteo et al. (2013b) have found that IR-bright PACS-detected LBGs at $z \sim 1$ are mostly located over the green valley, between the blue cloud and the red sequence of galaxies at their redshift.

Additionally, the $sBzK$ and UV-selected galaxies are located in a similar zone of the colour--mass diagram owing to the similarity between the two populations.

\begin{figure*}
\centering
\includegraphics[width=0.3\textwidth]{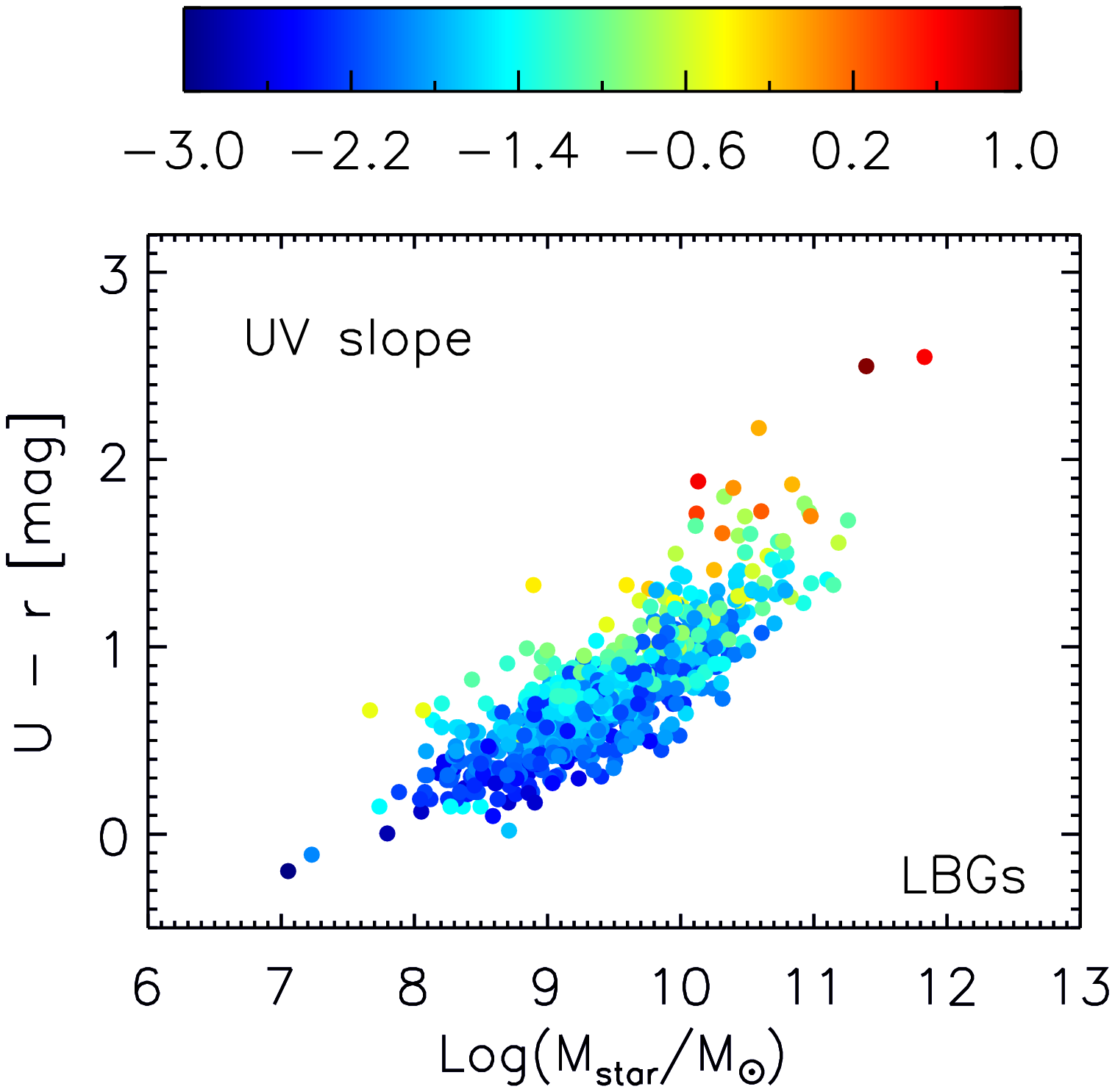}
\includegraphics[width=0.3\textwidth]{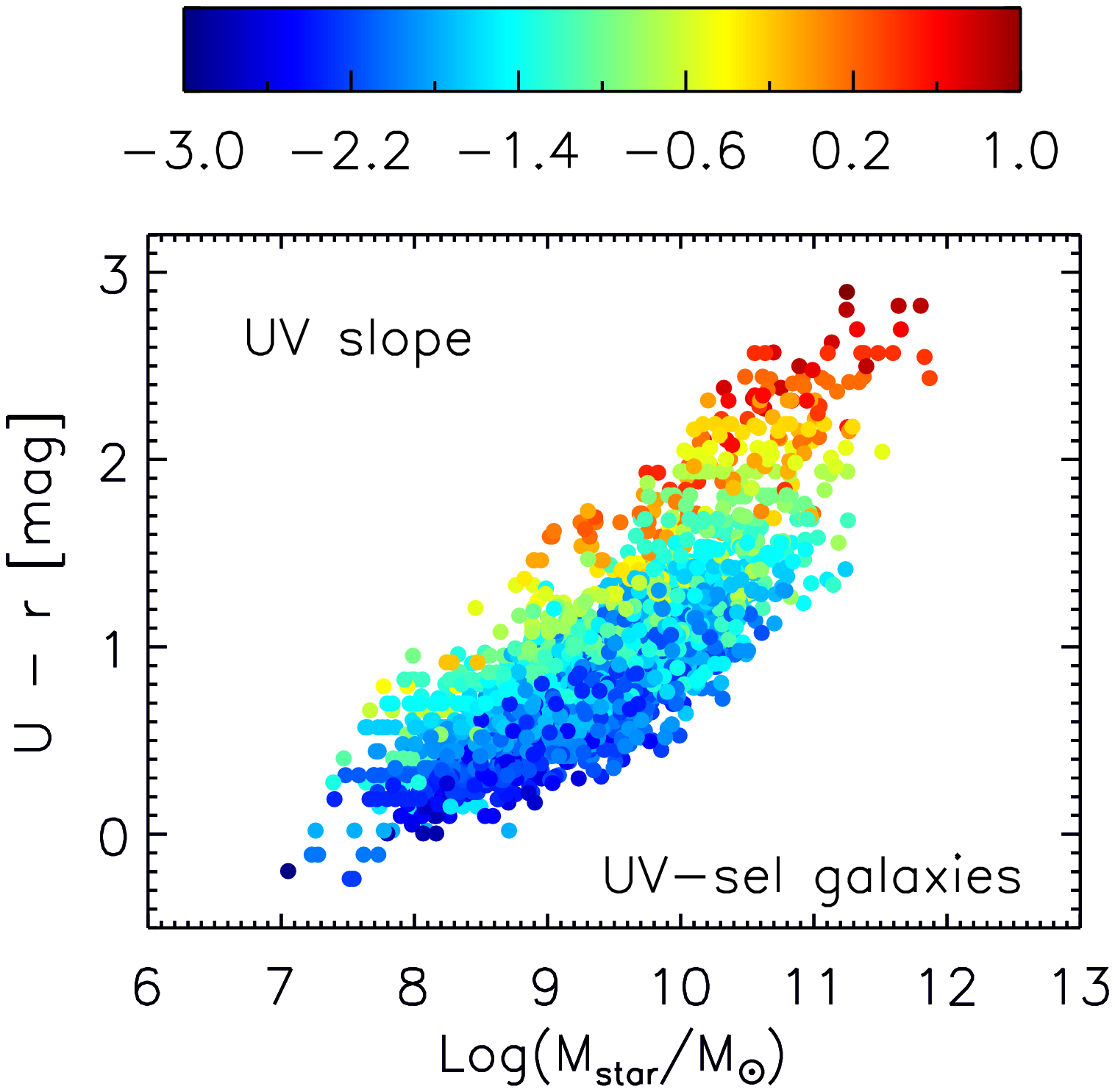}\\
\includegraphics[width=0.3\textwidth]{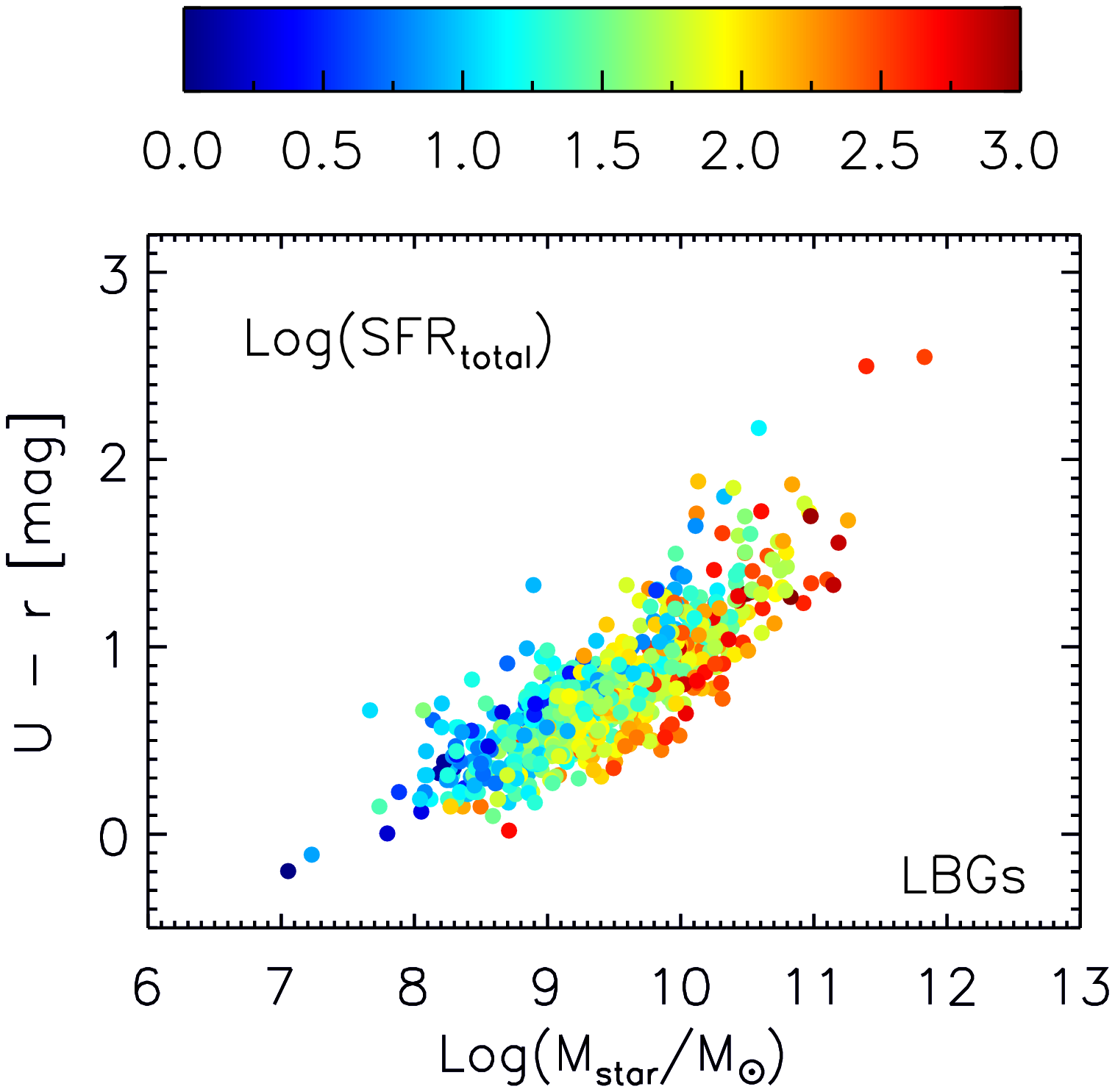}
\includegraphics[width=0.3\textwidth]{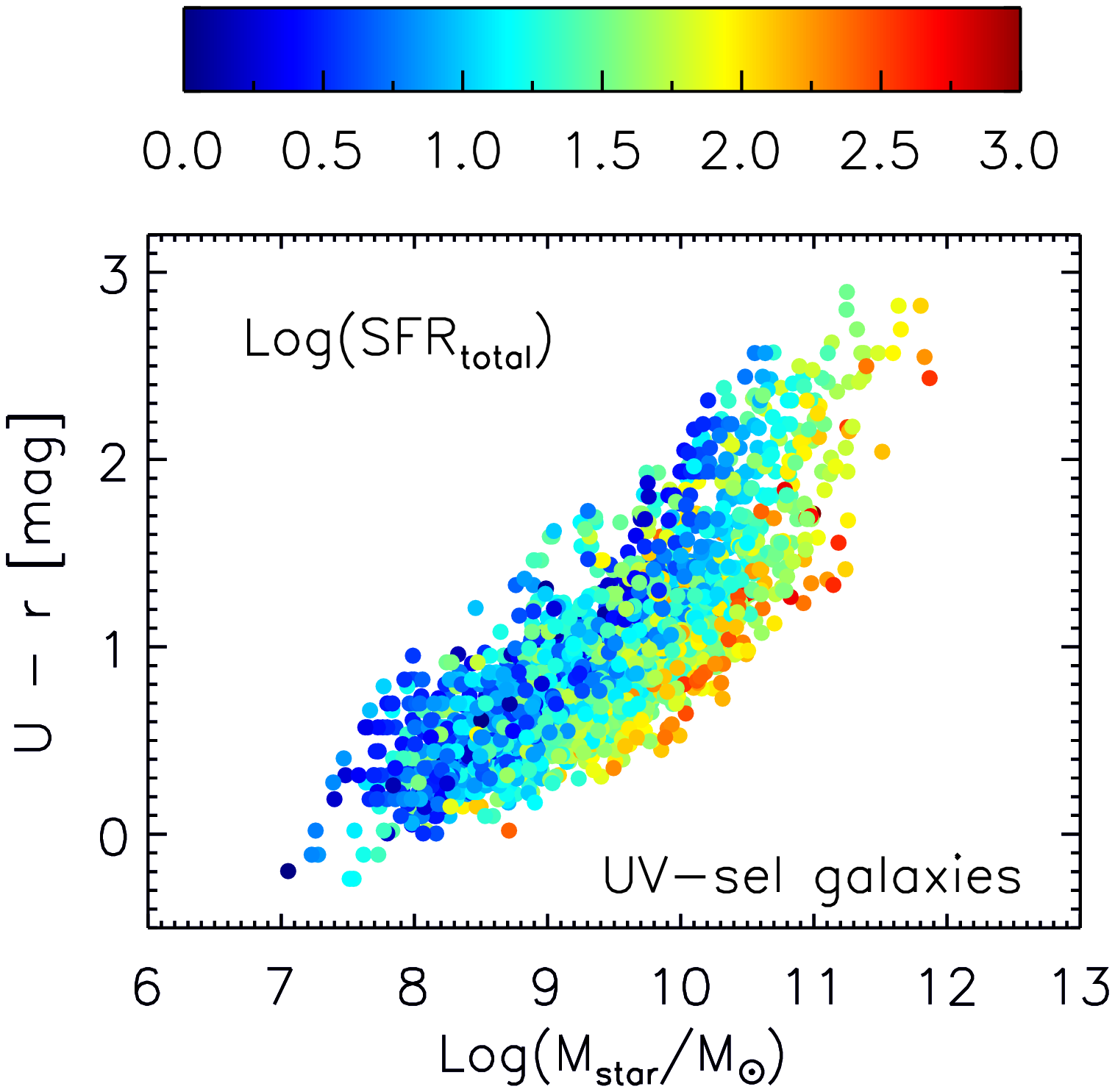}
\caption{Location of our selected LBGs {\emph{left}} and UV-selected galaxies (\emph{right}) in a colour versus stellar mass plane as a function of the values of the UV continuum slope (\emph{upper}) and the dust-corrected total SFR (\emph{bottom}). The values of the UV continuum slope and dust-corrected total SFR associated with each point are indicated with the colour bars situated at the top of each plot. In the interests of clarity, we refrain from plotting the location of $sBzK$ and $pBzK$ galaxies in these diagrams. The values of the dust-corrected total SFR are those obtained with the dust correction factors calculated with the \citet{Takeuchi2012} law.
              }
\label{color_color_SFR_beta}
\end{figure*}

Figure \ref{color_color_SFR_beta} shows the location of our LBGs and UV-selected galaxies in the colour vs stellar mass diagram as a function of the UV continuum slope and the dust-corrected total SFR. It can clearly be seen that the LBGs and UV-selected galaxies that tend towards the red sequence are those with the reddest values of the UV continuum slope. However, there is no correlation with the dust-corrected total SFR. What we do find is that the total SFR is higher for higher masses. This an indication of the existence of a main sequence (MS) at $z \sim 2$ for our selected galaxies (see Section \ref{sfrmass}).

\subsection{SFR vs stellar mass}\label{sfrmass}

\begin{figure*}
\centering
\includegraphics[width=0.45\textwidth]{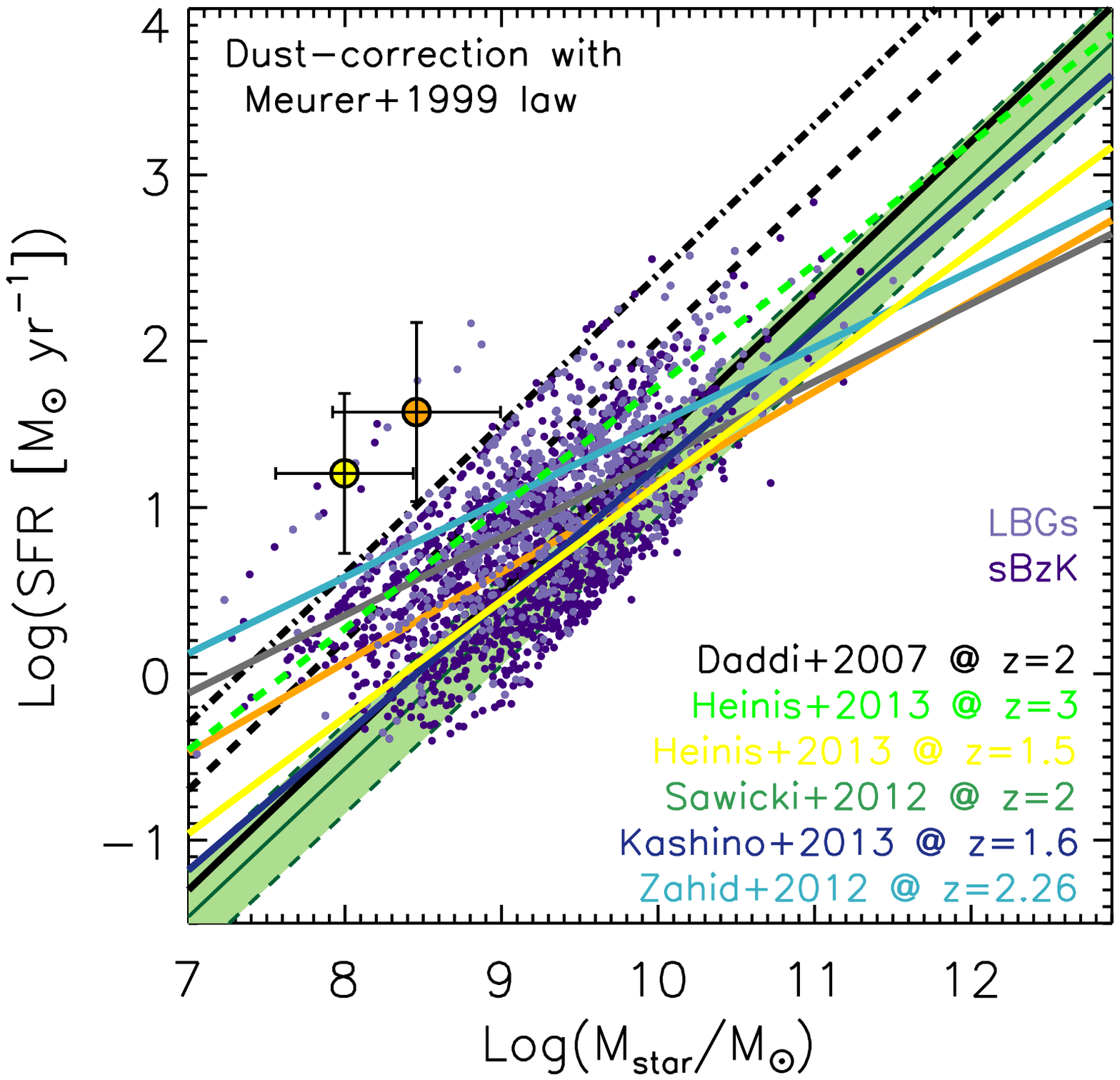}
\includegraphics[width=0.45\textwidth]{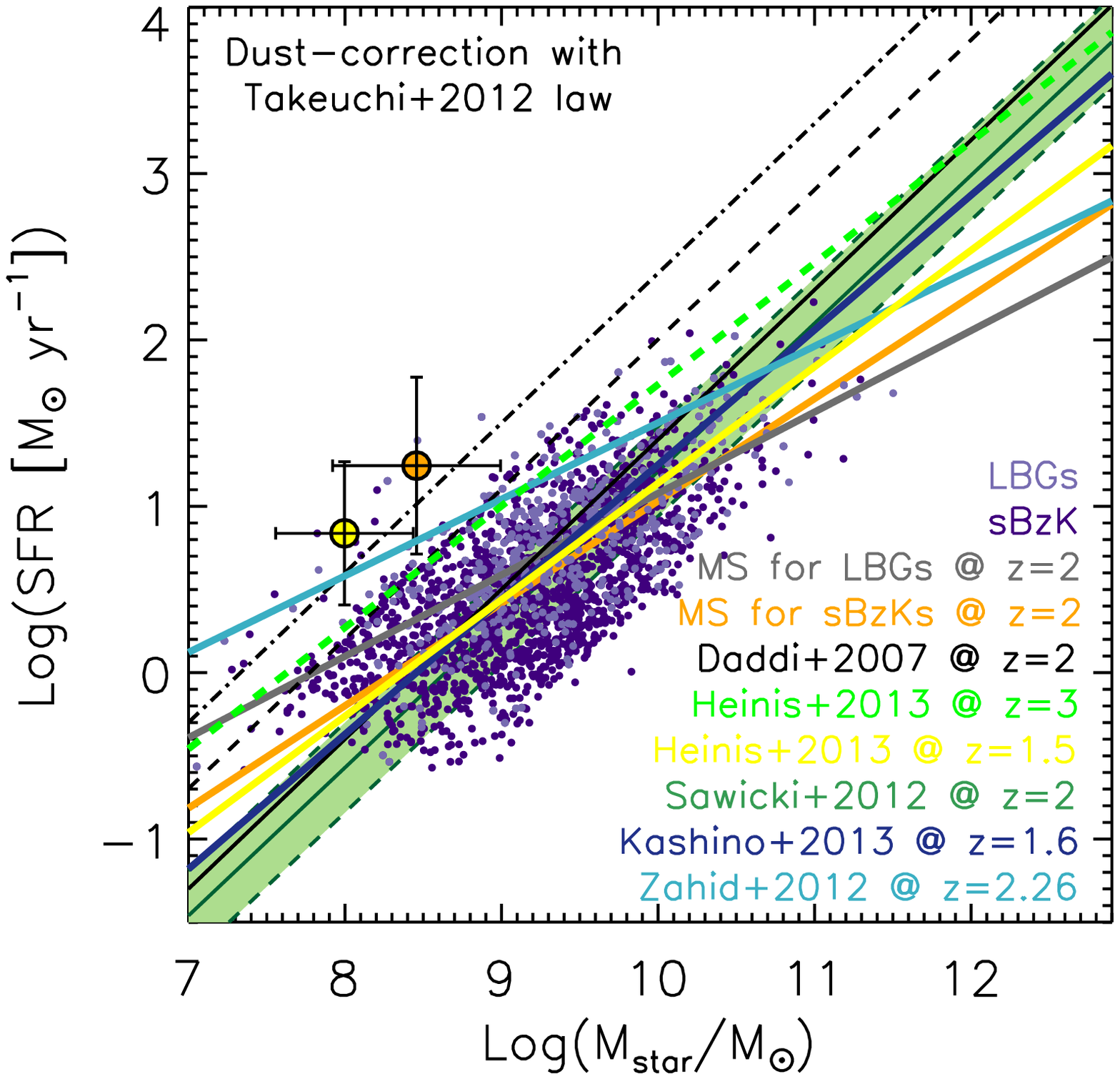}
\caption{SFR versus stellar mass plane for our selected LBGs (light purple dots) and $sBzK$ galaxies (dark purple dots) in the GOODS-S field. Two estimations of the dust-corrected total SFR are presented: the obtained with the \citet{Meurer1999} (\emph{left panel}) and \citet{Takeuchi2012} (\emph{right panel}) relations. A compilation of the MS for SF galaxies at different redshifts reported in several works is included \citep{Daddi2007,Heinis2013arXiv1310.3227H,Sawicki2012,Kashino2013arXiv1309.4774K,Zahid2012ApJ...757...54Z}. Dashed and dotted-dashed lines represent 4 and 10 times the \citet{Daddi2007} MS. The median locus of young (age $\leq$ 10 Myr) sources are indicated in each panel with a yellow filled dot (for LBGs) and a orange filled dot (for $sBzK$ galaxies). Their associated error bars represent the widths of the dust-corrected total SFR and stellar mass distributions. Grey and orange straight solid lines represent the fit to our LBGs and $sBzK$ galaxies (excluding age $\leq$ 10 Myr sources), respectively. 
              }
\label{sfr_mass_daddi}
\end{figure*}

A tight relation between the total SFR and stellar mass of galaxies has been found at a wide range of redshifts. This has allowed us to define an MS for SF galaxies at different redshifts \citep{Salim2007,Elbaz2007,Noeske2007,Daddi2007,Pannella2009,Rodighiero2010,Gonzalez2010,Karim2011,Elbaz2011,Sawicki2012,Salmi2012}. Galaxies in the MS are the reverse of the idea of 'starburst galaxies', which are those sources whose nature gives them higher sSFR values than MS galaxies for each stellar mass. The characteristic value of the sSFR for the MS of galaxies has been reported to change with redshift \citep{Elbaz2011}. Previous studies show that the sSFR increases with increasing redshift at all masses, and that the sSFR of massive galaxies is lower than that for less massive galaxies at any redshift \citep{Feulner2005,Erb2006,Dunne2009,Damen2009,Rodighiero2010}. Despite the number of studies analysing the relation between stellar mass and SFR, there is still some controversy, mostly regarding the slope of the sSFR-$M_*$ relation.

\cite{Daddi2007} found a correlation between SFR and stellar mass for massive 24 $\mu$m-detected $sBzK$ galaxies that has been traditionally used as a reference for the MS at $z \sim 2$. \cite{Sawicki2012} demonstrated the validity of the \cite{Daddi2007} relation for less massive and MIPS-undetected $sBzK$ galaxies. In this section we study the location of our selected SF galaxies at $z \sim 2$ in an SFR versus stellar mass diagram. In the left panel of Figure \ref{sfr_mass_daddi} we show the total SFR obtained by correcting the rest-frame UV luminosity with the dust correction factor obtained with the UV continuum slope and the application of the \cite{Meurer1999} IRX-$\beta$ relation. It can be seen that our galaxies are distributed around the \cite{Daddi2007} MS at $z \sim 2$ although with a significant spread. 

\begin{figure}
\centering
\includegraphics[width=0.45\textwidth]{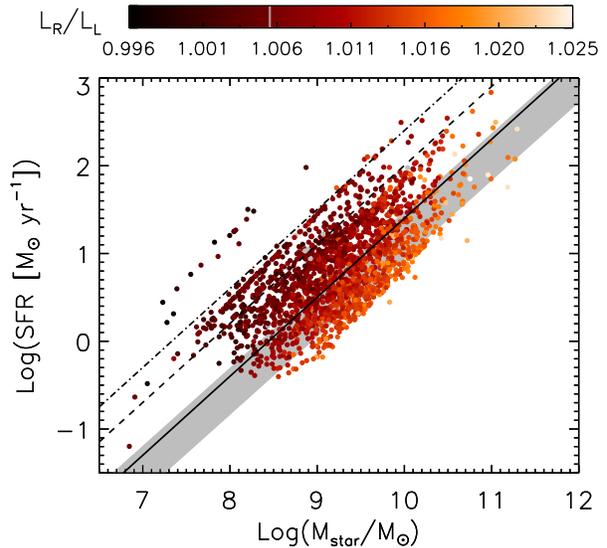}
\caption{Location in the SFR--mass diagram of our selected $sBzK$ galaxies in the GOODS-S field as a function of the Balmer break. The total SFR has been obtained by correcting the rest-frame UV luminosity with the dust attenuation derived with the UV continuum slope and the \citet{Meurer1999} relation. Black solid (dashed, dotted dashed) line is (4 times, 10 times) the MS of \citet{Daddi2007} at $z \sim 2$. The grey shaded zone represents the MS at $z \sim 2$ obtained in \citet{Sawicki2012}.
              }
\label{sfr_mass_edad}
\end{figure}

There is a subsample of LBGs and $sBzK$ galaxies whose location in the SFR vs stellar mass diagram is well above the MS at $z \sim 2$. These galaxies turn out to be very young sources whose SED-derived ages are lower than 10 Myr and represent 1.5\% of the total sample of LBGs and less than 1\% of the total sample of $sBzK$ galaxies in GOODS-S. We have plotted the median locus of these galaxies in the SFR--mass diagram (yellow filled dots for LBGs and orange filled dots for $sBzK$ galaxies). It can be seen that these young galaxies are only found in the low-mass end ($\log{\left( M_*/M_\odot \right)} \leq 8.5$) and are potentially outliers of the MS relation. It should be also taken into account that, owing to their young age, the \cite{Kennicutt1998} relation might not apply for these sources and the SFR plotted in Figure \ref{sfr_mass_daddi} might be underestimated. This behaviour for young galaxies was also reported in \cite{Sawicki2012}. In order to explore this issue further we represent in Figure \ref{sfr_mass_edad} the relation between SFR and stellar mass for our selected $sBzK$ galaxies in GOODS-S as a function of the amplitude of the Balmer break. We do not represent directly the age of the galaxy because the SED-derived ages tend to suffer from large uncertainties, mainly due to the degeneracy between age and SFH. Instead we use the Balmer break, which is an age indicator. As indicated in Section \ref{SED}, we measure the Balmer break for each galaxy from its best-fitted BC03 template as a ratio between luminosities and therefore, is similar to the one that would be obtained directly from the photometric data, since the best-fitted templates represent the shape of the observed SEDs of the galaxies. It can be seen that there is a clear correlation between the location of galaxies in the SFR-mass diagram and the amplitude of the Balmer break in the sense that the galaxies with the highest and lowest values of the Balmer break deviate more from the MS defined in \cite{Daddi2007} and \cite{Sawicki2012}. Additionally, the galaxies with the lowest values of the Balmer break are the ones that deviate most from the MS, located above five times the \cite{Daddi2007} relation. In this way, very young galaxies with nearly flat optical spectra are candidate interlopers of the MS of SF galaxies at $z \sim 2$.

By fitting a linear relation for our LBGs and $sBzK$ galaxies (without considering the young galaxies, which are probably interlopers of the MS) we obtain an MS with a slightly lower slope than the found in \cite{Daddi2007} or \cite{Sawicki2012}, but compatible with the results of \cite{Zahid2012ApJ...757...54Z}. However, over most of the mass range covered by our LBGs and $sBzK$ galaxies, our MS is compatible with the uncertainties in the \cite{Sawicki2012} relation. The differences in the slope of the MS obtained in different works are an indication that the definition of the MS is sensitive to the details of the SED-fitting procedures carried out to derive the stellar masses and also, and maybe more importantly, to the different dust-corrections factors adopted to estimate the dust-corrected total SFR.

Recently, \cite{Takeuchi2012} presented a correction to the \cite{Meurer1999} IRX-$\beta$ relation that takes into account the underestimation of the UV flux density of the galaxies caused by the small aperture of $IUE$ \citep[see also][]{Overzier2011}. When adopting the dust correction factor obtained with the new \cite{Takeuchi2012} relation (right panel of Figure \ref{sfr_mass_daddi}) our galaxies follow a MS with a similar slope but with a slightly lower zero point because, for each value of the UV continuum slope, the dust attenuation derived with the \cite{Takeuchi2012} relation is lower than the derived with the \cite{Meurer1999} one. This reinforce the previous finding that the definition of the MS for sources that are not individually detected in the FIR is dependent on the dust correction method employed to recover the total SFR of the galaxies studied (see also the discussion in \cite{Lee2013arXiv1310.0474L}). Stacking analysis in these cases might be a good alternative, although this is beyond the scope of this paper. Furthermore, as shown in \cite{deBarros2013}, the location of galaxies in the SFR-M$_*$ diagram is greatly dependent upon the SFH adopted in the elaboration of the BC03 templates used for determining the star formation rate and stellar mass.

\section{The FIR spectral energy distribution}\label{FIR_SED}

\begin{figure}

\centering
\includegraphics[width=0.22\textwidth]{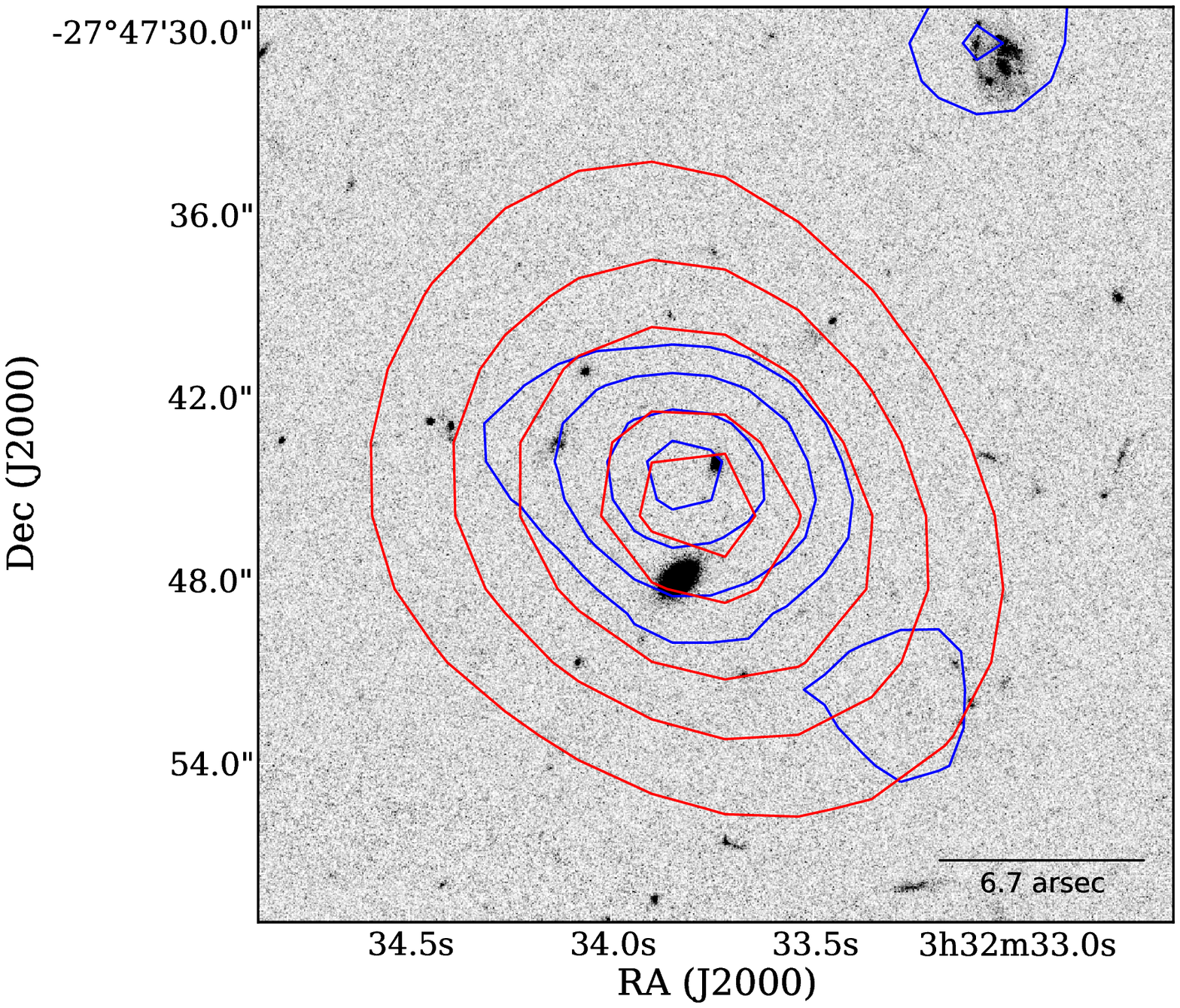}
\includegraphics[width=0.22\textwidth]{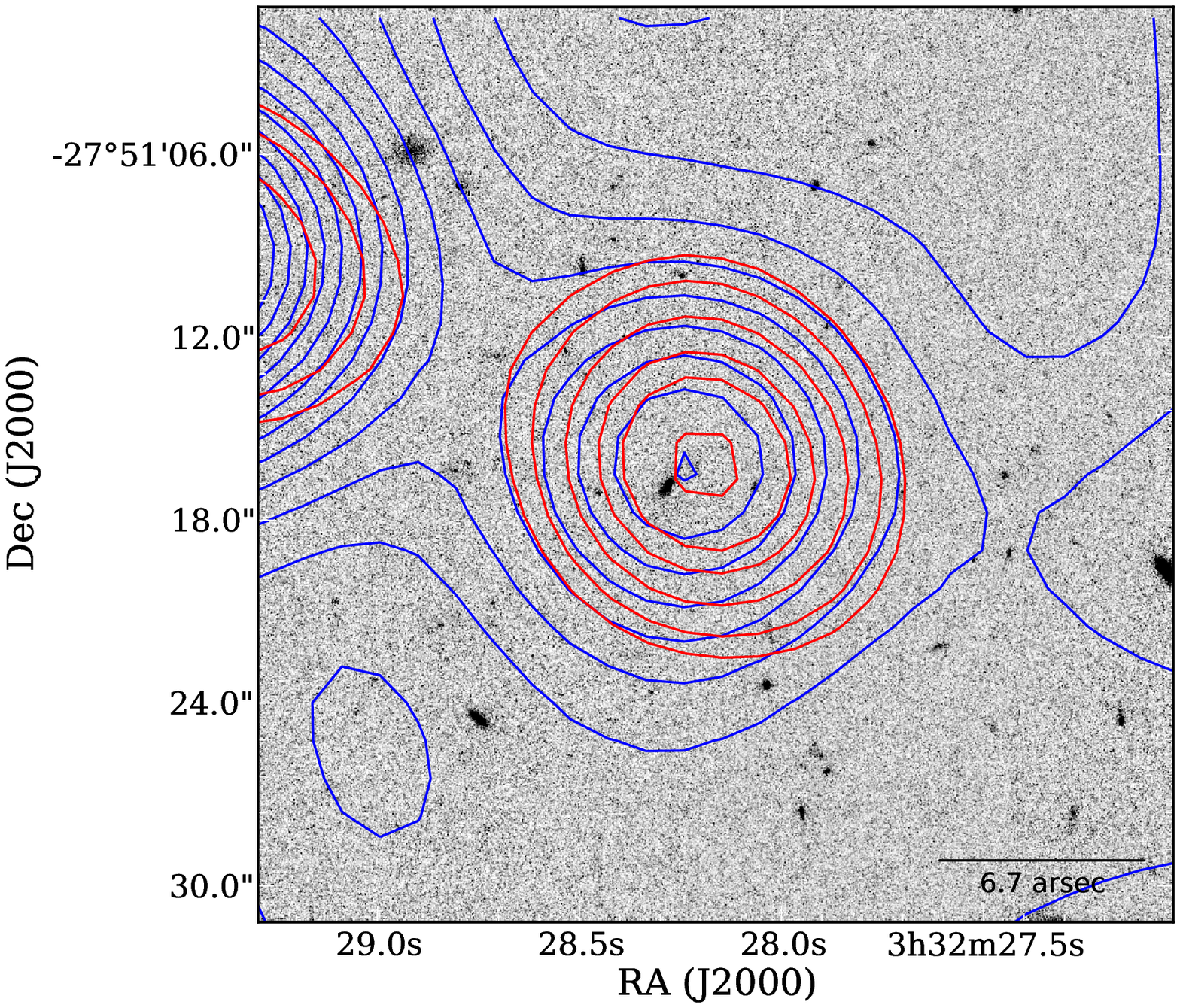}
\caption{ACS $z$-band cut-outs of two PACS-detected LBGs at $z \sim 2$ with good identification of the FIR counterpart. The LBG is located in the centre of each image. The MIPS-24 $\mu$m and PACS-100 $\mu$m contours are represented in blue and red, respectively. The PSF of the PACS-100 $\mu$m observations (6.7 arsecs) is indicated with a black horizontal bar in the bottom-left corner. These two cases represent the behaviour of the whole sample of PACS-detected galaxies with good identification of the FIR counterpart.
}
\label{confu}
\end{figure}

\begin{figure}
\centering
\includegraphics[width=0.22\textwidth]{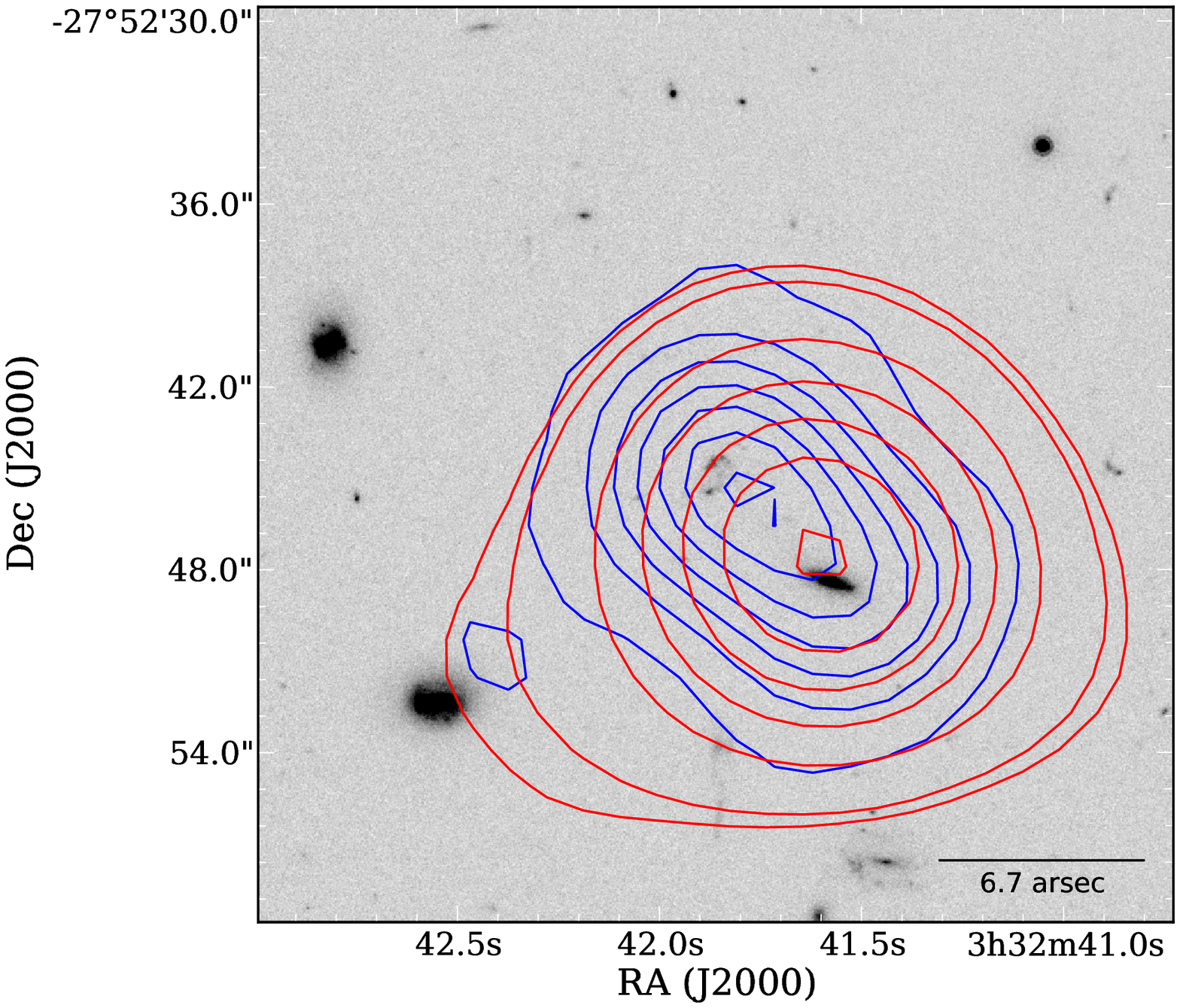}
\includegraphics[width=0.22\textwidth]{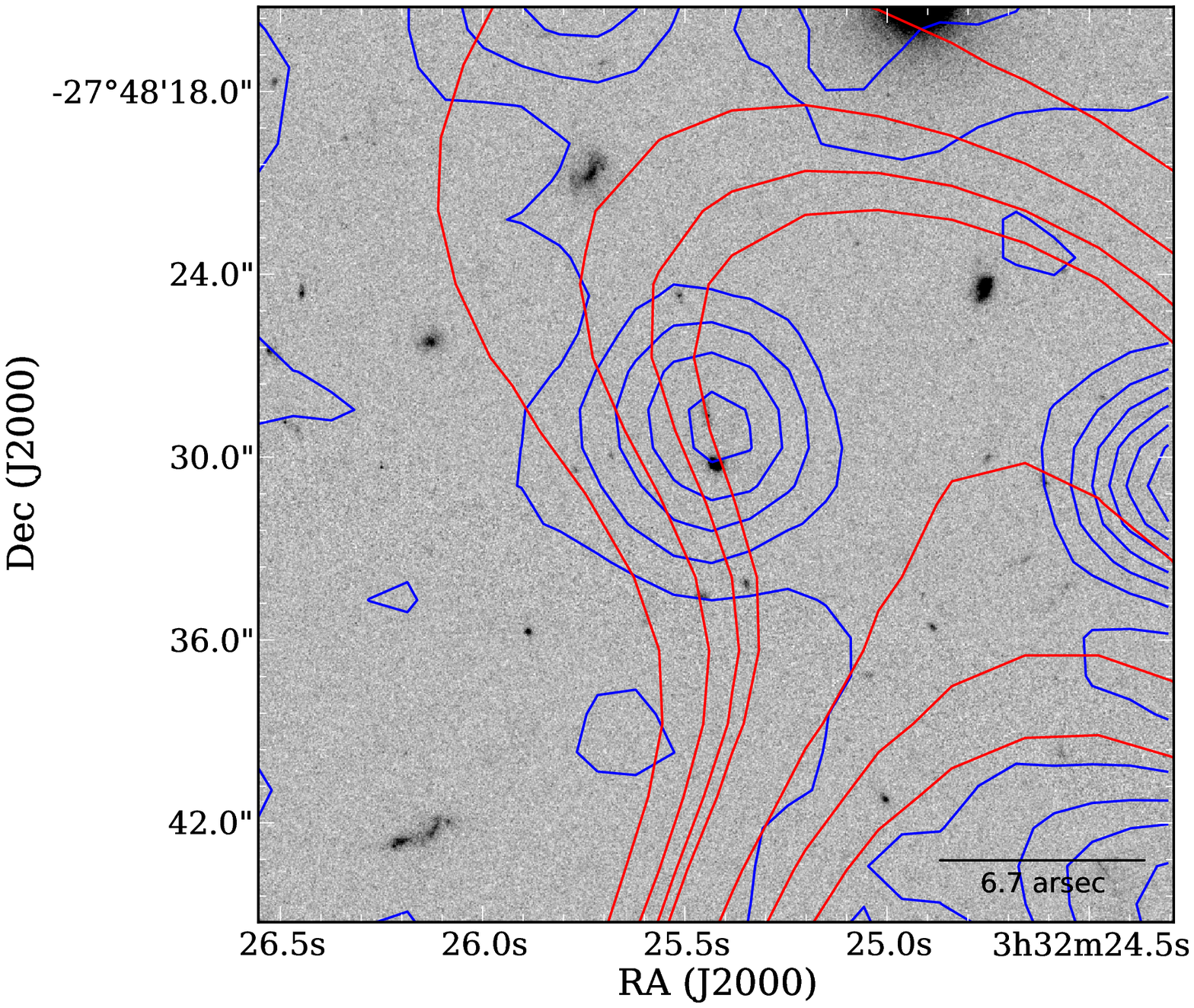}
\caption{ACS $z$-band cut-outs of two PACS-detected LBGs at $z \sim 2$ with bad identification of the FIR counterpart. The LBG is located in the centre of the image. The MIPS-24 $\mu$m and PACS-100 $\mu$m contours are represented in blue and red, respectively. The PSF of the PACS-100 $\mu$m observations (6.7 arsecs) is indicated with a black horizontal bar in the bottom-left corner. The left panel represents a case where the LBG has a clear MIPS-24 $\mu$m emission but the PACS flux is likely coming from a nearby source despite being less MIPS-24 $\mu$m luminous. The right panel represents a case where the LBG has a clear MIPS-24 $\mu$m detection but the PACS flux might be contaminated by the FIR emission of a nearby source, as indicated by its clean index parameter.
}
\label{confu_bad}
\end{figure}

So far, we have analysed the rest-frame UV to near-IR SED of our selected LBGs and $sBzK$ and UV-selected galaxies at $1.5 \lesssim z \lesssim 2.5$. This has allowed us to derive their SED-derived properties, such as age, amplitude of the Balmer break, dust attenuation, stellar mass, dust-corrected total SFR,  sSFR and UV continuum slope. We have also studied their location in the colour--M$_*$ diagram an the relation between their SFR and stellar mass. We now go further in wavelength coverage and analyse their FIR SED and FIR-derived physical properties by using deep FIR data taken from the GOODS-\emph{Herschel} project. Since it targeted only GOODS-N and GOODS-S, we focus in this section on those fields and discard COSMOS.

We look for possible PACS detections of our SF galaxies by using their optical positions as a reference and considering a matching radius of 1.2$''$. This is the typical blind pointing uncertainties of MIPS for the 24$\mu$m band. We consider a PACS-detected galaxy as one detected in at least either of the 100 $\mu$m or 160 $\mu$m bands. The number of FIR detections obtained when performing this simple match is shown in Table \ref{PACS_detections} in the column labelled 'All PACS detections'. The SPIRE detections for our LBGs and $sBzK$ galaxies will be reported in Section \ref{SPIREdetections}. When dealing with FIR data, and because of the large PSF of the observations, the source confusion becomes a major issue and two main problems have to be solved. First, one has to check that the identification of the FIR counterpart for a given optically selected sources is right. It is then necessary to check that the FIR fluxes of each identified source are not contaminated by the FIR emission of a close neighbour. The best way to ensure that we are identifying the FIR emission associated with a given optically selected sources properly is by plotting the FIR contours on top of an optical image (see Figures \ref{confu} and \ref{confu_bad}). These Figures show the FIR (MIPS in blue and PACS in red) contours for four initial candidates of PACS-detected LBGs at $z \sim 2$. It can be seen that the two sources shown in Figure \ref{confu} are correctly identified, and that the MIPS and PACS fluxes are very probably coming from each selected LBG. However, in the case presented in the left panel of Figure \ref{confu_bad} the PACS flux is probably coming from the companion source, despite the MIPS contours indicating that the LBG is brighter in MIPS. In the case presented in the right panel of Figure \ref{confu_bad}, the FIR identification of the LBG might be correct owing to the shape of the contours, but the FIR flux is likely to be contaminated by the emission of a nearby source. In order to address the problem of flux contamination, the GOODS-\emph{Herschel} catalogues provide an estimator, called the 'clean index', of the purity of the sources. This clean index indicates the number of sources ('bright neighbours') that might contaminate the flux in each of the PACS/SPIRE bands. In this study, we use only those PACS-detected galaxies whose MIPS/PACS contours indicate a correct FIR identification and whose clean index suggest that there are no bright neighbours in the PACS and SPIRE bands (\emph{clean sample}). The number of galaxies in the clean sample is also presented in Table \ref{PACS_detections}. The typical FIR contours of the sources in the clean sample are similar to those presented in Figure \ref{confu}. It should be pointed out that the clean index indicates the presence of one or some bright neighbours that might contaminate the flux of an FIR-detected source. However, the PSF-fitting method with prior positions employed might recover the FIR flux of a source properly even when a close neighbour is present. Despite this, to be conservative, we only use clean galaxies with no bright neighbours in PACS and SPIRE in our study.

The percentages of FIR-bright LBGs, $sBzK$, and UV-selected galaxies are very low. Despite this, those PACS detections indicate that there is a population of red and dusty SF galaxies at $1.5 \lesssim z \lesssim 2.5$ whose dust emission peak can be directly probed with \emph{Herschel}. \cite{Daddi2005} and \cite{Kurczynski2012} report MIPS-24$\mu$m detections for $sBzK$ galaxies $z \sim 2$. However, the extrapolation from MIPS-24$\mu$m to the total IR luminosity might suffer from significant uncertainties \citep{Elbaz2010,Nordon2010,Elbaz2011,Nordon2012}. A population of dusty and red LBGs also exists at lower \citep{Oteo2013_ALHAMBRA_PACS,Burgarella2011} and higher \citep{Oteo2013_z3} redshifts.

\begin{table*}
\caption{\label{PACS_detections}Number of PACS detections for each kind of SF galaxy studied in this work}
\centering
\begin{tabular}{lccccc}
\hline\hline

Galaxy & $N_{\rm total}$ & All PACS detections & Percentage & Clean PACS detections & Percentage\\

\hline

LBGs in GOODS-S & 681 & 17 & 2.4\%		&		9	& 1.3\%\\

LBGs in GOODS-N & 1300 & 60 & 4.6\%		&		39	& 3.0\%\\

\hline

$sBzK$ in GOODS-S & 2472 & 41 & 1.6\%	&		28	& 1.1\%\\

$sBzK$ in GOODS-N & 2192 & 99 & 4.6\%	&		61	& 2.7\%\\

\hline

UV-sel in GOODS-S & 2767  & 41 & 1.5\%	&		28	& 1.0\%\\

UV-sel in GOODS-N & 2581  & 103 & 3.9\%	&		63	& 2.4\%\\

\hline
\end{tabular}
\end{table*}













\begin{figure*}
\centering
\includegraphics[width=0.2\textwidth]{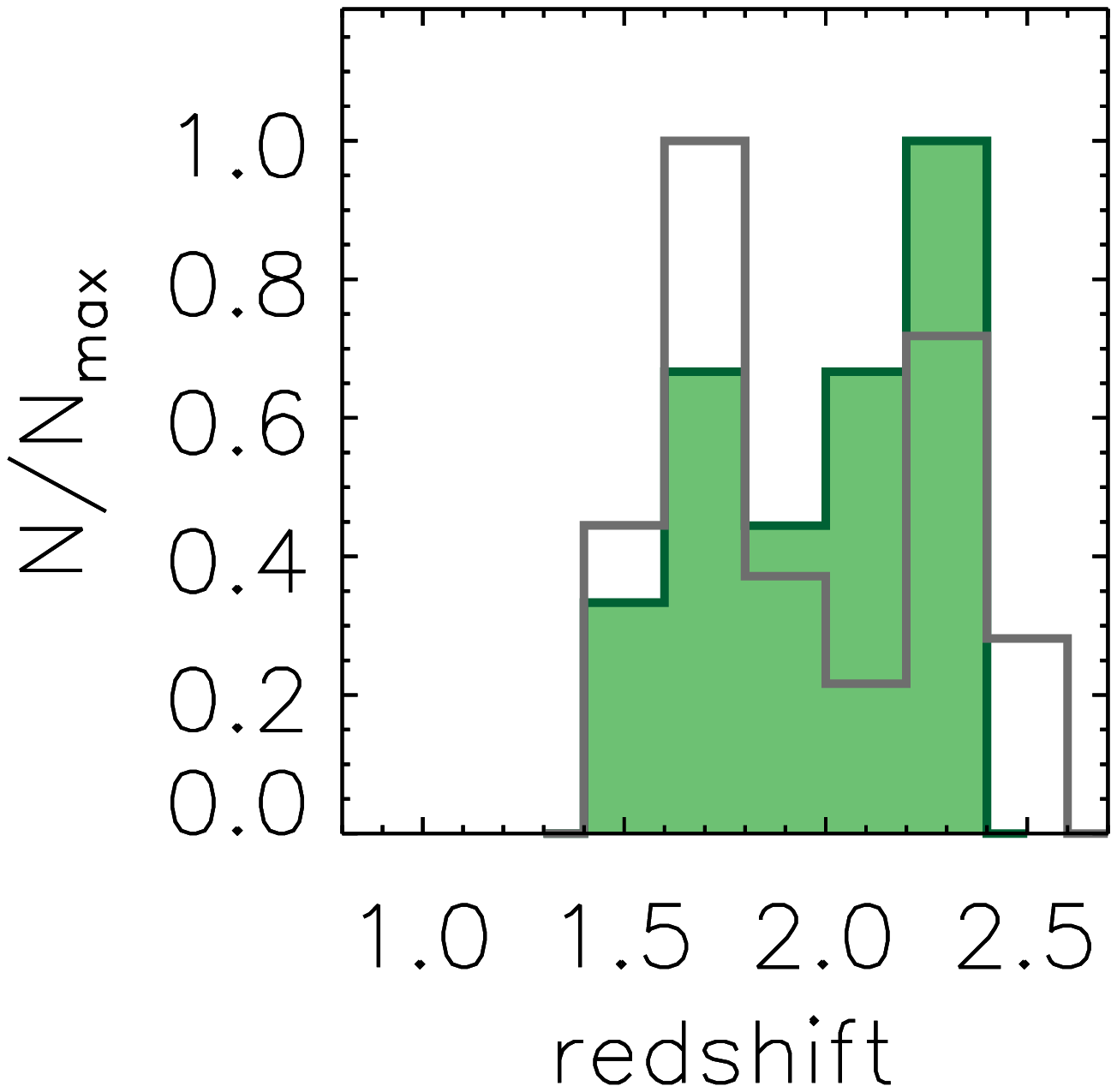}
\includegraphics[width=0.2\textwidth]{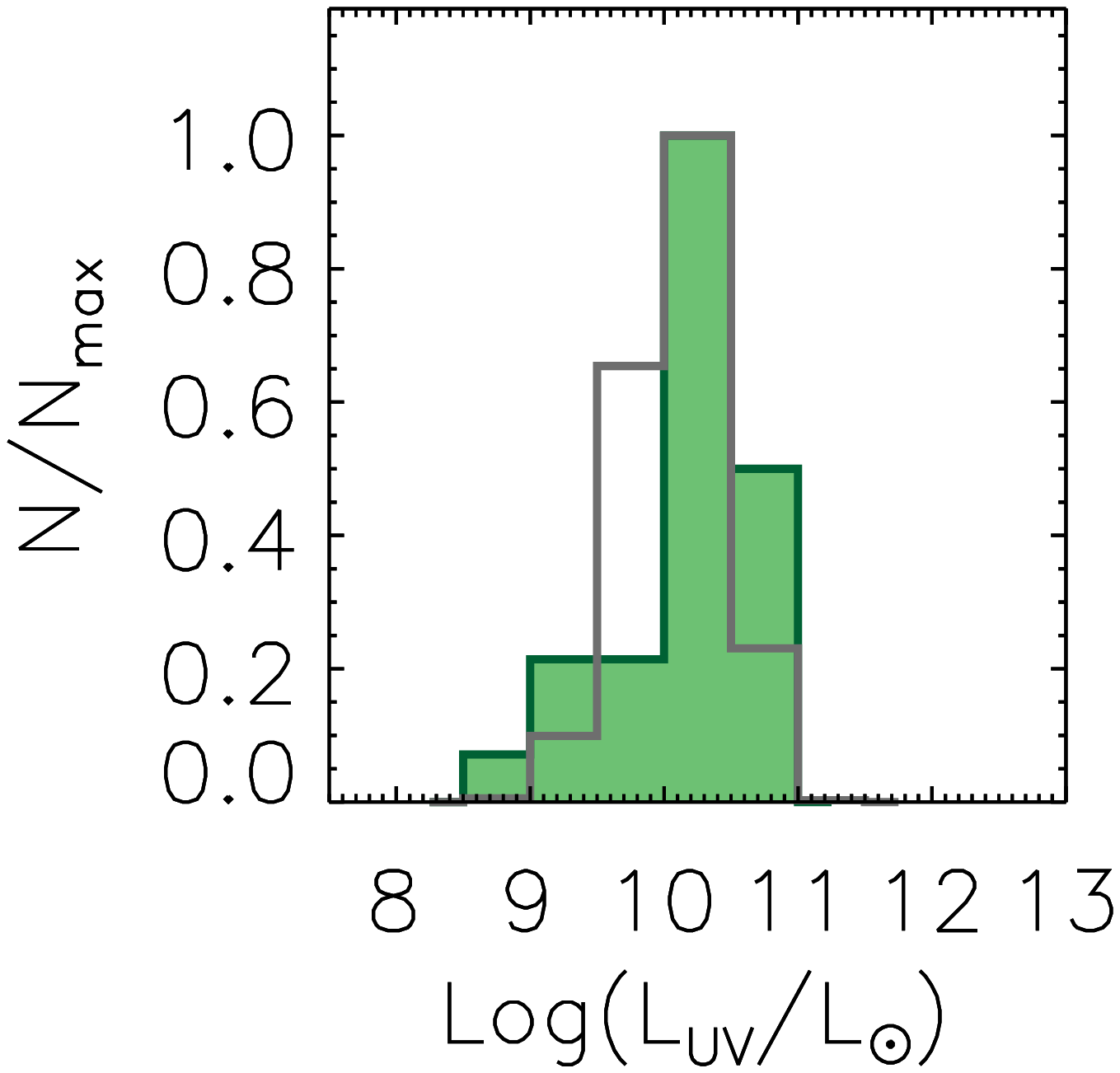}
\includegraphics[width=0.2\textwidth]{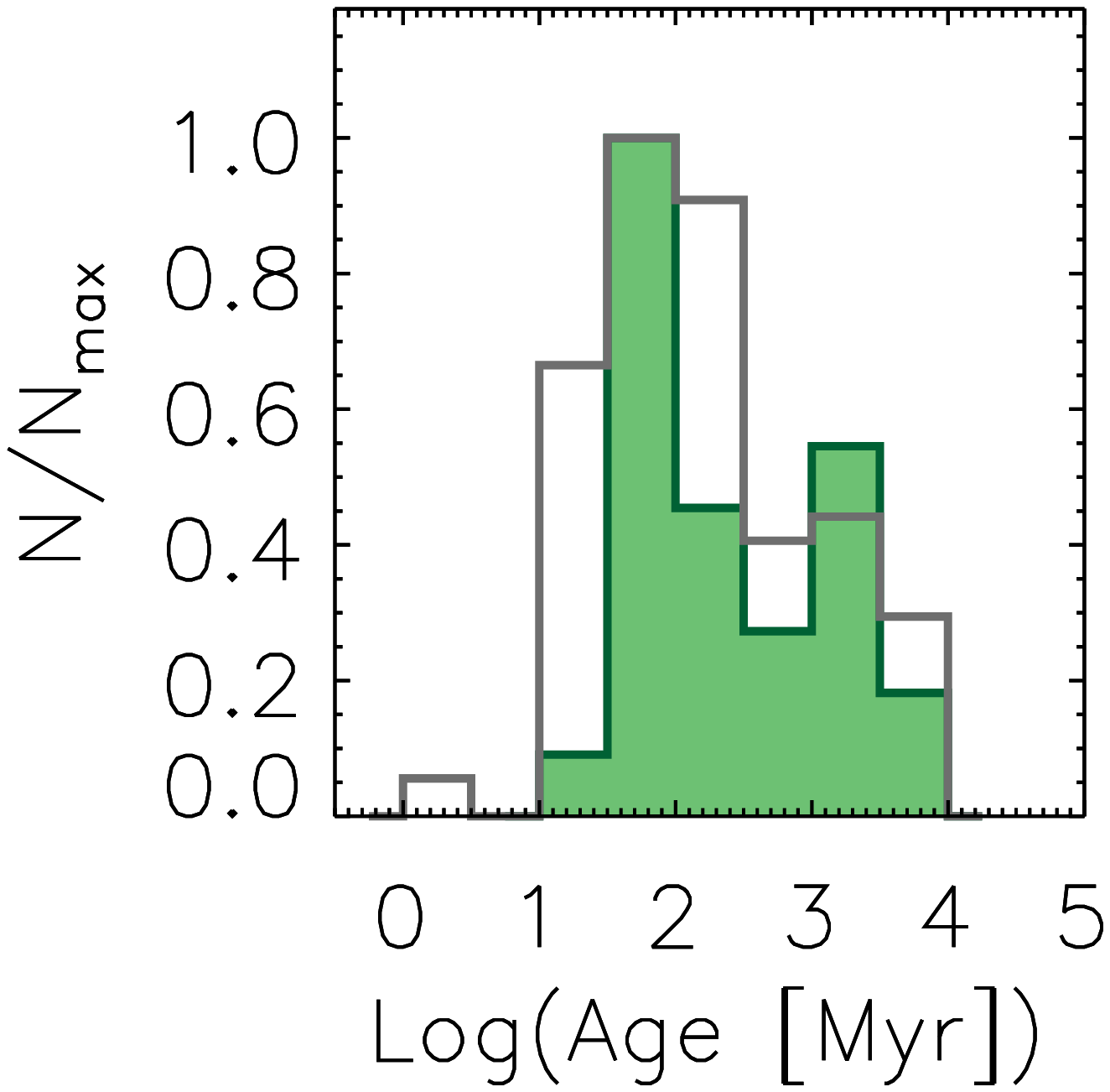} \\
\includegraphics[width=0.2\textwidth]{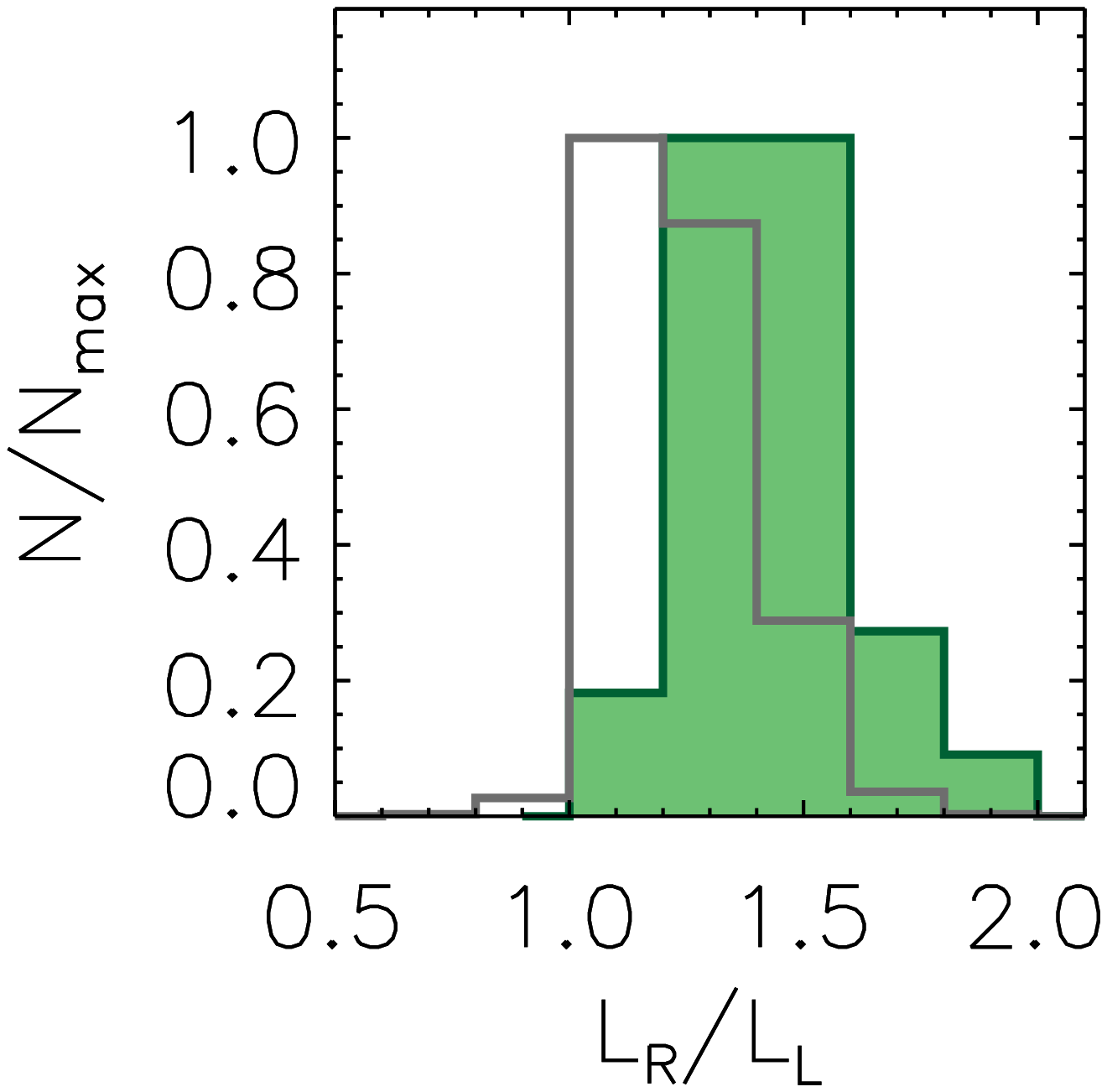}
\includegraphics[width=0.2\textwidth]{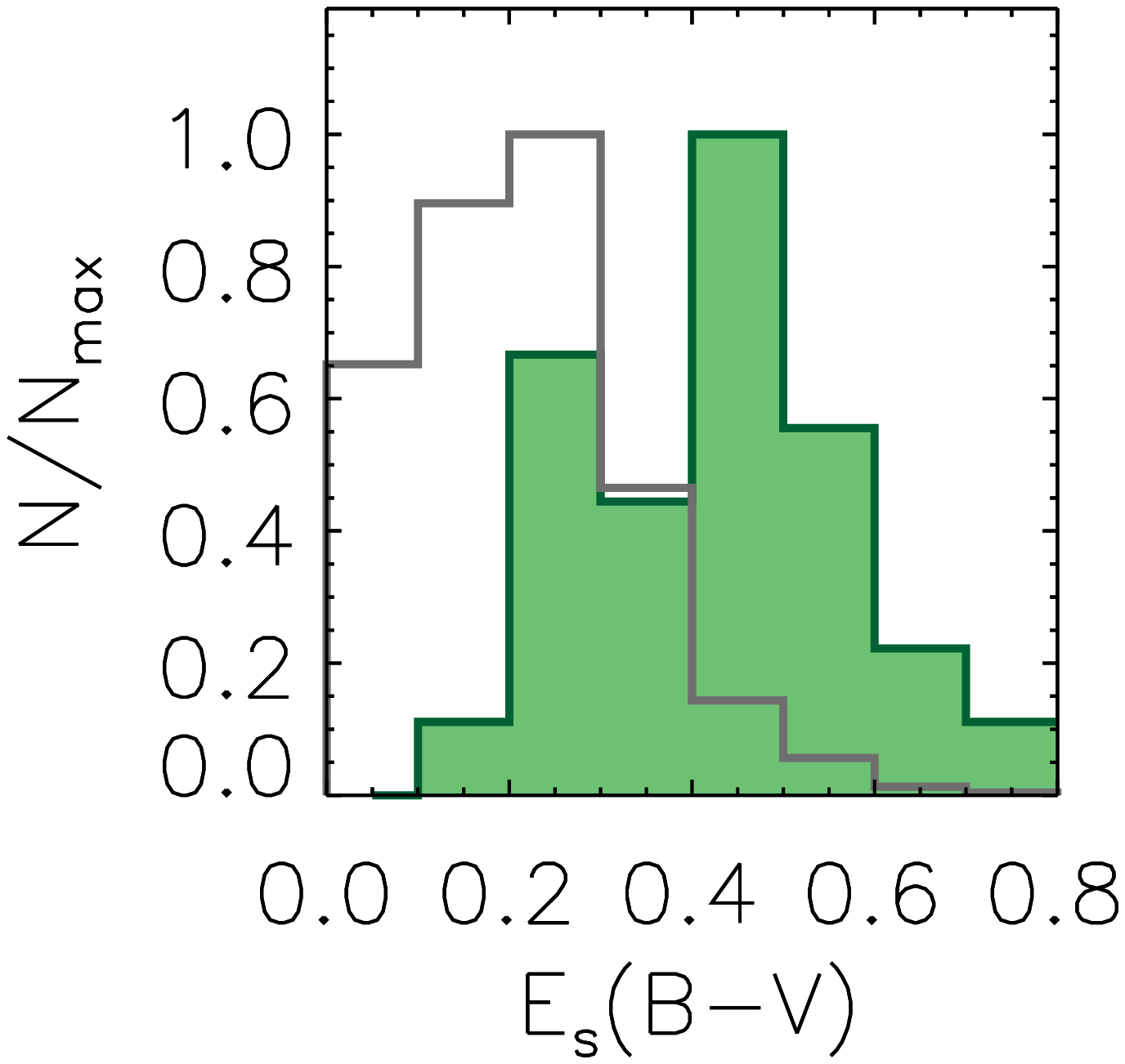}
\includegraphics[width=0.2\textwidth]{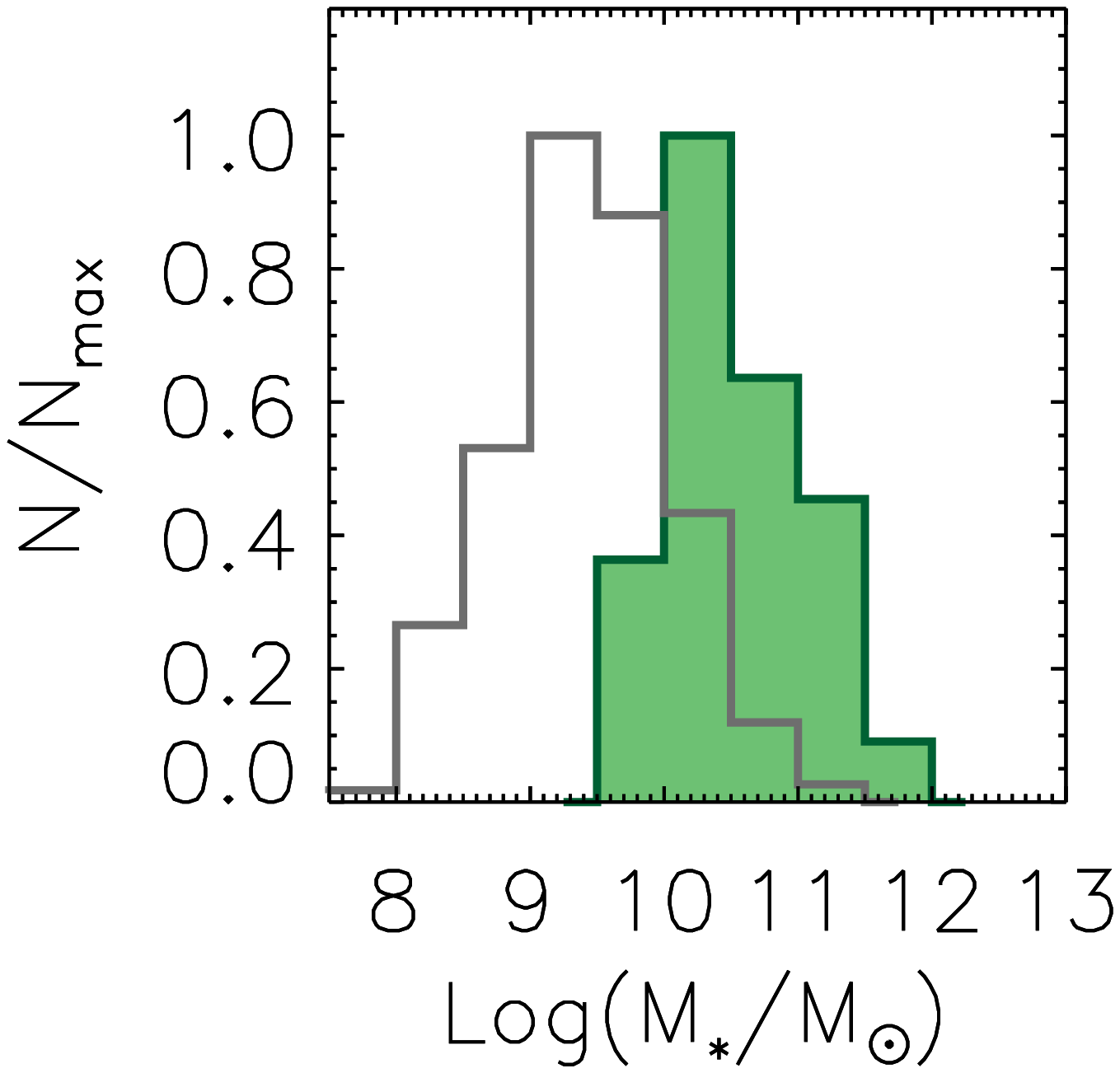} \\
\includegraphics[width=0.2\textwidth]{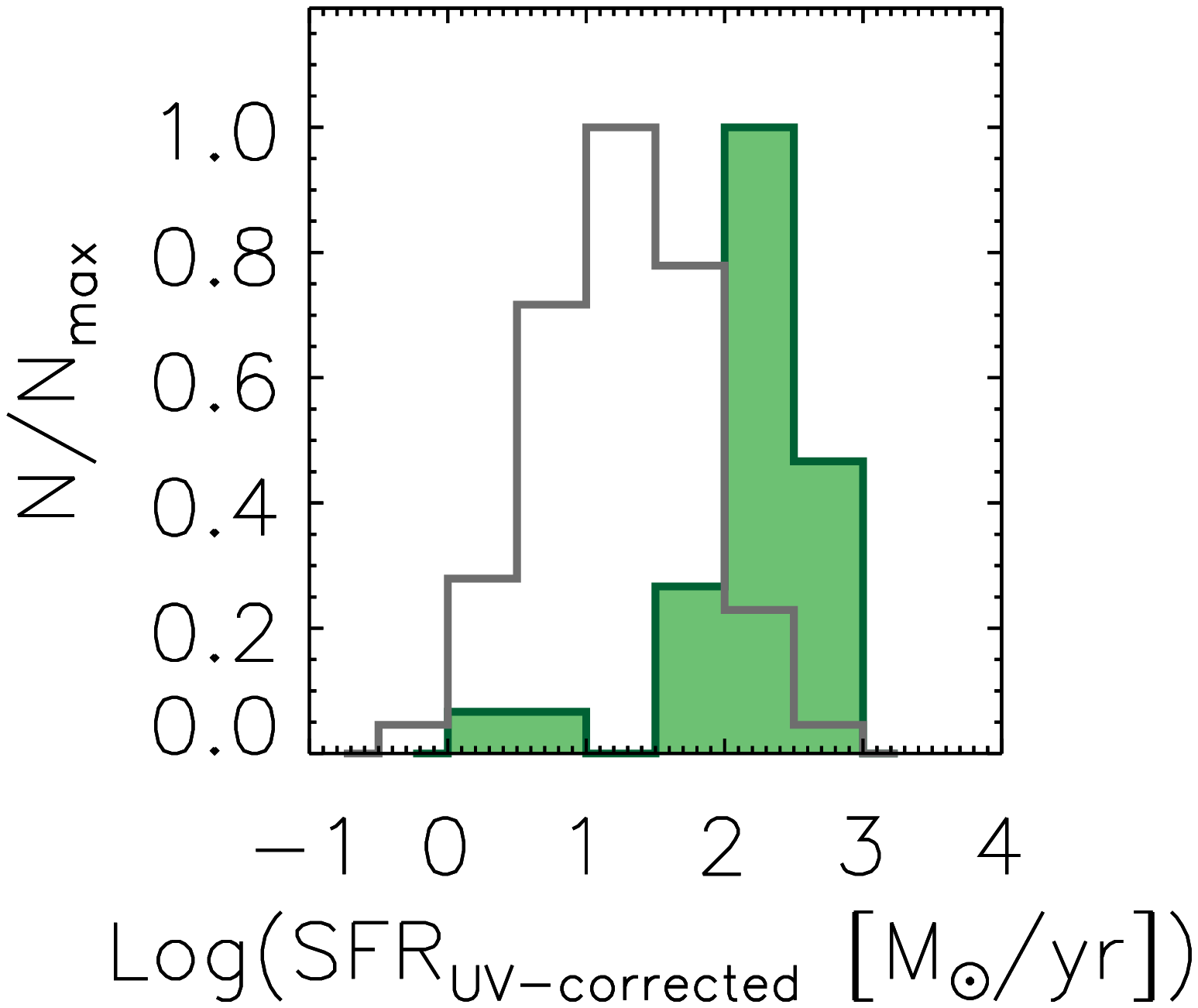}
\includegraphics[width=0.2\textwidth]{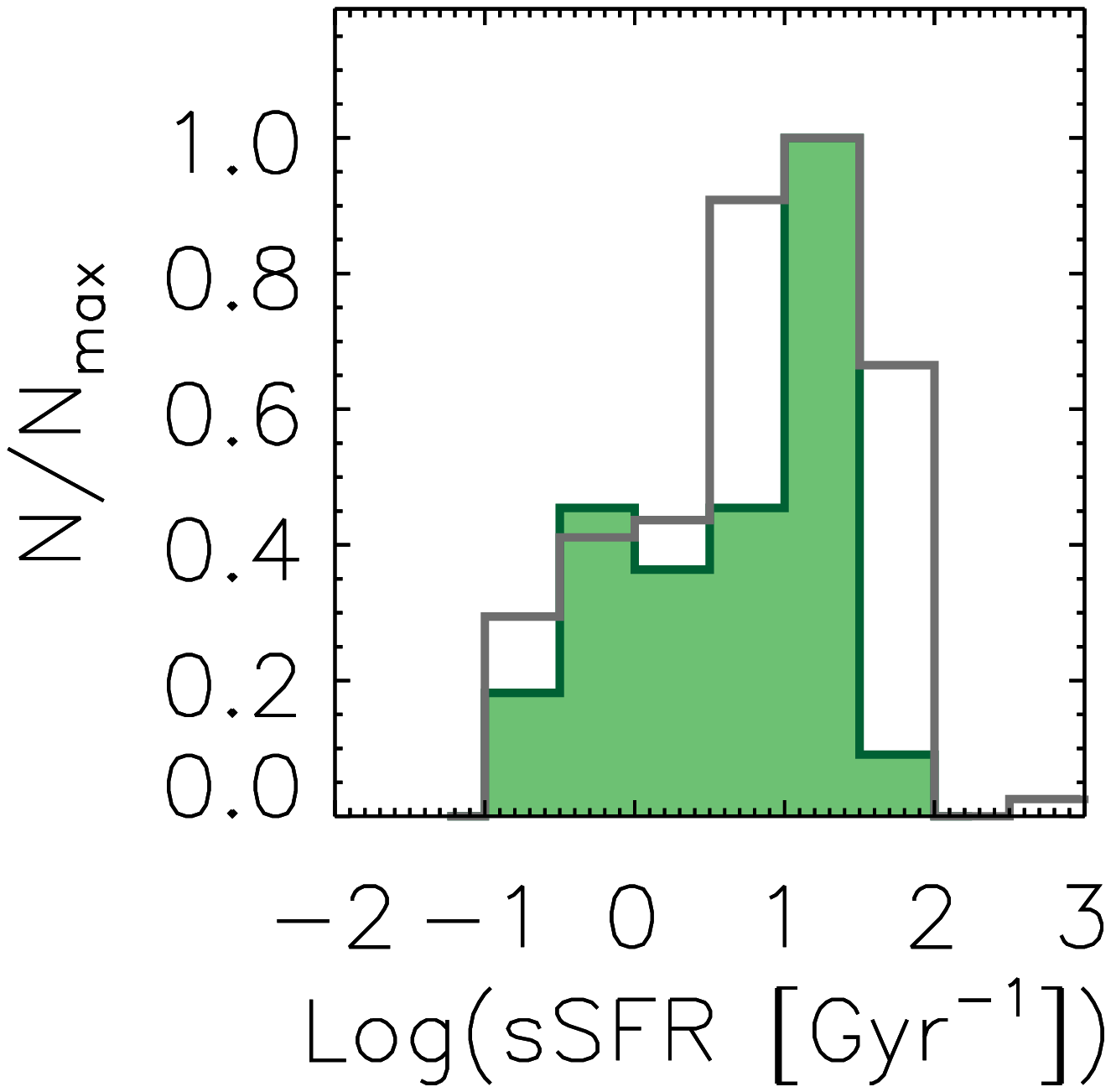}
\includegraphics[width=0.2\textwidth]{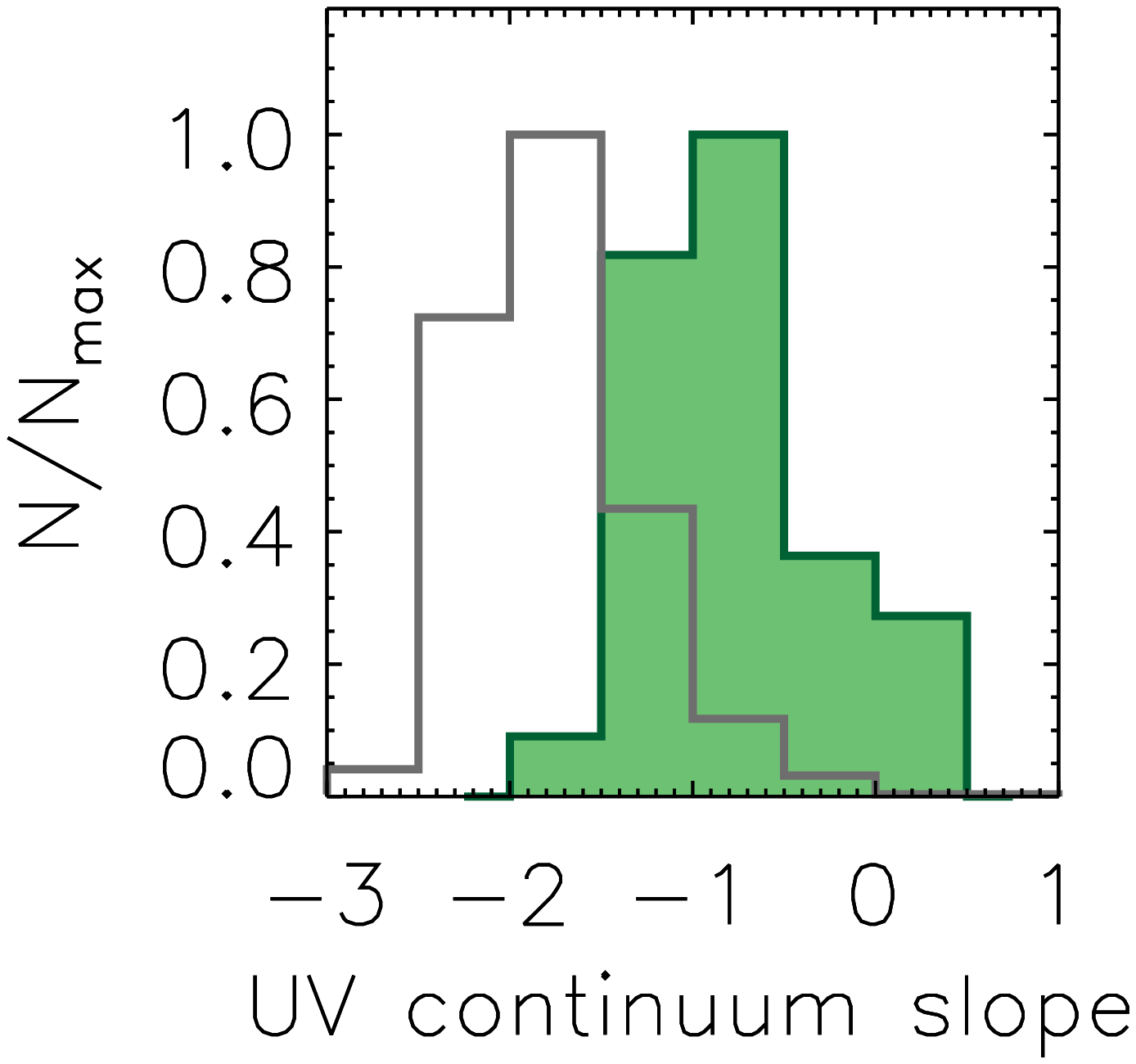}
\caption{Comparison of the best-fitted SED-derived photometric redshift, rest-frame UV luminosity, age, intensity of the Balmer break, dust attenuation, stellar mass, dust-corrected total SFR, specific SFR, and UV continuum slope for the PACS-detected (green shaded histograms) and PACS-undetected (grey histograms) LBGs. Histograms have been normalized to their maxima in order to clarify the representations. The templates used for the derivation of those properties are taken from the BC03 library and are built by considering a time-independent SFH and a constant value of metallicity $Z=0.2Z_\odot$. Only galaxies located in the GOODS-S field are considered in this plot.
              }
\label{properties_PACS}
\end{figure*}

\begin{table}

\caption{\label{PACS_detected} Median values of the physical properties shown in Figure \ref{properties_PACS} for PACS-detected and PACS-undetected LBGs in GOODS-N.}

\centering

\begin{tabular}{lccc}

\hline\hline

Property & PACS-detec. & PACS-undetec. & Prob K-S\\

\hline

$\log{\left(L_{\rm UV}/L_\odot\right)}$	&	10.3	&	10.1	& 0.01\\

Age [Myr]	&	171	&	111 & 0.50	\\

$L_{\rm R} / L_{\rm L}$	&	1.49	&	1.21 & $<$0.001	\\

$E_s(B-V)$	&	0.4	&	0.2	& $<$0.001\\

$\log{\left( M_*/M_\odot \right)}$	&	10.3	&	10.1	& $<$0.001\\

$SFR_{\rm total} \, [M_\odot \, {\rm yr}^{-1}]$	&	139.9	&	17.0     & $<$0.001	\\

sSFR [${\rm Gyr}^{-1}$]	&	6.21	&	9.01 & 0.21	\\

UV slope	&	-0.72	&	-1.81 & $<$0.001	\\

\hline
\end{tabular}
\end{table}

\subsection{Properties of the PACS-detected galaxies}\label{properties_PACS_detected_galaxies}

Figure \ref{properties_PACS} shows the distribution of the best-fitted SED-derived photometric redshift, rest-frame UV luminosities, age, amplitude of the Balmer break, dust attenuation, stellar mass, dust-corrected total SFR, sSFR and UV continuum slope for PACS-detected (green shaded histograms) and PACS-undetected (grey histograms) LBGs. Table \ref{PACS_detected} summarizes the median values of the distributions presented in Figure \ref{properties_PACS}. In order to compare the histograms analytically we have also performed a K-S test for each property, with the results shown in Table \ref{PACS_detected}. PACS-detected LBGs are slightly rest-frame UV-brighter, have a more intense Balmer break, are dustier and more massive, have higher values of the dust-corrected total SFR and exhibit redder UV continuum slopes than those PACS-undetected. A similar behavior is found for the galaxies in the GOODS-S field. The results of the K-S test indicate that those histograms are not driven by the same distribution. According to the K-S test, there is no significant difference in age or sSFR between PACS-detected and PACS-undetected LBGs. This behavior is the one expected because of the PACS limiting fluxes and is similar to some other found at higher and lower redshifts \citep{Oteo2013_z3,Oteo2013_ALHAMBRA_PACS}.

As indicated in Section \ref{SED}, the rest-frame UV luminosities of the galaxies studied here are derived by integrating their best-fitted BC03 templates with a top-hat filter centred in rest-frame 1500\AA. On the IR side, we obtain the total IR luminosities of our PACS-detected galaxies, $L_{\rm IR} [8-1000\, \mu {\rm m}]$, by carrying out SED fits with ZEBRA to their observed IRAC-8.0$\mu$m, MIPS-24$\mu$m, and PACS-100$\mu$m/PACS-160$\mu$m fluxes with \cite{Chary2001} (hereafter CE01) templates. Once the best-fitted templates are found they are shifted to the rest-frame and then integrated between 8 and 1000 $\mu$m. We have also fitted \cite{Dale2002} (hereafter DH02) templates to the FIR SED of our PACS-detected galaxies but, in general, we find better fits with the CE01 templates according to their $\chi^2_r$. Therefore, CE01 templates will be the ones employed throughout the work to obtain the total IR luminosity of the \emph{Herschel}-detected galaxies. Figure \ref{SEDs_SPIRE} shows four examples of the typical UV-to-FIR SEDs of our PACS-detected LBGs. The bolometric luminosity of the PACS-detected galaxies studied here are considered to be the sum of the UV and IR contributions: $L_{\rm bol} = L_{\rm UV} + L_{\rm IR}$. Once we know the rest-frame UV and total IR luminosities of our PACS-detected galaxies, their dust attenuation is obtained by employing the \cite{Buat2005} calibration and their total SFR is obtained with the \cite{Kennicutt1998} relations:

\begin{figure*}
\centering
\includegraphics[width=0.24\textwidth]{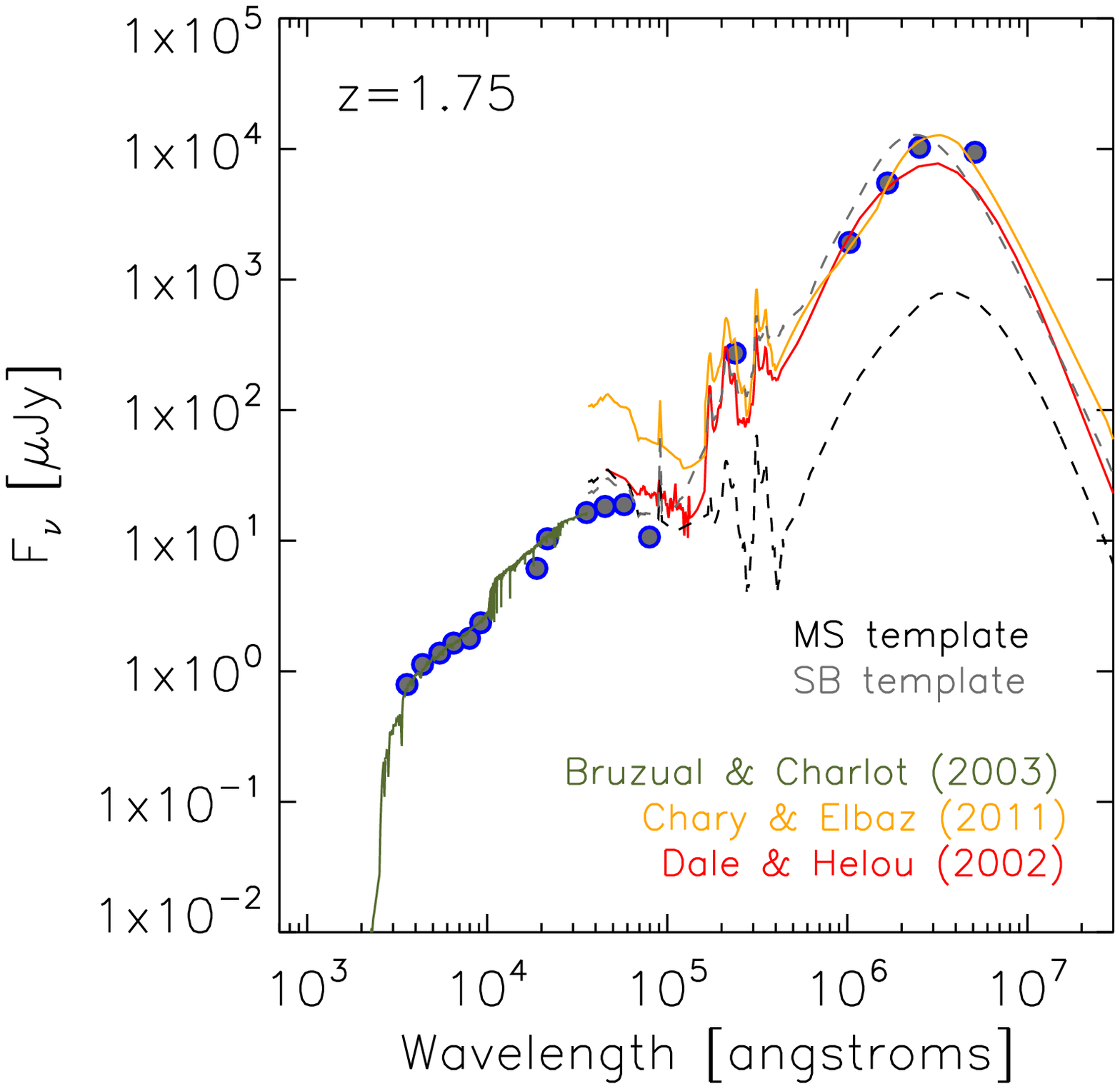}
\includegraphics[width=0.24\textwidth]{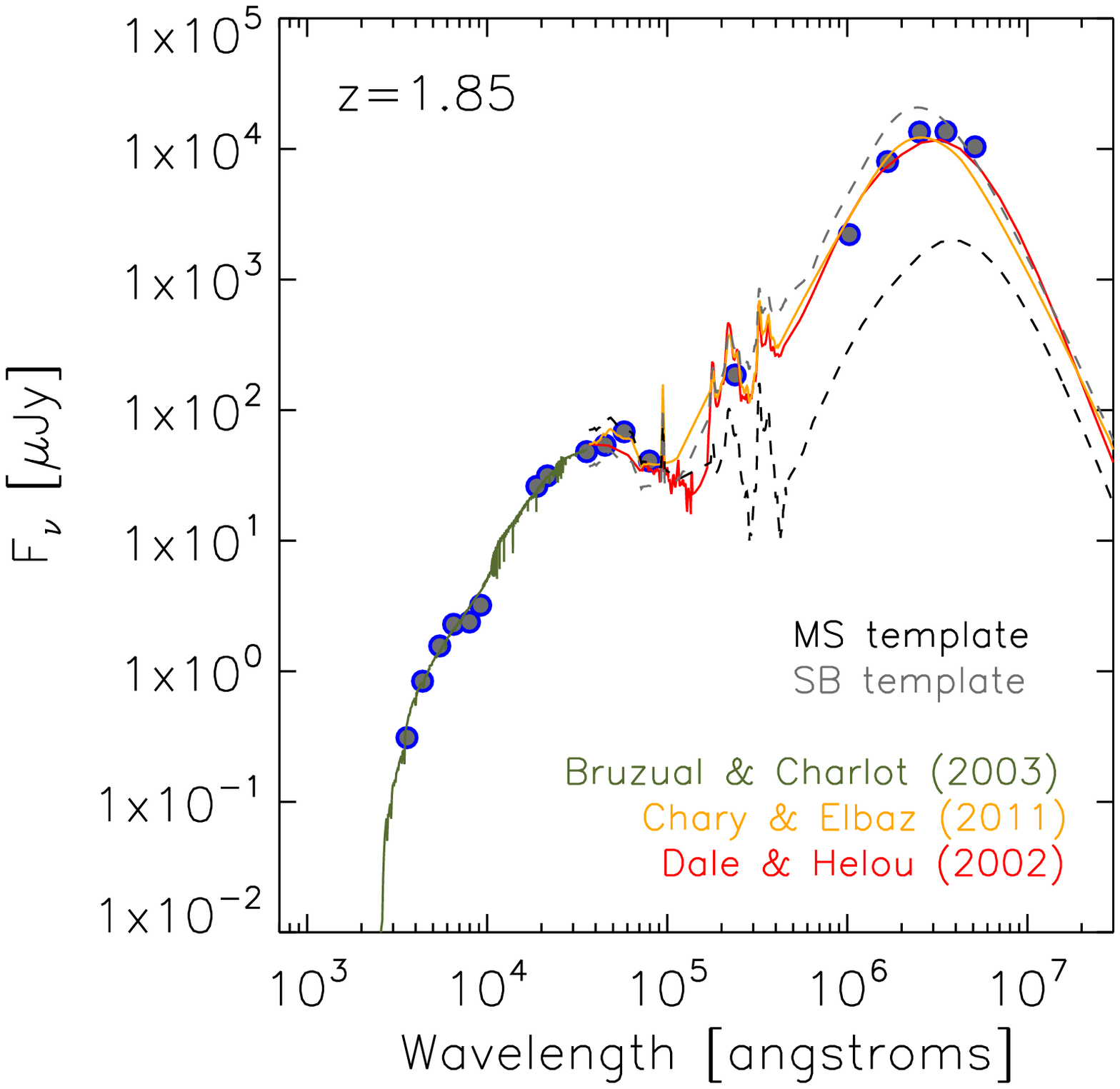}
\includegraphics[width=0.24\textwidth]{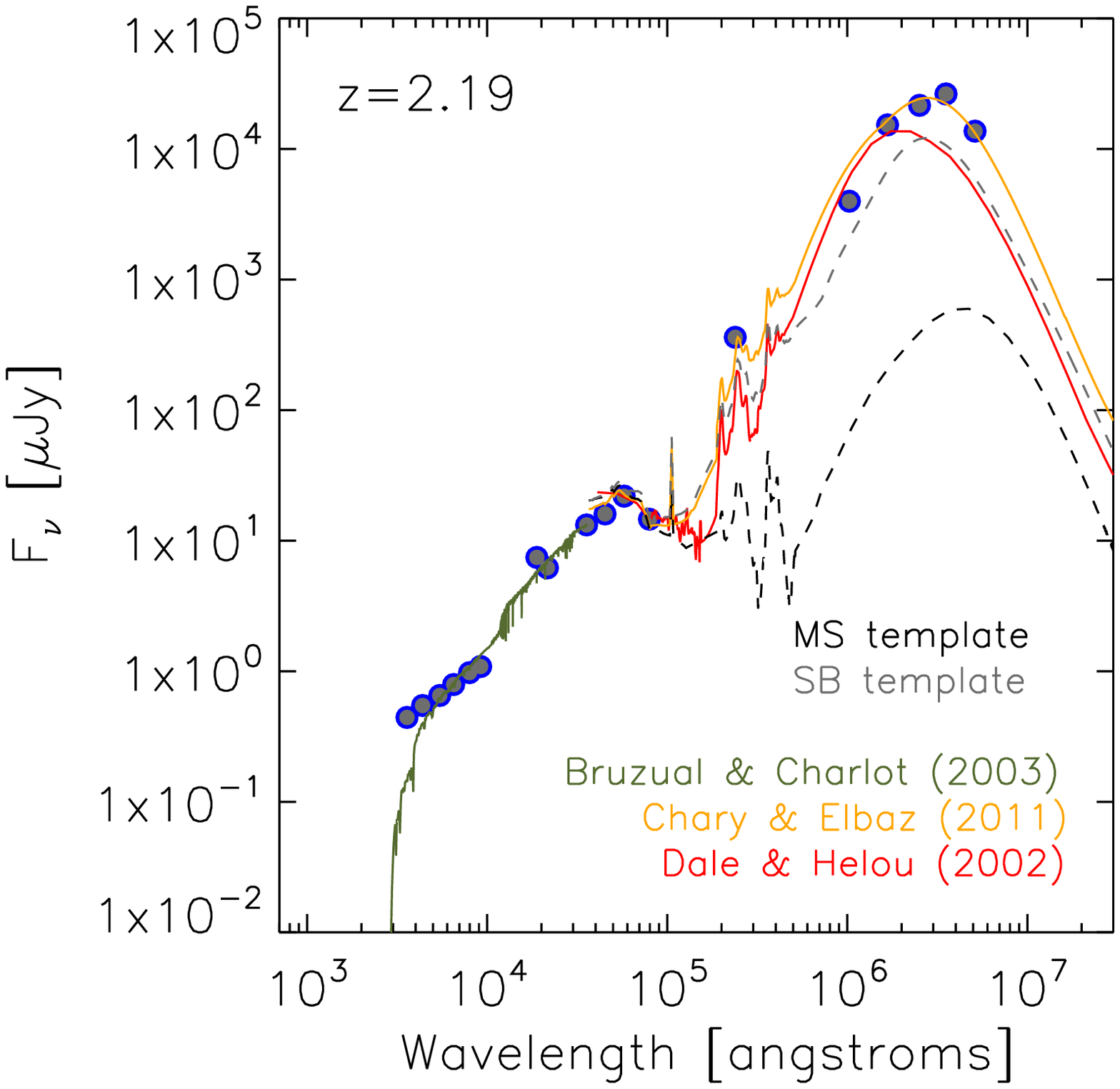}
\includegraphics[width=0.24\textwidth]{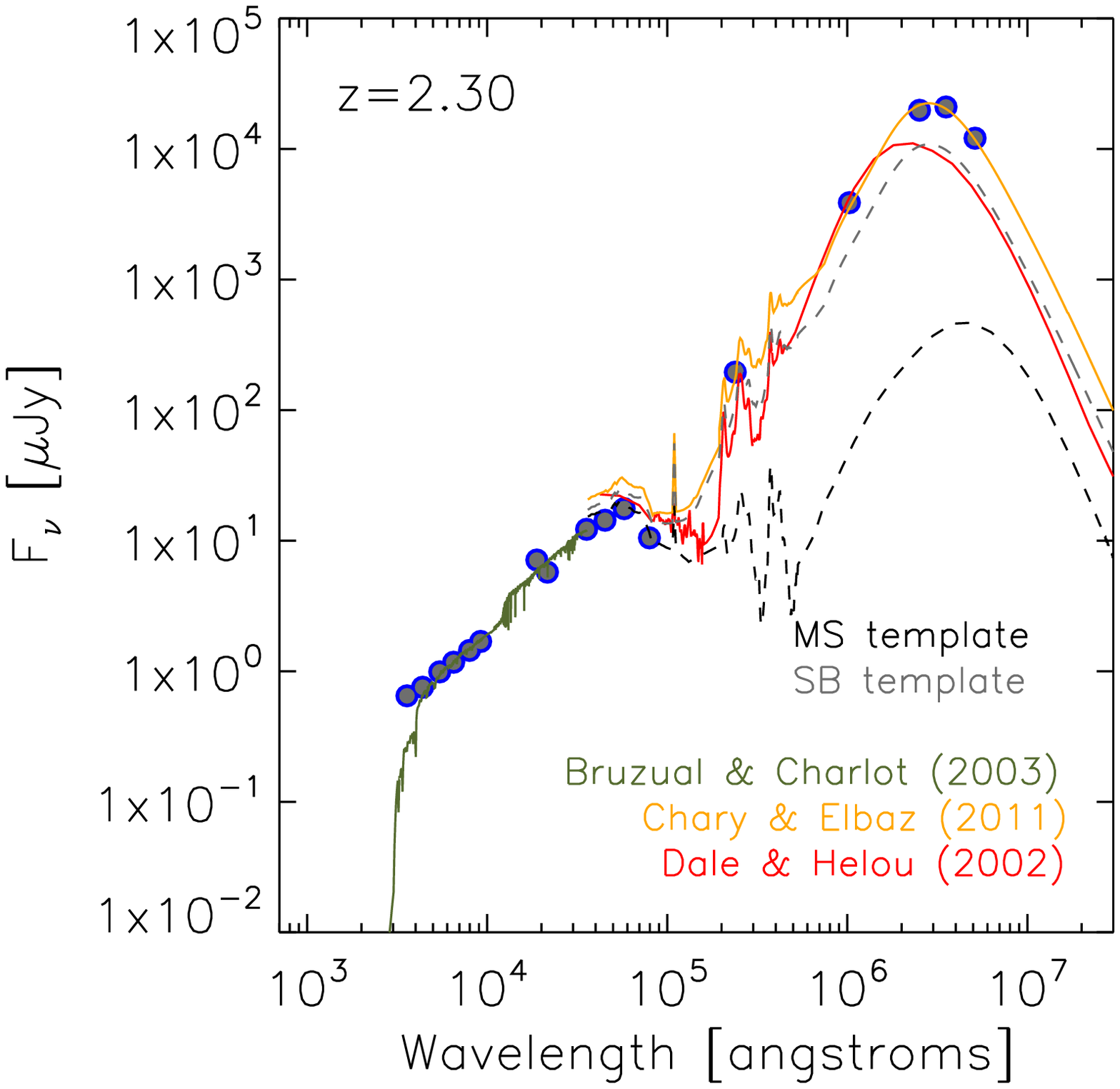}
\caption{UV-to-submm SEDs of the four SPIRE-500 $\mu$m detected LBGs located in the GOODS-N field. Green curves are the best-fitted \citet{Bruzual2003} templates to their GALEX to IRAC-8$\mu$m photometry. Orange and red curves are the best-fitted \citet{Chary2001} and \citet{Dale2002} templates to their MIPS-24$\mu$m-to-SPIRE-500$\mu$m fluxes. The best-fitted average IR SEDs for MS and SB galaxies defined in \citet{Elbaz2011} are also represented. The redshift of each source is indicated in each plot. These SED fits are representative of the whole sample of FIR-detected galaxies studied here.
              }
\label{SEDs_SPIRE}
\end{figure*}

\begin{equation}\label{A_NUV}
A_{FUV} = -0.0333x^3 + 0.3522x^2 + 1.1960x + 0.4967
\end{equation}

\noindent where $x=\log{\left(L_{\rm IR}/L_{\rm UV}\right)}$,

\begin{equation}\label{SFR_UV}
\textrm{SFR}_{\rm UV,uncorrected}[M_{\odot}\textrm{yr}^{-1}] = 1.4 \times 10^{-28}L_{1500}
\end{equation}

\begin{equation}\label{SFR_IR}
\textrm{SFR}_{\rm IR}[M_{\odot}\textrm{yr}^{-1}] = 4.5 \times 10^{-44}L_{\rm IR}
\end{equation}

\begin{equation}\label{SFR_total}
\textrm{SFR}_{\rm total} = \textrm{SFR}_{\rm UV, uncorrected} + \textrm{SFR}_{\rm IR}
\end{equation}

\noindent The calculation of the total SFR assumes that all the light absorbed by dust in the rest-frame UV is re-emitted in the FIR so that the Equation \ref{SFR_total} applies \citep[see for example][]{Magdis2010}.

\begin{figure*}
\centering
\includegraphics[width=0.2\textwidth]{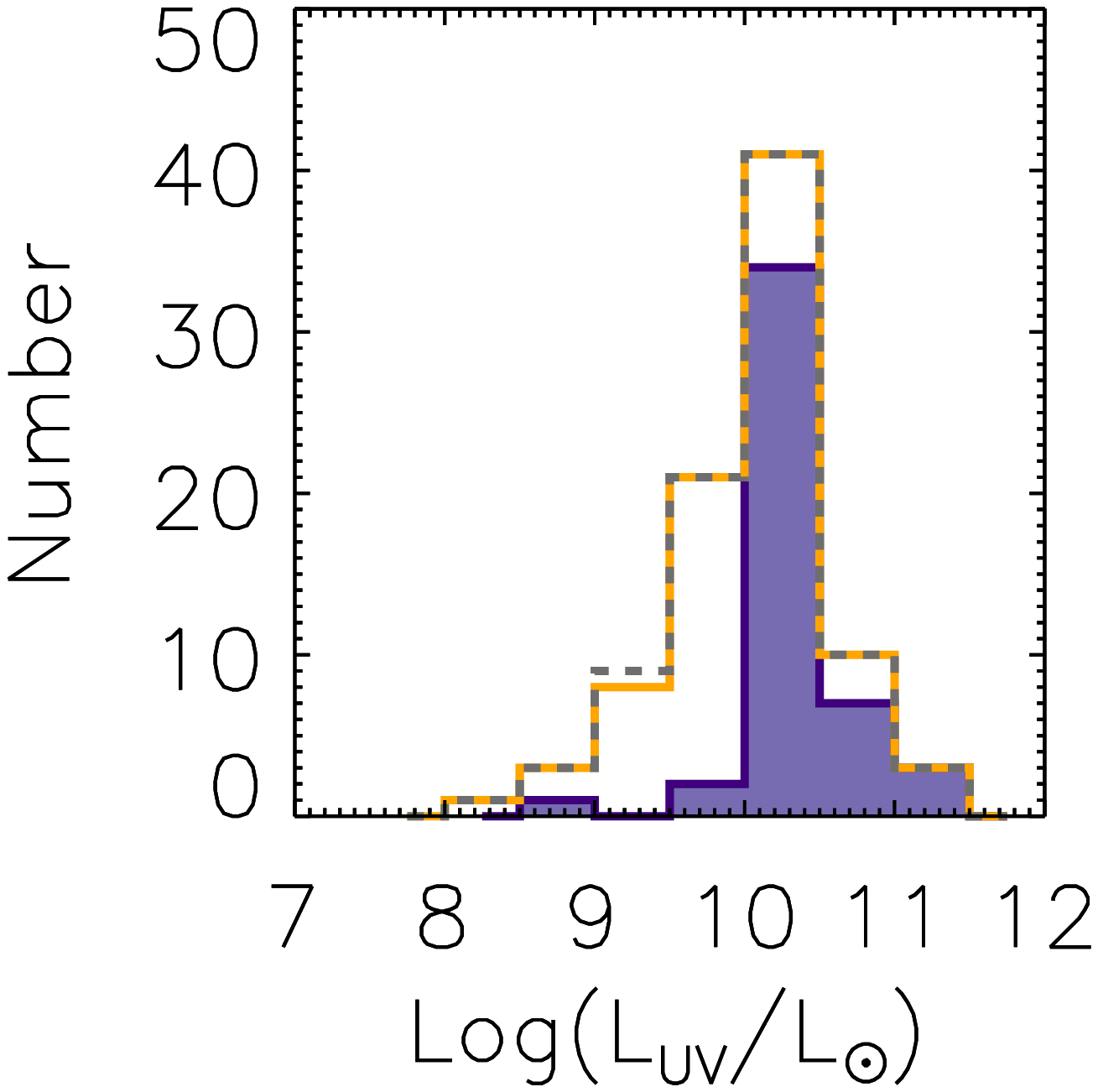}
\includegraphics[width=0.2\textwidth]{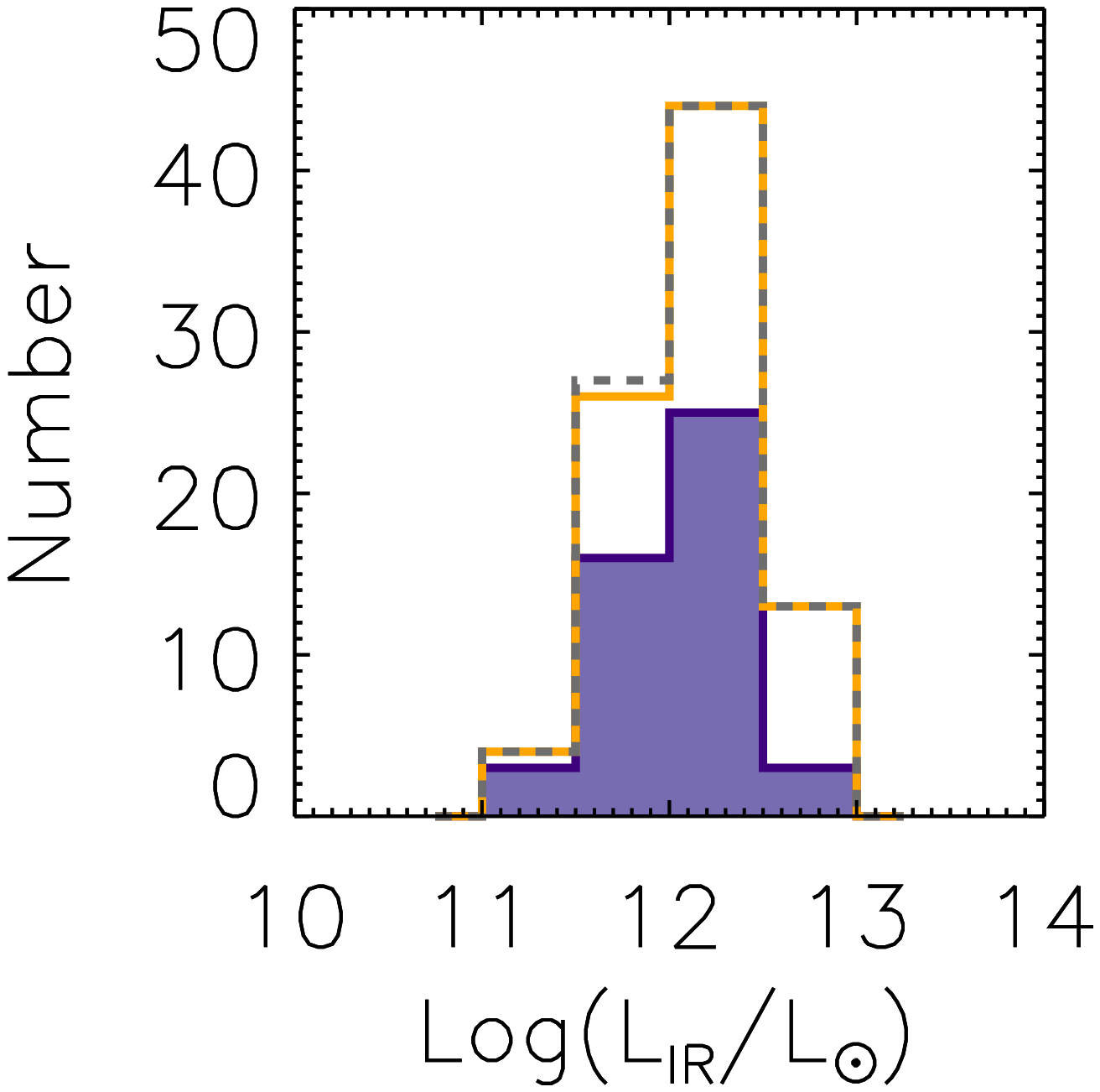}
\includegraphics[width=0.2\textwidth]{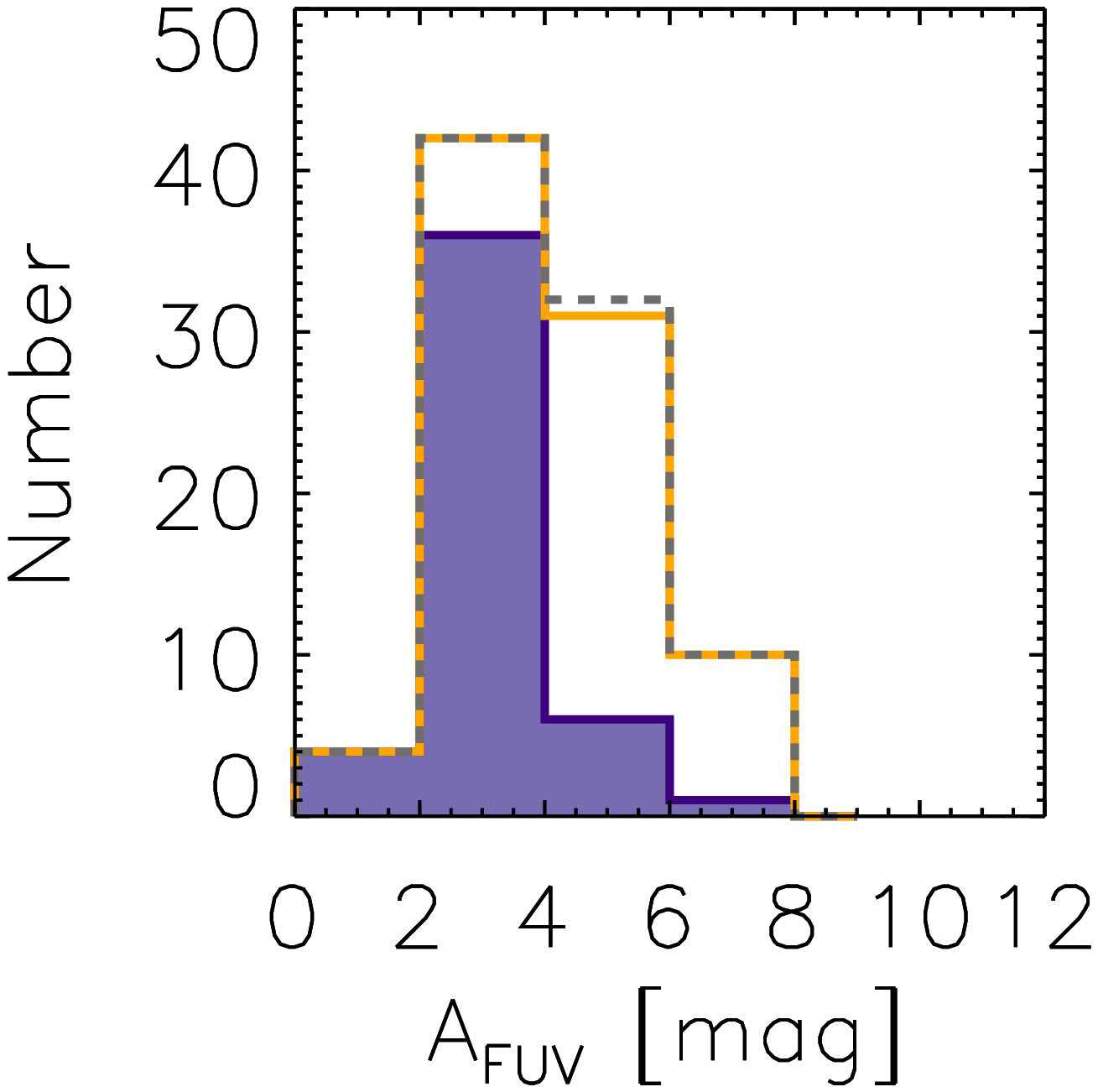} \\
\includegraphics[width=0.2\textwidth]{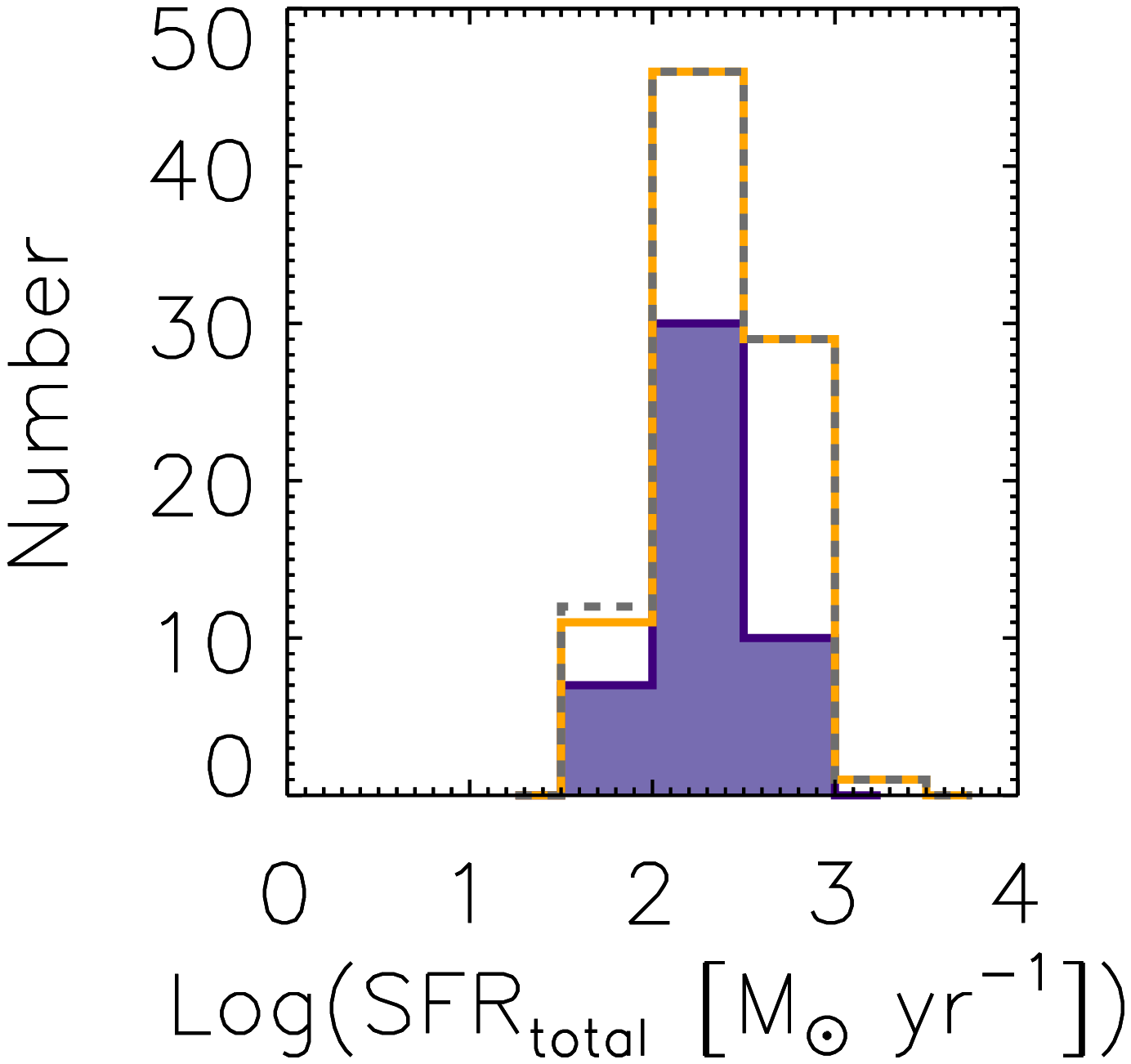}
\includegraphics[width=0.2\textwidth]{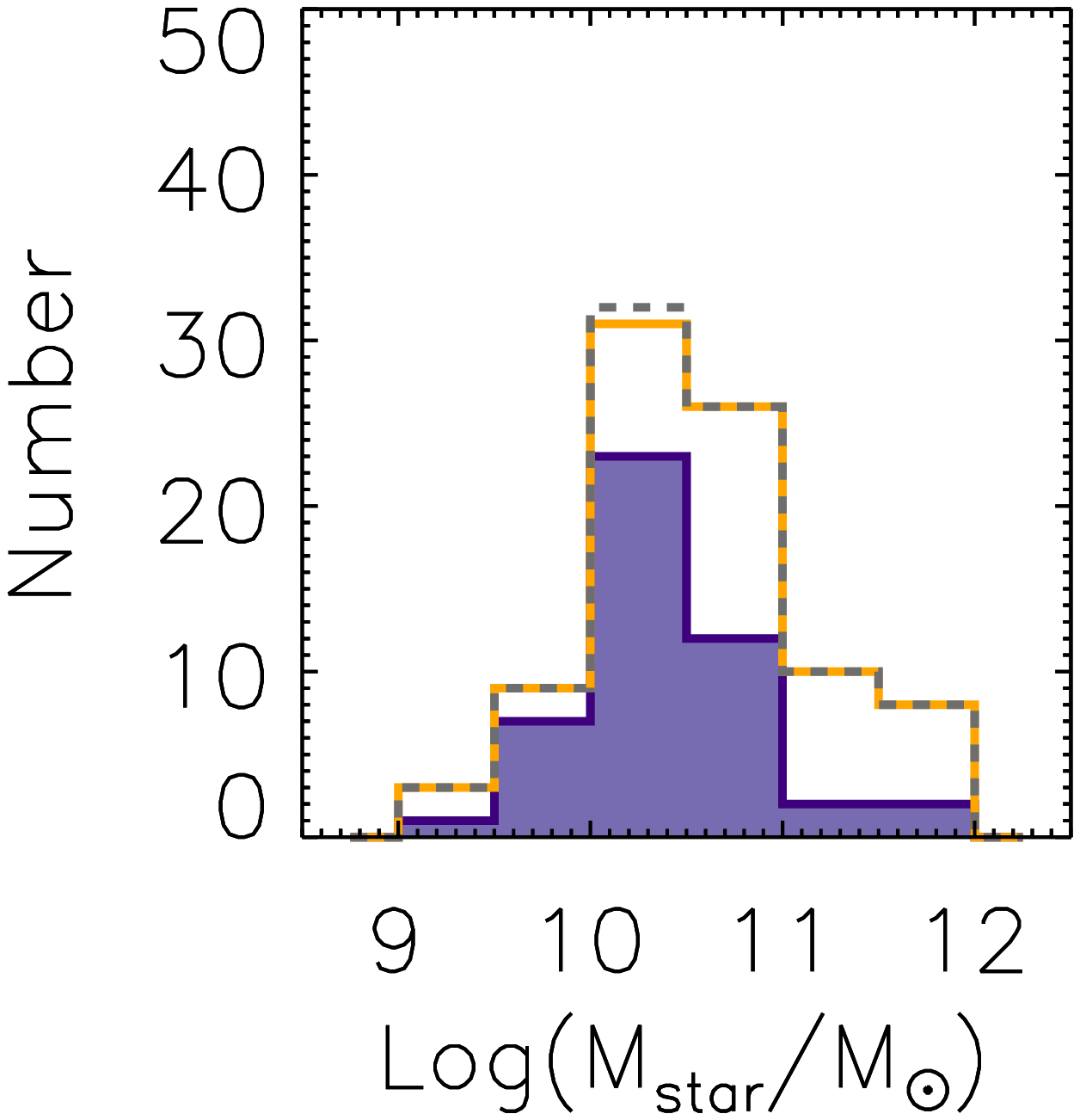}
\includegraphics[width=0.2\textwidth]{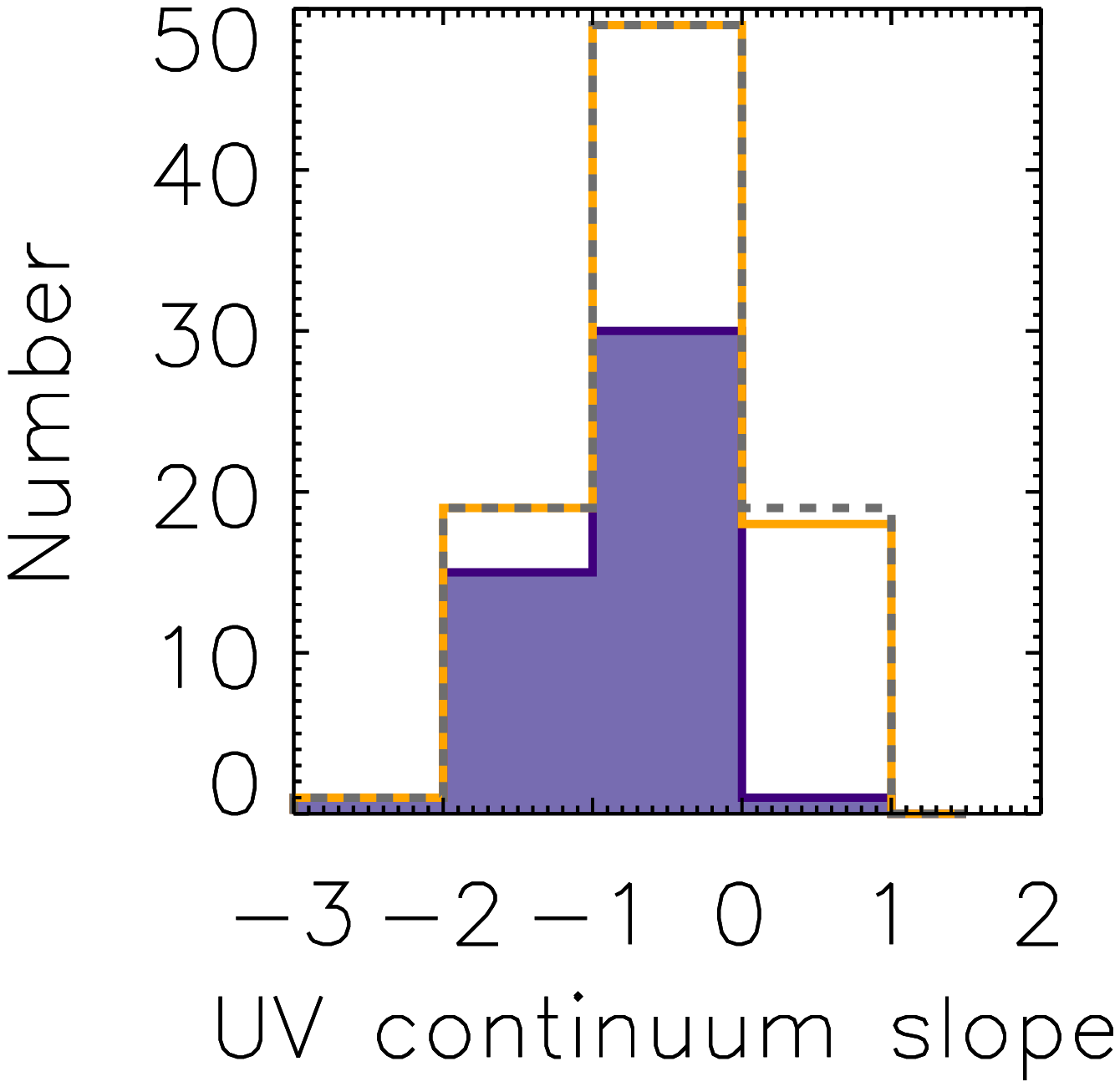}
\caption{Distributions of the rest-frame UV luminosity, total IR luminosity, dust attenuation, total SFR, stellar mass and UV continuum slope for our PACS-detected SF galaxies at $1.5 \lesssim z \lesssim 2.5$: LBGs with shaded purple histograms, $sBzK$ with solid empty orange histograms and UV-selected galaxies with grey empty dashed histograms. The dust attenuation has been obtained with the \citet{Buat2005} calibration and the total SFR from the \citet{Kennicutt1998} relations assuming $SFR_{\rm total} = SFR_{\rm UV} + SFR_{\rm IR}$.
              }
\label{PACS_comparison}
\end{figure*}

\cite{Elbaz2011} produced the average IR SED of MS and SB galaxies. They found that the typical MS IR SED has a broad dust emission peak centred around 90 $\mu$m and strong PAH features in emission. In contrast, the typical SB IR SED has a narrower dust emission peak and weak PAH emission. The broader dust emission peak of MS galaxies indicates a wider range of dust temperature than for SB galaxies. They also found that the average SB IR SED is very similar to the CE01 template associated with $L_{\rm IR} = 6 \times 10^{11} L_\odot$ and the average MS IR SED is closer to the CE01 template associated with $L_{\rm IR} = 4 \times 10^9 L_\odot$. We have studied which of the two SEDs (MS or SB) better  represents the IR SED of our PACS-detected galaxies. To this aim, we have rerun ZEBRA for each of the two templates and retain the values of the $\chi^2$ of the SED-fitting results. According to the $\chi^2_r$ values, we find that the SB IR SED is clearly a better representation of the typical IR SED of our PACS-detected galaxies than the MS IR SED. This is because for a given 100 $\mu$m fluxes, the SB IR SED predicts lower values of IRAC fluxes than the MS IR SED and, consequently, the MS IR SED is not able to reproduce the PACS-to-IRAC colours for our PACS-detected galaxies. In Figure \ref{SEDs_SPIRE} we over-plot the average MS and SB IR SEDs for each galaxy normalized to best-fit their observed FIR fluxes. It can be seen that in all cases the SB IR SED is better fitted than the MS one.

Figure \ref{PACS_comparison} shows the distributions of the rest-frame UV and total IR luminosities, dust attenuation, total SFR, stellar mass and UV continuum slope for our PACS-detected LBGs in GOODS-S and GOODS-N. The median values of the distributions of these parameters are shown in Table \ref{FIR_properties}. PACS-detected LBGs at $1.5 \lesssim z \lesssim 2.5$ have total IR luminosities which place them in the LIRG/ULIRG classes. It should be noted that we do not recover too many galaxies with $\log{\left( L_{\rm IR}/L_\odot \right)} <11.5$ owing to the depth of the PACS data employed. The only SPIRE-detected LBG at $1.5 \lesssim z \lesssim 2.5$ in \cite{Burgarella2011} also have a ULIRG nature. Oteo et al. (2013b) obtained the total IR luminosity of a sample of PACS-detected LBGs at $z \sim 1$ and found that all of them belong to the LIRG class. In that case, no PACS-detected LBGs at $z \sim 1$ with $\log{\left( L_{\rm IR}/L_\odot \right)} <11$ w
ere found, also as a consequence of the depth of the PACS data used. However, no PACS-detected LBGs with $\log{\left( L_{\rm IR}/L_\odot \right)} \geq 12$ at $z \sim 1$ were found either, where the observations are complete. The SPIRE-detected LBGs of \cite{Burgarella2011} at $0.8 \lesssim z \lesssim 1.2$ also have total IR luminosities compatible with their being LIRGs, with no ULIRG contribution. Therefore, comparing the results of Oteo et al (2013b) and \cite{Burgarella2011} with those found here, we conclude that at $1.5 \lesssim z \lesssim 2.5$ there is a population of  very dusty and red ULIRG-LBGs that has not been found at lower redshifts. This might represent an evolution of LBGs with redshift, at least in the brightest IR side. The median value of the UV+IR-derived total SFR for LBGs at $1.5 \lesssim z \lesssim 2.5$ is $SFR_{\rm total} = 249.9 M_\odot \, {\rm yr}^{-1}$, i.e.\ they are extreme SF galaxies. As a consequence of the difference in the total IR luminosity between LBGs at $z \sim 1$ and $z \sim 2$, the total SFR of the PACS-detected LBGs at $z \sim 2$ is much higher than those found at $z \sim 1$, suggesting an evolution in the tail of high values of the total SFR.

\begin{table*}
\caption{\label{FIR_properties}Median values of the magnitudes represented in Figure \ref{PACS_comparison} for each kind of galaxy.}
\centering
\begin{tabular}{|l|cc|ccc|}
\hline\hline

Property & LBGs  & $sBzK$  &  UV-sel \\

\hline

$\log{\left(L_{\rm UV}/L_\odot\right)}$		&	10.3		&	9.9 		& 	10.4	 		\\

$\log{\left(L_{\rm IR}/L_\odot\right)}$			&	12.1		&	12.3		& 	12.0			\\

$A_{\rm FUV}$ [mag]					&	3.7		&	5.29		&	3.4			\\

$SFR_{\rm total} \, [M_\odot \, {\rm yr^{-1}}]$	&	249.9	&	319.4	&	189.1		\\

$\log{\left(M_*/M_\odot\right)}$				&	10.3		&	10.6		&	10.3			\\

UV slope								&	-0.55		&	-0.26		&	-0.84				\\

\hline
\end{tabular}
\end{table*}

In Figure \ref{PACS_comparison} we also show the distributions for the studied PACS-detected $sBzK$ and PACS-detected UV-selected galaxies. The median values of the distribution of these parameters are also shown in Table \ref{FIR_properties}. PACS-detected $sBzK$ galaxies have a wider range of rest-frame UV luminosity reaching fainter values, have higher dust attenuation, are more massive, and have a more intense Balmer break than PACS-detected LBGs. This is a direct consequence of their difference in the rest-frame UV luminosity and the similarity in their total IR luminosity. The $BzK$ criterion employs a colour combination which is insensitive to dust attenuation, as  was shown in \cite{Daddi2004}; therefore, this technique is efficient at finding galaxies at $1.5 \lesssim z \lesssim 2.5$ more independently of their dust attenuation. However, the drop-out technique selects galaxies according to their rest-frame UV colours, and this implies that they are typically UV-bright galaxies with less obscured star formation. The ULIRG nature of some $sBzK$ galaxies has  already been reported in, for example, \cite{Daddi2005}. It should be noted that the distributions for $sBzK$ and UV-selected galaxies are very similar since only two UV-selected galaxies are not selected as $sBzK$ (see later in the text).

\begin{figure}
\centering
\includegraphics[width=0.45\textwidth]{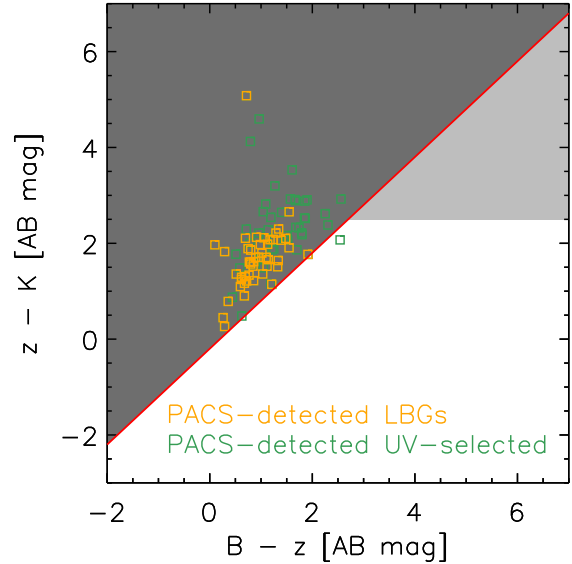}
\caption{Similar to Figure \ref{bzk_color}, but only PACS-detected LBGs and PACS-detected UV-selected galaxies are shown. We consider in this plot PACS-detected galaxies located in GOODS-N and GOODS-S.
              }
\label{bzk_color_PACS}
\end{figure}

\begin{figure}
\centering
\includegraphics[width=0.45\textwidth]{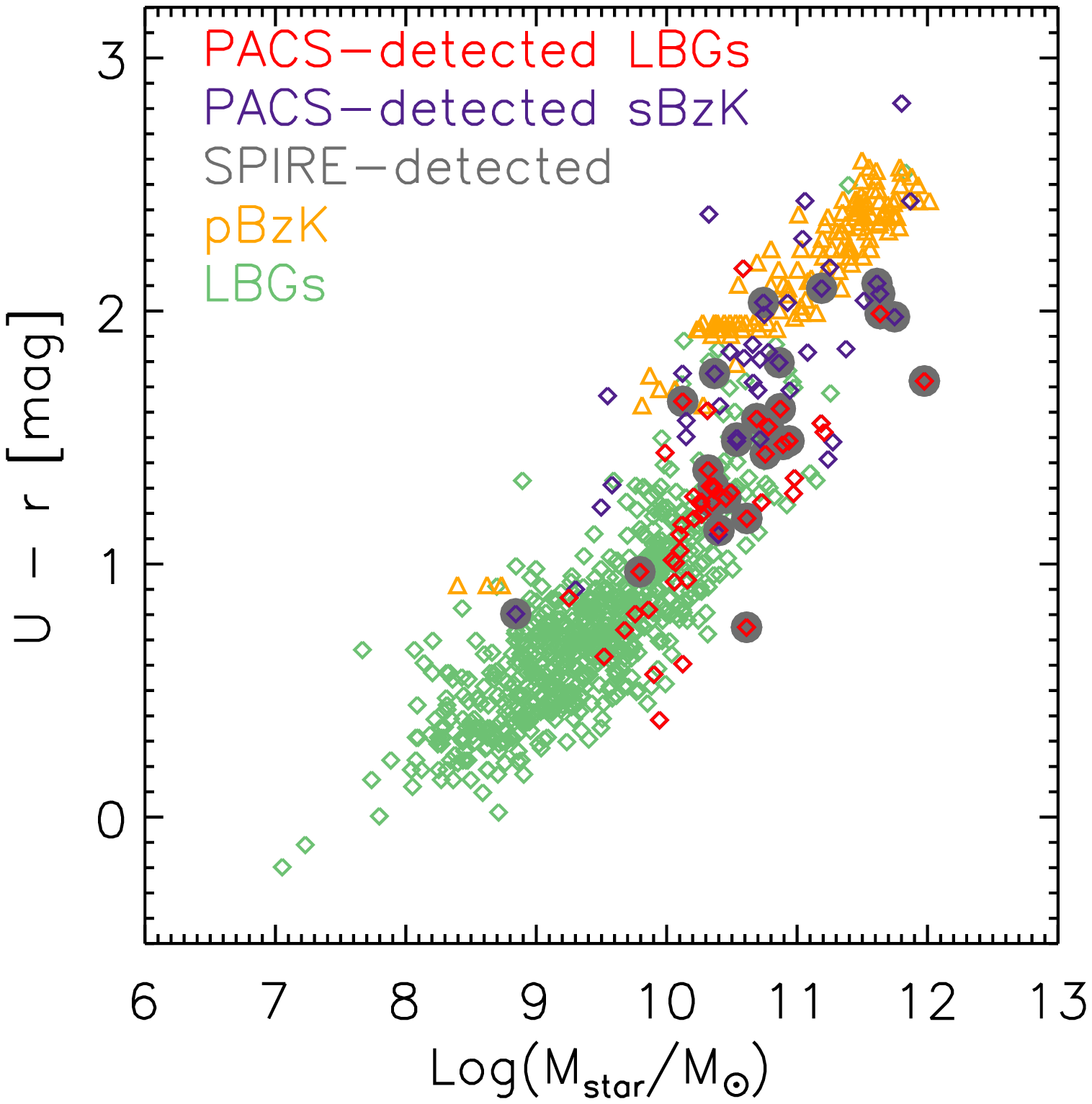}
\caption{Location of our PACS-detected galaxies (LBGs with red diamonds), $sBzK$ galaxies with purple diamonds, and UV-selected galaxies with grey diamonds) in a colour versus stellar mass diagram. SPIRE-detected sources (in any of the 250$\mu$m,  350$\mu$m, or 500$\mu$m bands) are indicated with large filled grey dots. For comparison, we also represent the whole population of LBGs segregated in the GOODS-S field (see Section \ref{lbgs}) regardless of their selection in PACS (green triangles). Additionally, with the aim of showing the location of the red sequence at $1.5 \lesssim z \lesssim 2.5$ in the diagram, we represent by orange triangles the sample of $pBzK$ galaxies shown in Figure \ref{color_color} (see Section \ref{bzk}).
              }
\label{color_color_PACS}
\end{figure}

In Section \ref{comparison} we studied the overlap between the galaxies selected through the Lyman break and $BzK$ criteria. We found that, even though most LBGs could have also been selected as $sBzK$ galaxies, only 25\% of the $BzK$ galaxies are also LBGs. Furthermore, compared to the UV-selected galaxies, we found that most of the UV-selected galaxies could also  have been selected as $BzK$ galaxies, but not many are LBGs. We now analyse the overlap between the samples of LBGs and $sBzK$ and UV-selected galaxies but only for those PACS-detected. In Figure \ref{bzk_color_PACS} we represent the colour-colour diagram utilized to look for $BzK$ galaxies (similar to that shown in Figure \ref{bzk_color}). We plot PACS-detected LBGs and PACS-detected UV-selected galaxies with orange and green open squares, respectively. It can be seen that there is almost a complete overlap between the samples of our PACS-detected galaxies (see also Table \ref{over}). Most PACS-detected LBGs and PACS-detected UV-selected galaxies are within the selection windows employed to look for $sBzK$ and   can thus be also selected as $sBzK$ galaxies. Therefore, according to Figures \ref{PACS_comparison} and \ref{bzk_color_PACS}, PACS-detected LBGs are a subsample of PACS-detected $sBzK$ galaxies with the highest values of rest-frame UV luminosity and lower amplitude of the Balmer break, dust attenuation and stellar mass. However, about 45\% of the PACS-detected $sBzK$ galaxies are not LBGs owing to their faintness in the UV continuum (see the different distributions of the rest-frame UV luminosity of PACS-detected LBGs and PACS-detected $sBzK$ galaxies in Figure \ref{PACS_comparison}). There are only two PACS-detected UV-selected galaxies that would have not been selected as $sBzK$ owing to their blue $z-K$ colour. Most UV-selected galaxies located below the $sBzK$ selection window in Figure \ref{bzk_color} have been missed when considering only PACS-detected galaxies since PACS selects the most massive sources and therefore with the brightest $K$ fluxes. All the PACS-detected $sBzK$ galaxies are detected in the B-band; therefore, according to the selection criterion of the UV-selected galaxies, all PACS-detected $sBzK$ galaxies are contained in the sample of PACS-detected UV-selected galaxies. As a consequence of all of this, PACS-detected $sBzK$ galaxies are perfectly suitable for studying the FIR emission of the general population of FIR-bright SF galaxies at $1.5 \lesssim z \lesssim 2.5$. This is not the case for LBGs since they are biased towards UV-brighter, less dusty and bluer galaxies.

\begin{table*}
\caption{\label{over} Number of PACS-detected galaxies in each field for each kind of SF galaxy and percentages of overlapping between the different samples.}
\centering
\begin{tabular}{ccc|ccc}
\hline\hline

 & GOODS-S & & & GOODS-N & \\

 \hline

LBGs & $sBzK$ & UV-sel & LBGs & $sBzK$ & UV-sel \\

9 & 28 & 28 & 39 & 61 & 63 \\

100\% are $sBzK$ & 39\% are LBGs & 39\% are LBGs & 97\% are $sBzK$ & 59\% are LBGs & 59\% are LBGs \\

 & & 100\% are $sBzK$ & &  & 96\% are $sBzK$ \\

\hline
\end{tabular}
\end{table*}

In Section \ref{CMD} we showed that most LBGs at $1.5 \lesssim z \lesssim 2.5$ are located in the blue cloud of galaxies at their redshift, and only the dustiest and oldest are shifted to the green valley or red sequence. According to this behaviour as a function of dust attenuation, it would be expected that PACS-detected LBGs are located in the green valley or the red sequence. To check this expectation we plot in Figure \ref{color_color_PACS} the location of our PACS-detected LBGs at $1.5 \lesssim z \lesssim 2.5$ in the same colour versus stellar mass plane as the shown in Figure \ref{color_color}. We also represent the location of PACS-undetected LBGs with green triangles and the location of $pBzK$ galaxies with orange triangles to guide the eye about the location of the red sequence of galaxies. It can be seen that, whereas PACS-undetected LBGs are located in the blue cloud, most PACS-detected LBGs have redder values of the $i-K$ colour, indicating that they are mostly located in the green valley. This behaviour is similar to that found in Oteo et al. (2013b) for PACS-detected LBGs at $z \sim 1$.

\begin{figure}
\centering
\includegraphics[width=0.45\textwidth]{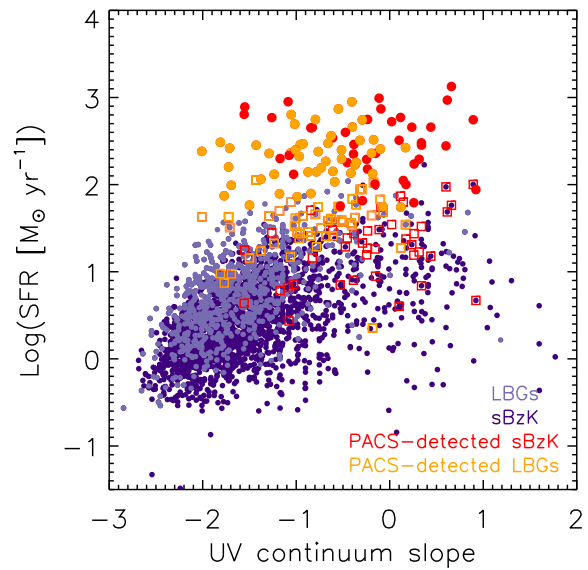}
\caption{Relation between the total SFR and the UV continuum slope for PACS-detected LBGs and $sBzK$ galaxies (orange and red symbols) and for the whole population of LBGs and $sBzK$ regardless their detection in PACS. For PACS-detected sources, we have distinguished the total SFR estimated from the dust correction factors obtained with the \citet{Takeuchi2012} IRX-$\beta$ relation (open symbols) and the total SFR derived with direct UV and IR detections, ${\rm SFR_{\rm total} = SFR_{\rm UV} + SFR_{\rm IR}}$. For the whole population of LBGs and $sBzK$ galaxies, the total SFR is estimated with the \citet{Takeuchi2012} IRX-$\beta$ relation.              }
\label{sfr_beta_PACS}
\end{figure}

Figure \ref{sfr_beta_PACS} shows the relation between the dust-corrected total SFR and the UV continuum slope for the whole population of LBGs and $sBzK$ galaxies and for those PACS-detected. The dust-corrected total SFR of the whole population of LBGs and $sBzK$ galaxies has been derived with the dust attenuation factor obtained with the \cite{Takeuchi2012} IRX-$\beta$ relation. For PACS-detected galaxies we distinguish the total SFR estimated through the \cite{Takeuchi2012} IRX-$\beta$ relation (open symbols) and the total SFR obtained with direct detections in the UV and IR, ${\rm SFR_{\rm total} = SFR_{\rm UV} + SFR_{\rm IR}}$ (filled symbols). It can be seen that, as a general trend, redder galaxies in the UV continuum have a higher total SFR. The location of the PACS-detected galaxies whose total SFR has been obtained with the \cite{Takeuchi2012} dust correction indicates that they are separated from the cloud where most points (the bulk of the galaxies) are located. This indicates that PACS-detected galaxies, as well as representing a low percentage of the whole population of SF galaxies in the redshift desert, also have properties/trends that are not similar to the rest of the population of SF galaxies. 

Figure \ref{sfr_beta_PACS} indicates that the total SFR obtained with the \cite{Takeuchi2012} IRX-$\beta$ relation is quite underestimated since, for a given UV continuum slope, the values are much lower than those derived with the more accurate ${\rm SFR_{\rm total} = SFR_{\rm UV} + SFR_{\rm IR}}$ determinations. The underestimation holds when applying the \cite{Meurer1999} relation for recovering the total SFR. This suggests that the total SFR of PACS-detected galaxies at $z \sim 2$ cannot be recovered from UV/optical-based methods (see also Section \ref{irxbpacs}). This is similar to the results found in \cite{Oteo2013_z3}, where it is reported that the total SFR of PACS-detected LBGs at $z \sim 3$ cannot be recovered with the dust attenuation derived from SED-fitting with BC03 templates, or with the dust correction factors obtained from the \cite{Meurer1999} or \cite{Takeuchi2012} IRX-$\beta$ relations. This highlights the importance of using individual FIR detections to constrain the FIR SED (total SFR and dust attenuation, for example) of high-redshift IR-bright galaxies.

\subsection{Relation between dust attenuation and UV continuum slope at $z \sim 2$}\label{irxbpacs}

\begin{figure*}
\centering
\includegraphics[width=0.45\textwidth]{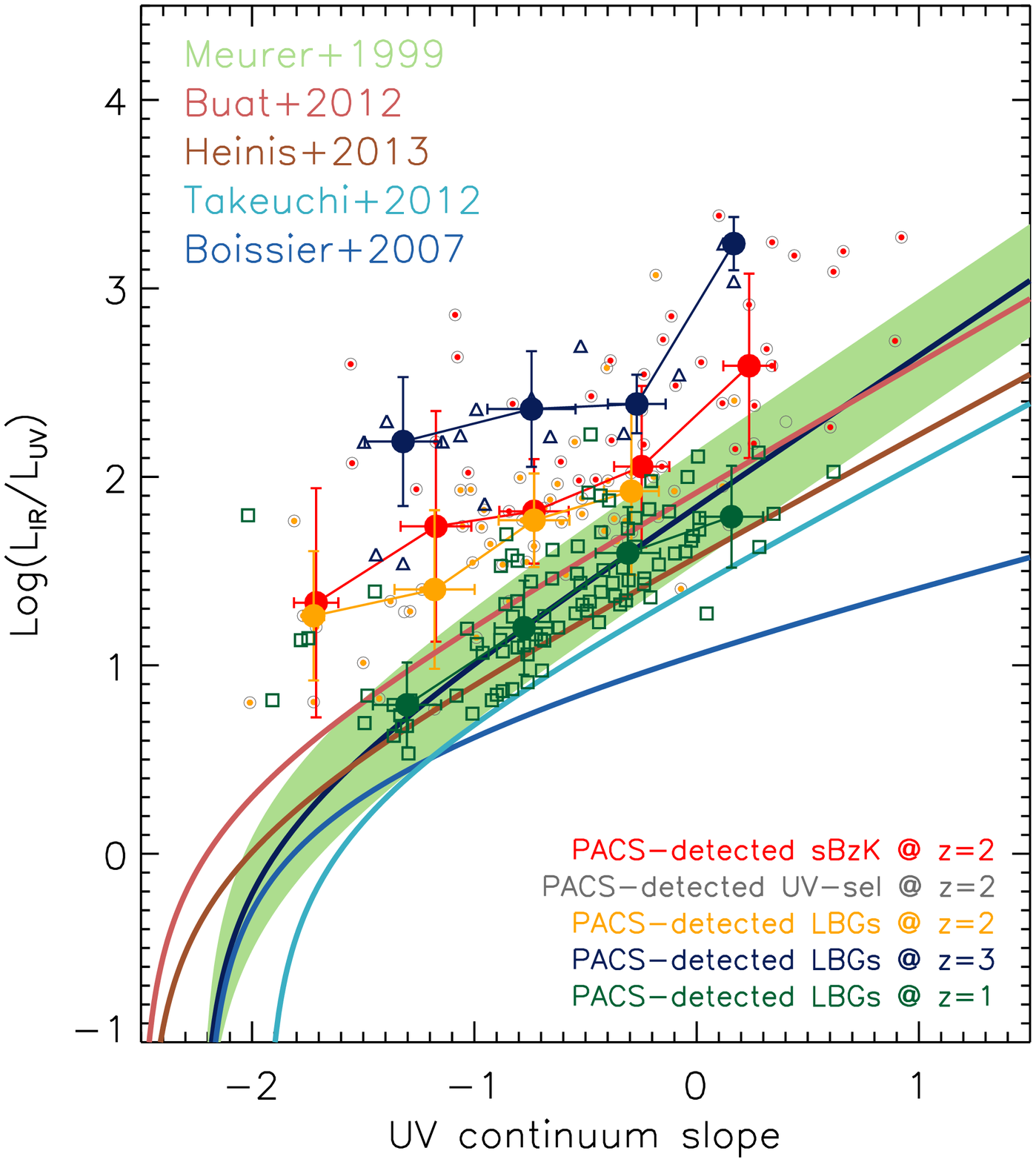}
\includegraphics[width=0.45\textwidth]{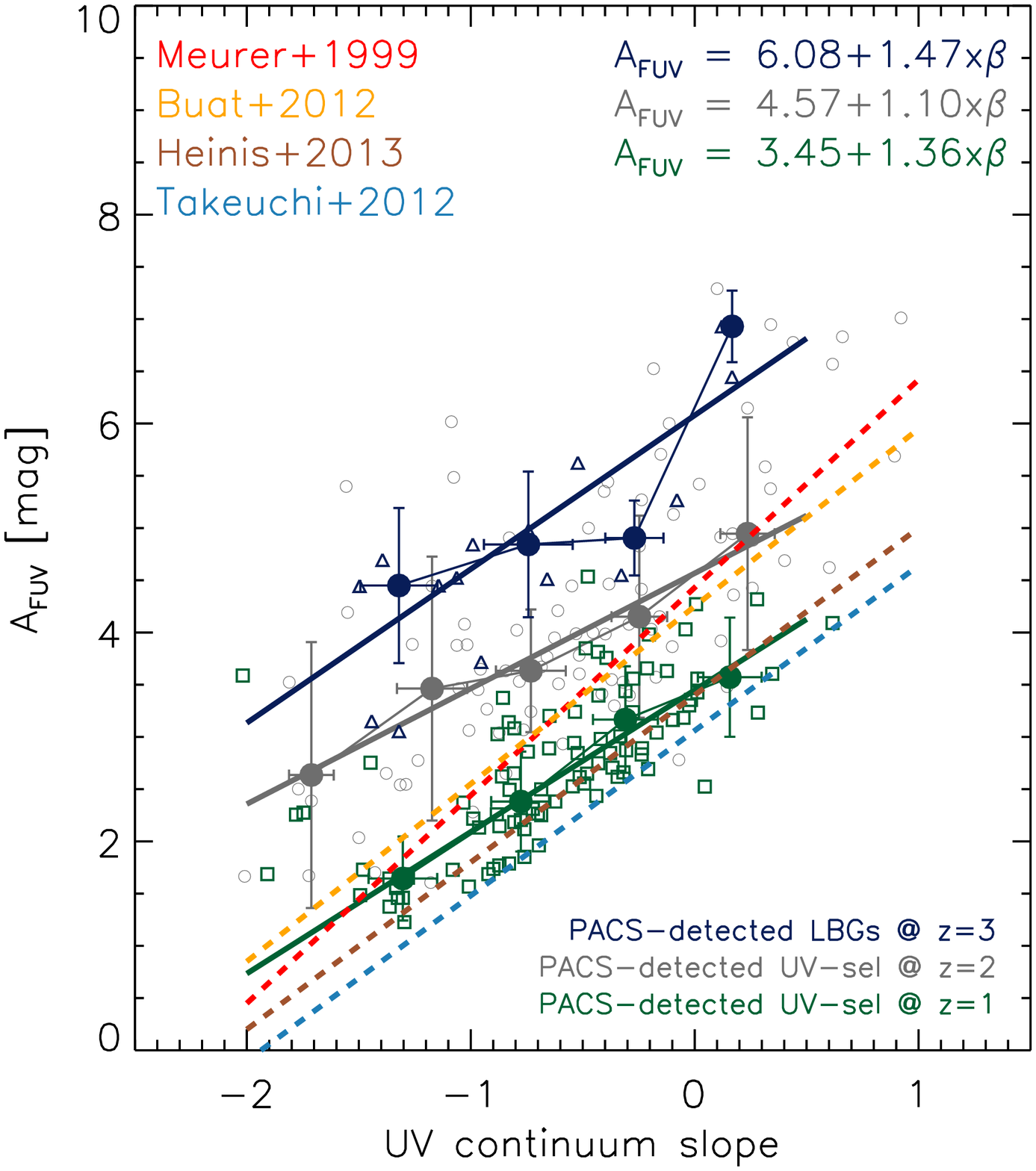}
\caption{\emph{Left}: Location in an IRX-$\beta$ diagram of our PACS-detected LBGs (small orange dots), PACS-detected $sBzK$ galaxies (small red dots), and PACS-detected UV-selected galaxies (open grey dots). The large dots, in the same colour code, represent the median value of the IRX ratio in different bins of the UV continuum slope. The error bars represent the width of the distributions. We also include a sample of PACS-detected LBGs at $z \sim 3$ from \citet{Oteo2013_z3} with open blue triangles and a sample of PACS-detected LBGs at $z \sim 1$ from \citet{Oteo2013a} with green open squares. The IRX-$\beta$ relations reported by \citet{Meurer1999,Buat2012,Heinis2013,Takeuchi2012}, and \citet{Boissier2007} are represented with the colour code indicated in the upper-left legend. The green shaded region represents the uncertainties of the \citet{Meurer1999} relation. \emph{Right}: Dust attenuation as a function of the UV continuum slope for PACS-detected LBGs at $z \sim 3$ (blue symbols) and PACS-detected UV-selected galaxies at $z \sim 2$ (grey symbols) and $z \sim 1$ (green symbols) taken from this work, \citet{Oteo2013_ALHAMBRA_PACS} and \citet{Oteo2013_z3}. We also overplot the relation of \citet{Meurer1999}, \citet{Buat2012}, \citet{Heinis2013} and \citet{Takeuchi2012}, as indicated in the upper-left legend. The solid lines are linear fits to the median points for each sample. The analytical expressions are presented in the same colour code.
              }
\label{dust_beta_PACS}
\end{figure*}

The location of our PACS-detected galaxies in an IRX-$\beta$ diagram \citep{Buat2012} can give us information about the validity of the dust correction factors that are obtained from the UV continuum slope and the application of a IRX-$\beta$ relation \citep{Meurer1999,Overzier2011,Takeuchi2012}. In this Section we explore how our PACS-detected galaxies at $z \sim 2$ are distributed in an IRX-$\beta$ diagram (Figure \ref{dust_beta_PACS}). Our PACS-detected LBGs, $sBzK$ and UV-selected galaxies at $z \sim 2$ are represented with filled orange, filled red and open grey dots, respectively. Along with the point, we also represent the IRX-$\beta$ relations of \cite{Meurer1999}, \cite{Buat2012}, \cite{Heinis2013}, \cite{Takeuchi2012} and \cite{Boissier2007}. It can be clearly seen that most of our PACS-detected galaxies at $z \sim 2$ are above the aforementioned IRX-$\beta$ relations. Therefore, the application of any of those relations would underestimate the dust attenuation of our PACS-detected sources, and consequently their total SFR, similarly to what  happens at $z \sim 3$ \citep{Oteo2013_z3}. At a similar redshift than our PACS-detected galaxies at $z \sim 2$, the \cite{Heinis2013} relation gives lower values of dust attenuation for a given UV continuum slope probably because of the stacking analysis carried out in that work which allows the authors to study the FIR emission of less IR-luminous galaxies.

In order to explore the location of PACS-detected LBGs in the IRX-$\beta$ diagram as a function of redshift, we also represent in Figure \ref{dust_beta_PACS} two samples of PACS-detected LBGs at $z \sim 1$ and $z \sim 3$ taken from \cite{Oteo2013_ALHAMBRA_PACS} and \cite{Oteo2013_z3}, respectively. First, it can be seen that the location of LBGs at $z \sim 1$ is compatible with the relation of \cite{Buat2012} and \cite{Meurer1999}. The agreement with the \cite{Buat2012} relation is mainly due to the similar selection criteria of the sources in both works since most of the PACS-detected UV-selected galaxies in \cite{Buat2012} are around $z \sim 1$. However, the \cite{Meurer1999} relation was derived for local galaxies and it is still valid at $z \sim 1$. 

Comparing the location of LBGs in the IRX-$\beta$ diagram at $z \sim 1$, $z \sim 2$ and $z \sim 3$, it can be seen that the upper envelope of the locus of PACS-detected LBGs at $z \sim 3$ is associated with higher dust attenuation than at $z \sim 2$ and at $z \sim 1$. To further explore this issue and to generalize to general populations of UV-selected galaxies instead of LBGs we represent in the right panel of Figure \ref{dust_beta_PACS} the dust attenuation of PACS-detected UV-selected galaxies at $z \sim 1$ and $z \sim 2$ and PACS-detected LBGs at $z \sim 3$ as a function of their UV continuum slope. It should be observed that the plot shown in the right panel is very similar to the one in the left panel. The only difference is the quantity represented in the y-axis. However, we have separated them in order to clarify the explanation. The UV-selected galaxies at $z \sim 2$ have been taken from this work. The selection criterion for UV-selected galaxies at $z \sim 1$ is the same as the one employed in \cite{Oteo2013_ALHAMBRA_PACS}, i.e.\ galaxies detected in the GALEX NUV band and PACS (any band), with photometric redshifts within $0.8 \lesssim z \lesssim 1.2$, and no X-ray counterpart. This criterion has been applied to the multiwavelength catalogs employed in this work in the GOODS-S and GOODS-N fields. Furthermore, we have included in the sample the PACS-detected UV-selected galaxies at $z \sim 1$ of \cite{Oteo2013_ALHAMBRA_PACS} in the COSMOS field. We have performed linear fits to the median points corresponding to the PACS-detected sources at different redshifts, obtaining:

\begin{equation}
	A_{\rm FUV} = 3.45 + 1.36 \times \beta \qquad \qquad (z \sim 1)
\end{equation}

\begin{equation}
	A_{\rm FUV} = 4.57 + 1.10 \times \beta \qquad \qquad (z \sim 2)
\end{equation}

\begin{equation}
	A_{\rm FUV} = 6.08 + 1.47 \times \beta \qquad \qquad (z \sim 3)
\end{equation}

These relations imply that, for a given UV continuum slope, the dustiest UV-selected galaxies at higher redshifts are more attenuated. This might be considered as an evolution of the FIR emission of UV-selected galaxies with redshift. Interestingly, PACS-detected UV-selected at $z \sim 1$, $z \sim 2$ and PACS-detected LBGs $z \sim 3$ have a UV continuum slope spanning a similar range despite their large difference in dust attenuation.

\subsection{SFR-mass for PACS-detected galaxies}

\begin{figure}
\centering
\includegraphics[width=0.45\textwidth]{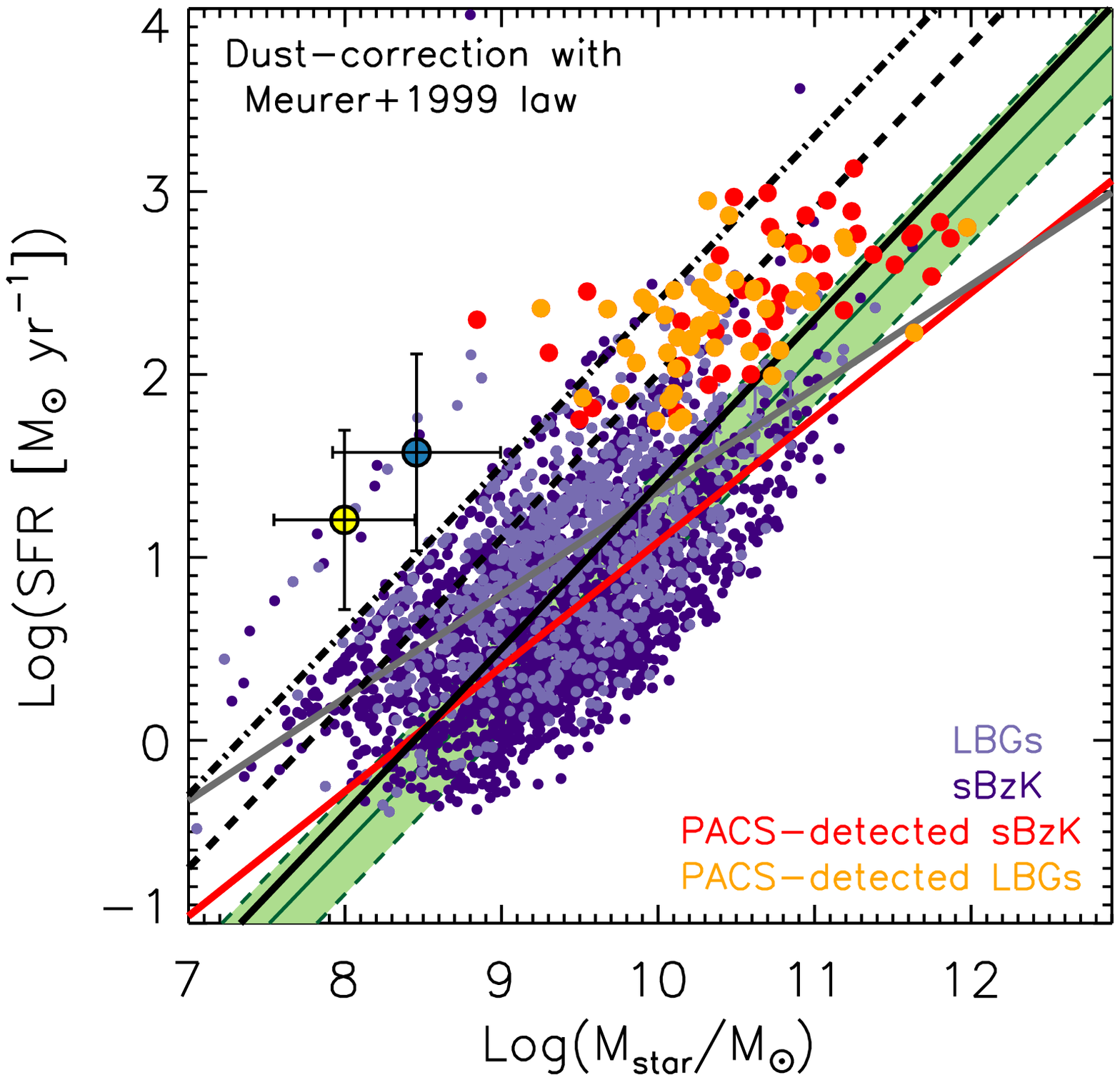}
\caption{SFR versus stellar mass plane for our PACS-detected LBGs (orange symbols) and $sBzK$ galaxies (red symbols) in GOODS-N and GOODS-S. Their total SFR has been derived with the direct UV and IR detections, ${\rm SFR_{total} = SFR_{UV}+SFR_{IR}}$. The black solid line is the main sequence (MS) of star-forming galaxies at $z \sim 2$ derived in \citet{Daddi2007} for MIPS-detected $sBzK$ galaxies. Dashed and dotted-dashed lines represent 4 and 10 times the \citet{Daddi2007} MS. For comparison, we also plot the points associated with LBGs and $sBzK$ galaxies regardless of their detection in PACS, whose total SFR have been estimated with the dust correction factors obtained with the \citet{Meurer1999} IRX-$\beta$ relation.
              }
\label{sfr_mass_daddi_PACS}
\end{figure}

Figure \ref{sfr_mass_daddi_PACS} shows the location in an SFR--mass diagram of our PACS-detected LBGs (orange symbols) and $sBzK$ galaxies (red symbols). For comparison, we also plot the location of our LBGs and $sBzK$ regardless their PACS-detection with light and dark purple symbols, respectively, as in the left panel of Figure \ref{sfr_mass_daddi}. For PACS-detected galaxies, the total SFR are determined with direct UV and IR detections, ${\rm SFR_{total} = SFR_{UV}+SFR_{IR}}$, whereas for PACS-undetected galaxies we estimate the total SFR with the dust correction factors obtained with the \cite{Meurer1999} IRX-$\beta$ relation. The UV+IR-derived total SFR of our PACS-detected galaxies indicates that they do not follow the \cite{Daddi2007} MS, but there are mostly above (see also the discussion in \cite{Lee2013arXiv1310.0474L}); they are therefore expected to have a starburst nature rather than being normal SF galaxies \citep{Elbaz2011}. This is in agreement with the fact that the IR SED of our PACS-detected galaxies resembles the SB IR SED defined in \cite{Elbaz2011}.

\subsection{Dust attenuation and stellar mass}

\begin{figure*}
\centering
\includegraphics[width=0.8\textwidth]{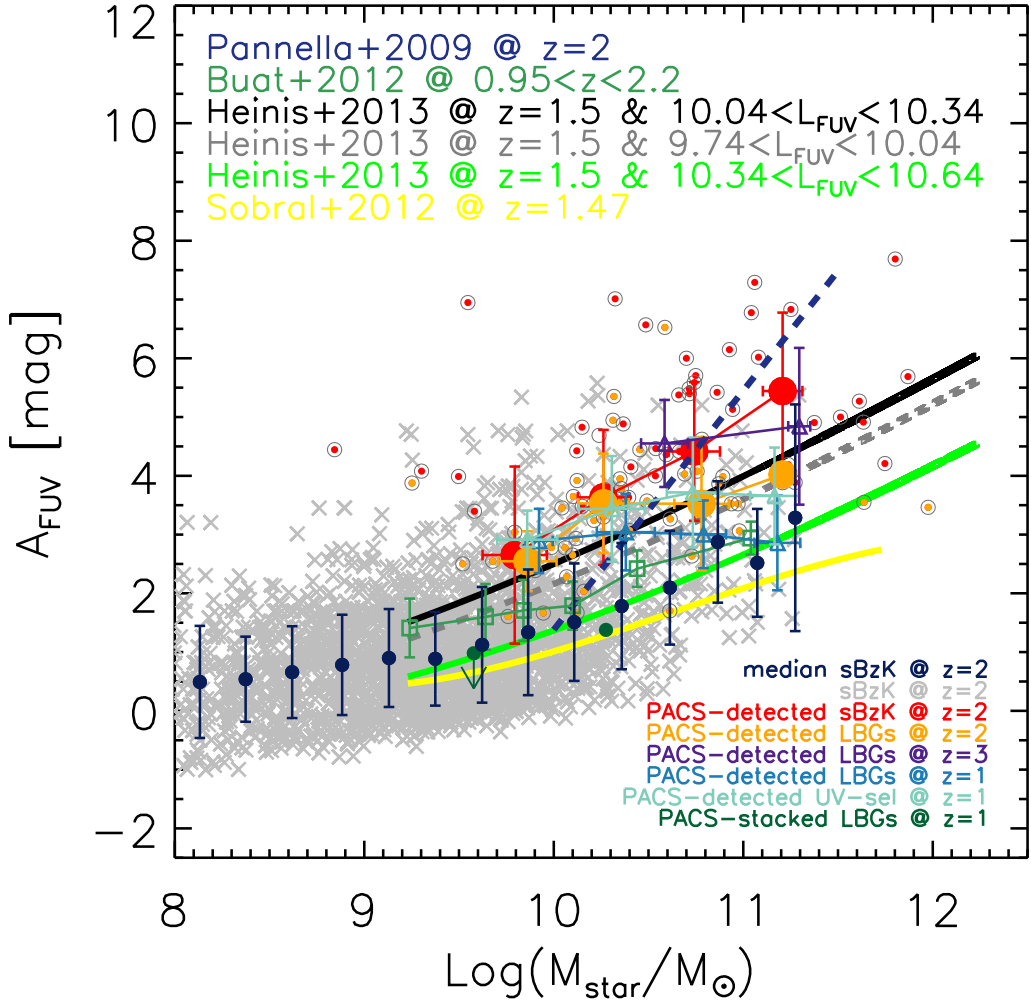}
\caption{Relation between the dust attenuation and the stellar mass for our PACS-detected LBGs (small orange dots), $sBzK$ (small red dots), and UV-selected galaxies (open grey dots). The large dots, in the same colour code, represent the median value of the dust attenuation in different stellar mass bins. The error bars represent the width of the distributions. For comparison, we also represent the dust attenuation versus stellar mass for out whole sample of $sBzK$ in the GOODS-S field with grey crosses (blue points indicate median values for each bin of stellar mass and blue bars are the widths of the distributions). We also represent a sample of PACS-detected LBGs at $z \sim 1$ \citep{Oteo2013_ALHAMBRA_PACS} and a sample of PACS-detected LBGs at $z \sim 3$ \citep{Oteo2013_z3}. PACS-stacked LBGs at $z \sim 1$ are represented by black diamonds, with the arrow indicating an upper limit in the dust attenuation when no stacked flux is recovered. The relations between the dust attenuation and stellar mass derived in \citet{Pannella2009,Buat2012,Sobral2012MNRAS.420.1926S}, and Heinis et al. (2013, in preparation) are represented in the colour code included in the plot. In all the cases, the stellar mass have been renormalized to a common Salpeter IMF.
              }
\label{dust_masa_PACS}
\end{figure*}

Some previous studies have reported a correlation between stellar mass and dust attenuation in a wide range of redshifts \citep{Martin2007,IglesiasParamo2007,Pannella2009,Buat2009,Buat2012}. Most of them agrees that more massive galaxies are more attenuated \citep{Garn2010MNRAS.409..421G}. Additionally, \cite{Sobral2012MNRAS.420.1926S} found evidences of no evolution in the dust-mass relation with redshift, at least up to $z \sim 1.47$. In this Section we analyse the relation between these two quantities for our samples of PACS-detected LBGs, $sBzK$ and UV-selected galaxies at $z \sim 2$ (see Figure \ref{dust_masa_PACS}). We only find a correlation for PACS-detected $sBzK$ and PACS-detected UV-selected galaxies: for $\log{\left( M_*/M_\odot \right)} > 9.8$, more massive galaxies are more attenuated. The narrow range of values of stellar mass and dust attenuation for PACS-detected LBGs prevents us from obtaining a significant relation between dust attenuation and stellar mass. The correlation between dust attenuation and stellar mass is also seen in the whole sample of $sBzK$ galaxies regardless their PACS detection, with scatter of the relation being higher at the high-mass end.

\begin{figure*}
\centering
\includegraphics[width=0.8\textwidth]{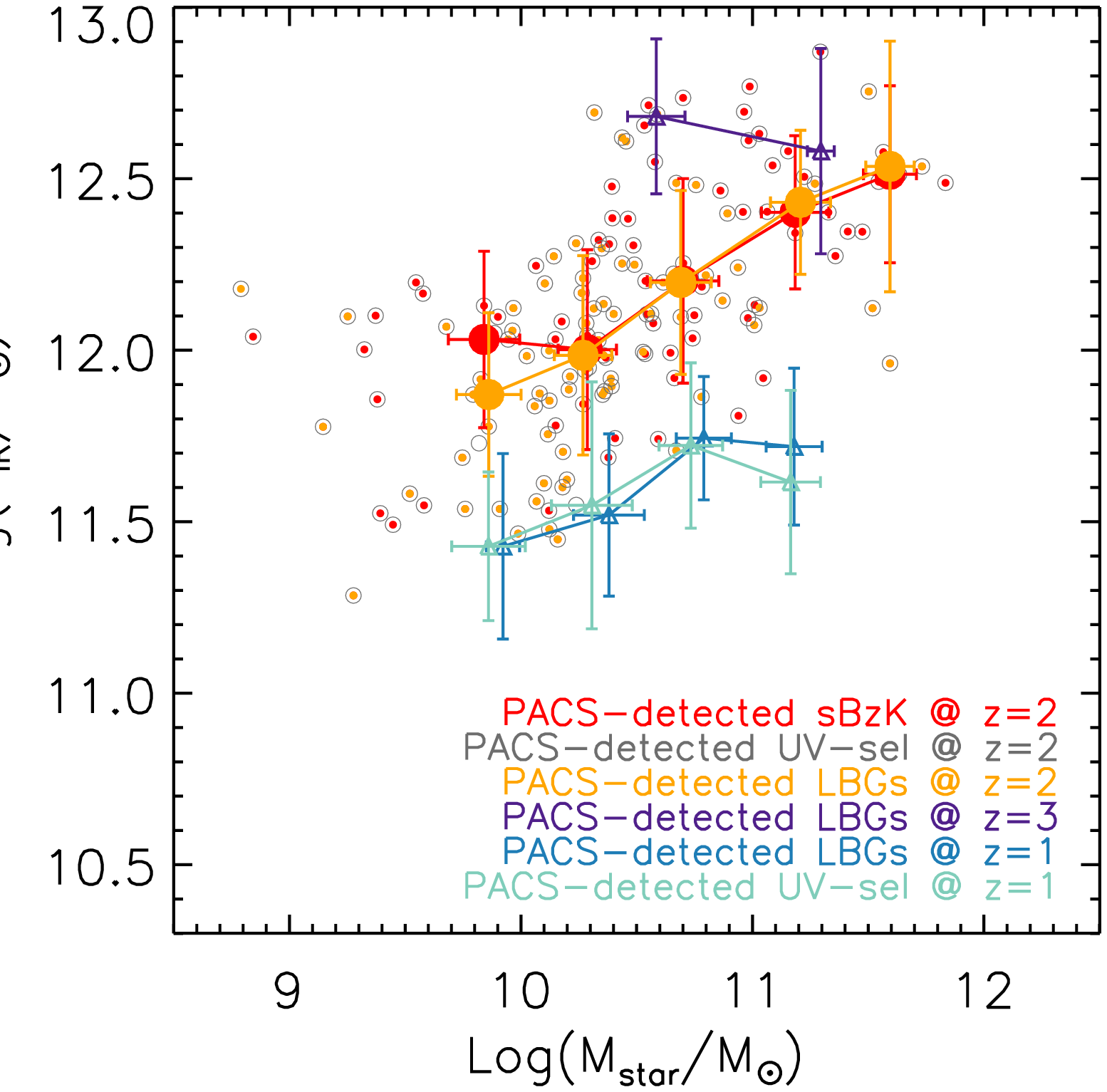}
\caption{Relation between the total IR luminosity and the stellar mass for our PACS-detected LBGs (small orange dots), $sBzK$ (small red dots) and UV-selected galaxies (open grey dots). The large dots, in the same colour code, represent the median value of the total IR luminosity in different bins of the stellar mass. The error bars represent the width of the distributions. We also represent a sample of PACS-detected LBGs at $z \sim 1$ \citep{Oteo2013_ALHAMBRA_PACS} and a sample of PACS-detected LBGs at $z \sim 3$ \citep{Oteo2013_z3}.
              }
\label{lir_masa_PACS}
\end{figure*}

In Figure \ref{dust_masa_PACS} we also compare our results with those reported in previous studies. It should be noted at this point that some of the differences found between our and other works can be due to the methodology followed to derive the stellar mass and the dust attenuation of the galaxies under study. We have converted all the stellar masses to a common Salpeter IMF, the one employed in this work. \cite{Pannella2009} derived a relation between dust attenuation and stellar mass by using stacking analysis in 1.4 GHz for a sample of $sBzK$ galaxies at $z \sim 2$ and assuming the local relation between radio and total IR luminosity. They obtained a relation that provides, for the most massive galaxies in our sample of PACS-detected $sBzK$ galaxies, similar dust attenuation for a given stellar mass, although the slope of their relation is slightly higher.


We also plot in Figure \ref{dust_masa_PACS} the results of Heinis et al.\ (2013, submitted). In that paper, the dust attenuation is obtained by performing stacking analysis in SPIRE bands in three samples of UV-selected galaxies at $z \sim 1.5$. They obtain, in agreement with \cite{Buat2012}, that the relation between the dust attenuation and stellar mass depends upon the rest-frame UV luminosity. We represent the relations of Heinis et al.\ (2013) in their three ranges of rest-frame UV luminosity translating their their stellar masses derived with a Chabrier IMF into a Salpeter IMF. The median rest-frame UV luminosity of our PACS-detected $sBzK$ galaxies (considering those in GOODS-S and GOODS-N at the same time) is $\log{\left( L_{\rm UV}/L_\odot \right)} = 10.16$. The Heinis et al.\ (2013) relation associated with this luminosity is represented with a black solid curve in Figure \ref{dust_masa_PACS}. It can be seen that our points are above the Heinis et al. relation. This is because we only consider PACS-detected sources; therefore, for a given stellar mass we only select the dustiest galaxies. However, Heinis et al.\ employ a stacking analysis that enables them to find less attenuated galaxies for a given stellar mass. The median value of the rest-frame UV luminosity of our PACS-detected LBGs is $\log{\left( L_{\rm UV}/L_\odot \right)} = 10.37$, which corresponds with the Heinis et al.\ relation shown in green. Again, our points are above the Heinis et al. relation for the same reasons above.


The relation between dust attenuation and stellar mass at $z \sim 1.47$ derived in \cite{Sobral2012MNRAS.420.1926S} is also represented in Figure \ref{dust_masa_PACS}. Again, our points are above the Sobral et al. relation since we only work with PACS-detected galaxies, while Sobral et al. use a large sample of emission line galaxies selected by their H$\alpha$ emission. Interestingly, the dust--mass relation derived in \cite{Sobral2012MNRAS.420.1926S} and \cite{Buat2012} are in very good agreement with the relation for our whole sample of $sBzK$ samples at $z \sim 2$ regardless their PACS detection. This indicates that the dust--mass relation did not evolve from $z \sim 2.5$ down to $z \sim 1.47$. Since the Sobral et al. relation is very similar to that found in the local universe \citep{Garn2010MNRAS.409..421G}, we find evidence that the dust--mass relation has not evolve with redshift, at least for $z \lesssim 2.5$.

We also represent in figure \ref{dust_masa_PACS} the location of a sample of PACS-detected UV-selected galaxies at $z \sim 1$ taken from \cite{Oteo2013_ALHAMBRA_PACS} and at $z \sim 3$ from \cite{Oteo2013_z3}. Comparing the samples at $z \sim 1$, $z \sim 2$, and $z \sim 3$ we do not find significant evidence of evolution on the FIR either. In most cases the median values at the three redshifts are within the widths of the distributions. This contrasts with the clear evolution found in the IRX-$\beta$ diagram in Section \ref{irxbpacs}. Therefore, the dust-mass relation does not seem to be a useful relation to constrain the evolution of the dust properties of galaxies at any redshift.

As shown in Figure \ref{lir_masa_PACS} there is also a relation between the stellar mass and the total IR luminosity of the PACS-detected SF galaxies at $z \sim 2$: more massive galaxies have stronger IR emission. In this case there is almost no difference between the trends for PACS-detected LBGs and PACS-detected $sBzK$ and UV-selected galaxies due to their similar selection in the FIR. Since all the galaxies are within a similar redshift range and are detected by PACS under the same limiting fluxes, all the galaxies have a similar range of total IR luminosity for a given stellar mass. Additionally, as found in Section \ref{properties_PACS_detected_galaxies}, the stellar masses of PACS-detected LBGs and $sBzK$ galaxies span a similar range. Again, it can be seen that there is an evolutionary trend with redshift in the sense that, for a given stellar mass, PACS-detected LBGs at higher redshifts have stronger FIR emission. The most significant change is between $z \sim 1$ and $z \sim 2$, where the total IR luminosity changes by more than 0.5 dex for a given stellar mass. No significant evolution is found between $z \sim 2$ and $z \sim 3$. Since the total IR luminosity is a very good proxy of the total SFR for IR-bright galaxies, the relation between the total IR luminosity and the stellar mass is directly related to the correlation between SFR and stellar mass \citep{Daddi2007}. In the same way, the evolution of the FIR emission for a given stellar mass is directly related to the correlation between total SFR and stellar mass and its evolution with redshift \citep[see for example][]{Elbaz2011} and also to the evolution of the sSFR with redshift \citep[see for example][]{Rodighiero2010}.

\subsection{IR luminosity functions and contribution to the cosmic star-formation density}

Obtaining the IR luminosity function of our PACS-detected galaxies allows us to determine their contribution to the cosmic SFR density (SFRD) of the universe at their redshift. We compute the IR luminosity function of our PACS-detected galaxies by using the $1/V_{\rm max}$ accessible volume technique, as in many previous studies \citep{Chapman2005,Gruppioni2010,Casey2012_z2,Casey2012_redshift,Roseboom2012}:

\begin{equation}
	\phi \left( L \right) \Delta L = \sum_{i=0}^n \frac{1}{V_{{\rm max},i}}
\end{equation}

\begin{figure*}
\centering
\includegraphics[width=0.45\textwidth]{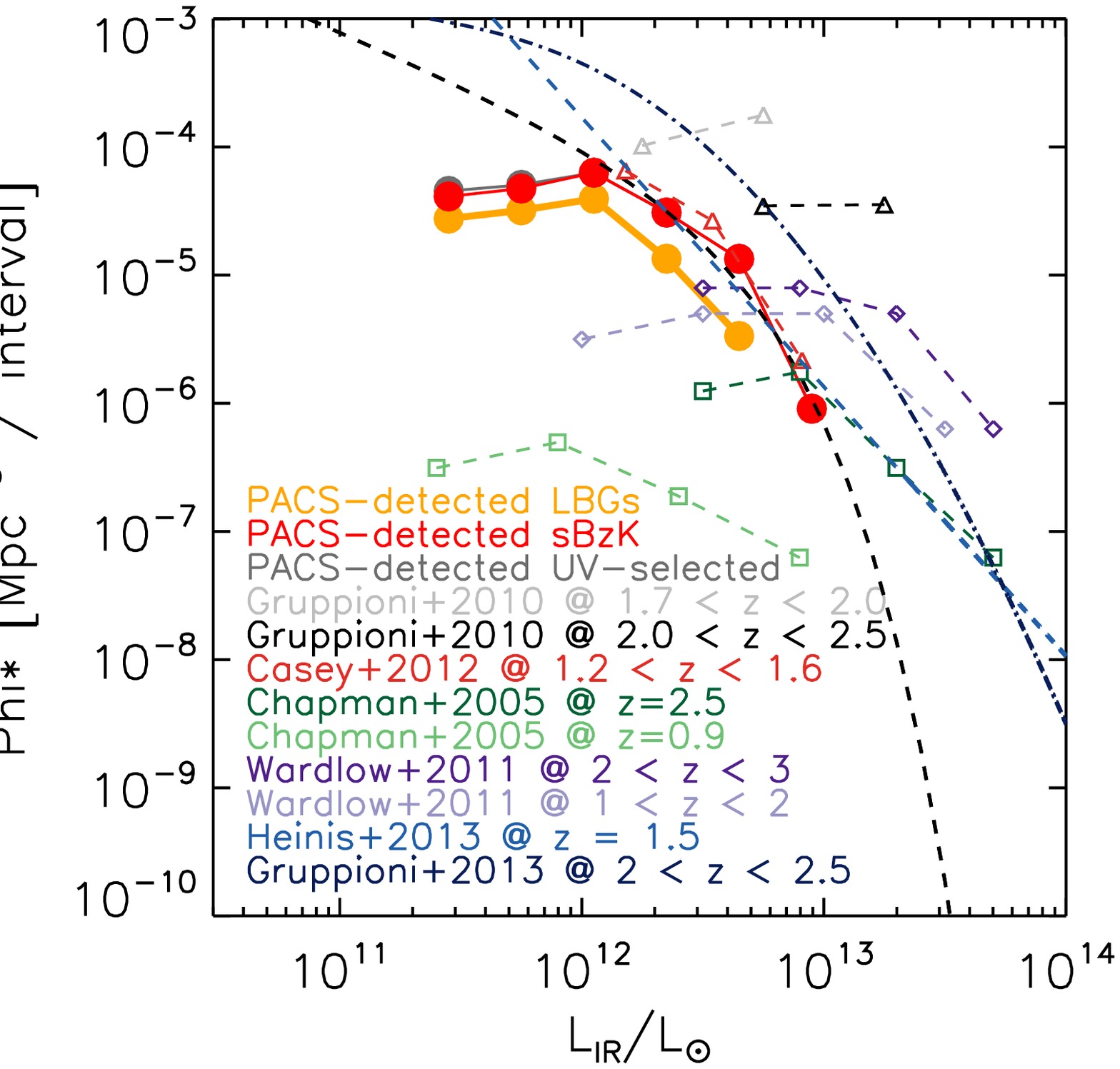}
\includegraphics[width=0.45\textwidth]{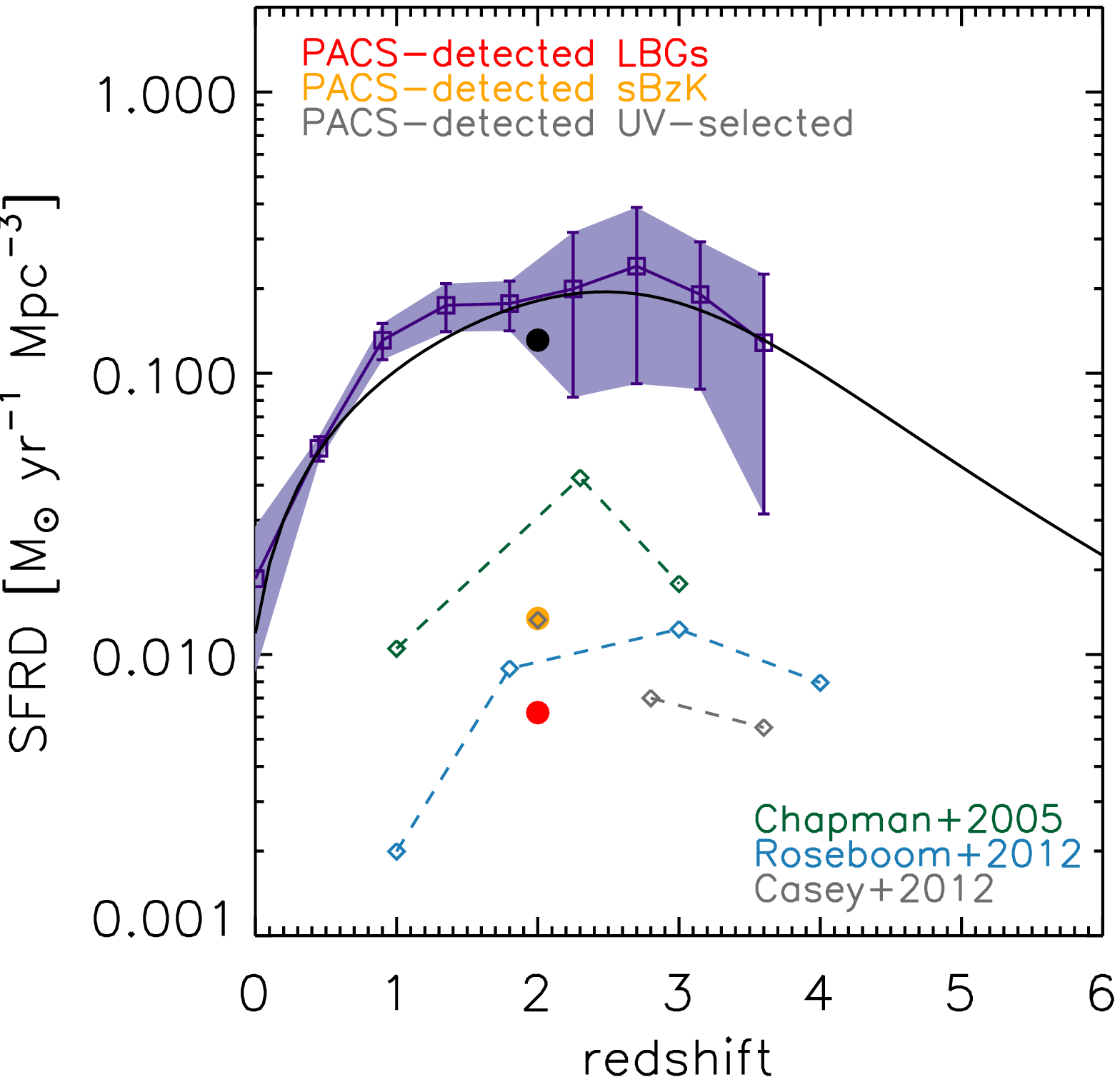}
\caption{\emph{Left plot}: IR luminosity functions for our PACS-detected LBGs (red points), $sBzK$ (grey points) and UV-selected (orange points) galaxies. We also show the luminosity functions reported in \citet{Gruppioni2010}, \citet{Chapman2005}, \citet{Wardlow2011}, \citet{Heinis2013}, \citet{Gruppioni2013}, and \citet{Casey2012_redshift} for samples of PACS-detected, SPIRE-detected and sub-mm galaxies at different redshift ranges, as indicated by the colour legend. Some \citet{Schechter1976} functions associated with different parameters are over-plotted to guide the eye regarding the shape of the luminosity functions of each type of galaxy. \emph{Right plot}: Contribution of our PACS-detected LBGs (red point), $sBzK$ (grey open diamond), and UV-selected (orange dot) galaxies to the cosmic SFR density (SFRD). We also plot with a black curve the redshift evolution of the SFH of the universe defined in \citet{Hopkins2006} and with a purple shaded zone the redshift evolution of the total (UV+IR) SFRD reported in \citet{Burgarella2013}. The evolution of the SFR density for some samples of submm and FIR-detected galaxies \citep{Chapman2005,Wardlow2011,Roseboom2012,Casey2012_z2}, are also shown for comparison.
              }
\label{LF}
\end{figure*}

\noindent where $\phi \left( L \right) \Delta L$ is the number density of sources whose total IR luminosities are within $[L,L+\Delta L]$ and $\phi \left( L \right)$ is given in units of ${\rm Mpc}^{-3} \log L^{-1}$. The parameter $V_{{\rm max},i}$ is the comoving volume where the $i$th source could be located according to its luminosity $L_i$ and the depth of the FIR survey employed to determine the total IR luminosity. The luminosity of each object in combination with the limiting IR luminosity of the survey at the redshift of the source determines the maximum redshift where the galaxy could have been located for being detected in the survey. This maximum redshift is then translated into the maximum comoving volume. This method assumes a homogeneous distribution of objects in space, and any spatial clustering in the sample will distort the shape of the luminosity function. Therefore, the procedure is sensitive to cosmic variance \citep{Weiss2009}. For the calculation of the limiting IR luminosity for the redshift of each source we employ the PACS-160$\mu$m limiting flux and carry out extrapolations from PACS-160$\mu$m to the total IR luminosity with CE01 templates \citep{Elbaz2010,Elbaz2011}. We adopt the PACS-160$\mu$m channel instead of the PACS-100$\mu$m one since the former provides lower values of the limiting luminosity at each redshift and therefore higher values of the maximum redshift where each galaxy could be located. The IR LF for PACS-detected LBGs, $sBzK$, and UV-selected galaxies are shown in the left plot of Figure \ref{LF}. For each kind of PACS-detected galaxy we try to fit Schechter functions of the form $\phi = \phi^* x^\alpha e^{-x}$, where $x=L/L^*$. In this process, $\phi^*$, $\alpha$, and $L^*$ are the three parameters to be fitted. For each type of galaxy we only fit those points of the LF that are not greatly affected by incompleteness, i.e.\ those with $\log{\left( L_{\rm IR}/L_\odot \right)} > 12$. In the case of PACS-detected $sBzK$ and UV-selected galaxies there are enough galaxies which meet that criterion, but this is not the case for PACS-detected LBGs. In this way, we do Schechter fits only for PACS-detected $sBzK$ and UV-selected galaxies. The depth of the PACS observations employed here, despite being one of the deepest FIR data sets available to date, does not allow us properly to constrain the $\alpha$ parameter. For this reason, \emph{we adopt a fixed value $\alpha = -1.8$} and leave $\phi^*$ and $L^*$ free. As a result, we obtain for both PACS-detected $sBzK$ and UV-selected galaxies values of $\log{\left( L_*/L_\odot \right)} = 12.47$ and $\log \phi^* = -3.75$. The result for both kind of galaxies is the same since the two PACS-detected UV-selected galaxies that are not selected as $sBzK$ have $\log{\left( L_{\rm IR}/L_\odot \right)} < 12$.

Figure \ref{LF} shows the IR LFs of our PACS-detected LBGs, $sBzK$ and UV-selected galaxies. We also represent in the IR luminosity functions found in \cite{Chapman2005}, \cite{Wardlow2011}, \cite{Casey2012_z2} and \cite{Gruppioni2010} for PACS-detected, SPIRE-detected and sub-mm galaxies at different redshifts. It can be seen that the LFs for PACS-detected $sBzK$ and PACS-detected UV-selected galaxies tend to follow the \cite{Schechter1976} functions for the points associated with the brightest luminosities, where the completeness of the observations is high. For each total IR luminosity bin the density of PACS-detected LBGs is lower than that for $sBzK$ and UV-selected galaxies owing to the lower number of PACS detections. The IR LF of our PACS-detected $sBzK$ and UV-selected galaxies is similar to that for the SPIRE-detected galaxies of \cite{Casey2012_redshift} at $1.2 < z < 1.6$. Those IR LF are also compatible with the LF of the submm galaxies of \cite{Chapman2005} at $z \sim 2.5$ in the overlapping luminosity range. Furthermore, the bright end of the FIR luminosity function of UV-selected galaxies at $z \sim 1.5$ obtained in \cite{Heinis2013} from stacking analysis in SPIRE bands is in very good agreement with our determinations for PACS-detected $sBzK$ and UV-selected galaxies. The IR LFs of our PACS-detected galaxies are dissimilar to that associated with sub-mm galaxies at their redshift \citep{Wardlow2011} due to the different number density of galaxies in each sample. The IR LF of \cite{Wardlow2011} are also shifted towards higher values of the total IR luminosity (have higher values of $L^*$) than our IR LFs owing to the difference in the selection criterion of their sources. Finally, the IR LF given by \cite{Gruppioni2013} gives higher values of the density for each total IR luminosity likely because it includes the contribution of both SF galaxies and AGN. See \cite{Gruppioni2013} for details of the contribution of galaxies of different natures to the IR luminosity function.

The right planel of Figure \ref{LF} shows the contribution of our PACS-detected galaxies to the cosmic SFRD of the universe at their redshift. In order to convert the our LF into SFR functions we employ the \cite{Kennicutt1998} relation shown in Equation \ref{SFR_IR}. The SFRD values for each kind of galaxy are then obtained by integrating each SFR function over the luminosity range where our galaxies are located: we consider the integration as the raw sum of the SFR weighted by the number density of sources in each luminosity bin. This is an alternative procedure to the traditional one, which would be to integrate the analytical expression of the IR LF over the whole range of total IR luminosities. However, due to limitation produced by the depths of the FIR observations employed here we cannot constrain the faint-end slope of the IR LF and, consequently, this leads to an uncertainty in the SFRD measurements. This effect is stronger at higher redshifts, since only the bright end of the IR LF can be sampled, leaving the faint end completely unconstrained. This is true for any FIR data sets that could be employed, since the FIR detections are strongly biased towards very bright IR sources. Only stacking analysis in FIR bands could help us to find the properties of the IR LF of IR-faint galaxies \citep{Heinis2013}.

Compared to the evolution of the cosmic SFRD defined in \cite{Hopkins2006}, the contributions of PACS-detected LBGs, $sBzK$ and UV-selected galaxies are 7\%, 12\% and 12\%, respectively. We also plot in the right plot of Figure \ref{LF} the redshift evolution of the FIR-bright galaxies of \cite{Chapman2005}, \cite{Roseboom2012} and \cite{Casey2012_z2}. The contribution of our PACS-detected galaxies to the cosmic SFRD is slightly higher than that for the mm-detected galaxies of \cite{Roseboom2012} but lower than the contribution of the submm galaxies of \cite{Chapman2005} in the same redshift range. The percentages found for our PACS-detected galaxies indicate that they play an important role in the SFH of the universe and are consequently important in understanding the formation and evolution of galaxies over cosmic time. Finally, we also represent by a black dot  in the right panel of Figure \ref{LF} the SFRD corresponding to the integration of the Schechter function of the sample of PACS-detected UV-selected galaxies over the entire range of IR luminosity. This integration would include both detected and undetected galaxies in the FIR. This point would correspond to the total SFRD of the bulk of SF galaxies at $z \sim 2$ and is in agreement with the \cite{Hopkins2006} and \cite{Burgarella2013} curves, although with the difference that the rest-frame UV LF is not considered in our work.

\subsection{SPIRE sub-mm detections}\label{SPIREdetections}

Within the GOODS-\emph{Herschel} project, the GOODS-N field has beed also observed in SPIRE-250$\mu$m, SPIRE-350$\mu$m and SPIRE-500$\mu$m. In principle, these additional FIR data are not needed to obtain the total IR luminosity of our FIR-bright sources, but reveal a population of submm-detected LBGs. We find that among the sample of PACS-detected (LBGs, $sBzK$ galaxies, UV-selected galaxies) in GOODS-N, (15, 24, 24) are detected in SPIRE-250$\mu$m, (10, 14, 15) are detected in SPIRE-350$\mu$m, and (4, 4, 4) are detected in SPIRE-500$\mu$m, respectively. The UV-to-FIR SEDs of the four LBGs detected in SPIRE-500 $\mu$m are shown in Figure \ref{SEDs_SPIRE}. The reported SPIRE detections increase the sample of SPIRE-detected LBGs at $1.5 \lesssim z \lesssim 2.5$ of \cite{Burgarella2011}, where only one SPIRE detection (in SPIRE-250$\mu$m and SPIRE-350$\mu$m) was reported. Additionally, apart from the higher number of detections in SPIRE-250$\mu$m and SPIRE-350$\mu$m, the four SPIRE-500$\mu$m-detected LBGs at $1.5 \lesssim z \lesssim 2.5$ represent the first sample of sub-mm emission in LBGs in that redshift range.

We find no significant difference between the SPIRE-detected LBGs and the population of PACS-detected LBGs. Our SPIRE-detected LBGs at $z \sim 2$ have a median SED-derived dust attenuation of $E_s(B-V) = 0.5$, median total IR luminosity $\log{L_{IR}/L_\odot} = 12.2$, median stellar mass of $\log{M_*/M_\odot} = 10.6$, median age of 71 Myr, ${\rm SFR_{total}} = 284 \, M_\odot {\rm yr}^{-1}$, and median UV slope of $\beta = -0.41$. Similar values are obtained for $sBzK$ and UV-selected galaxies. We represent in Figure \ref{color_color_PACS} the location of the SPIRE-detected galaxies with large filled grey dots and find no significant difference in their locus either. The similarity between the PACS-detected and SPIRE-detected LBGs could be due to either the low number of galaxies individually detected in the FIR, which does not allow statistically significant results, or to the insensitivity of the SED-derived and FIR-derived properties to the detection rate as a function of FIR wavelength. We have not computed the dust temperature of our PACS and SPIRE-detected sources owing to the lack of submm data \citep{Pope2005}. PACS and SPIRE detections alone do not allow accurate dust temperate determinations at the redshift of our galaxies since the red wing of the dust emission peak is not very well constrained. Because of their brightness in the SPIRE bands and their non-detection in redder submm bands, we speculate that SPIRE-detected LBGs are the bridging population between submm galaxies and LBGs.

The above-reported SPIRE-detected sources are all detected in all the PACS bands. Moreover, there are five LBGs (all meeting the $sBzK$ selection criterion) and eight $sBzK$ galaxies (which are also UV-selected galaxies) that are detected in all of the SPIRE bands but are undetected in PACS. The existence of this population was expected since at $z \sim 2$ the SPIRE bands sample the maximum of the dust emission peak whereas the PACS fluxes are expected to be lower. The main problem in the analysis is source confusion. The jump in spatial resolution between MIPS-24$\mu$m and SPIRE bands is very high; consequently, the observed SPIRE source could be affected by the emission of close ($\lesssim 20''$) galaxies.







\section{Conclusions}\label{conclu}

Throughout this work we have studied the UV-to-FIR SED of a sample of LBGs, $sBzK$ and UV-selected galaxies at $1.5 \leq z \leq 2.5$ in the COSMOS, GOODS-N and GOODS-S fields. We have divided the study into two main parts. First, we have analysed the similarities and differences of the rest-frame UV-to-near-IR SEDs of LBGs, $sBzK$ and UV-selected galaxies at $z \sim 2$ with the aim of having a better comprehension about the populations that segregate the different selection criteria that look for SF galaxies in the redshift desert. Secondly, we have reported the PACS and SPIRE detections of a subsample of those galaxies located in the GOODS-S and GOODS-N fields by using data from the GOODS-\emph{Herschel} project. These FIR measurements are essential for accurately deriving, for example, their total IR luminosity, dust attenuation and total SFR without the uncertainties that the SED-fitting procedure with BC03 templates introduces. The main conclusions of the work can be summarized as follows:

\begin{enumerate}

	\item We have compiled a sample of 3207, 681 and 1300 LBGs at $1.5 \leq z \leq 2.5$ in COSMOS, GOODS-S and GOODS-N fields. Additionally, in order to compare the properties of LBGs at $z \sim 2$ with those for other SF galaxies at that redshift, we have segregated a sample of 9539, 2472 and 2192 star-forming $BzK$ galaxies and 8100, 2767 and 2581 UV-selected galaxies in COSMOS, GOODS-S and GOODS-N fields, respectively. The $sBzK$ galaxies have been selected through application of the classical \cite{Daddi2004} criterion, and the UV-selected galaxies have seen isolated imposing a detection in the $B$-band (rest-frame UV at $z \sim 2$) and restricting their photometric redshifts to $1.5 \leq z \leq 2.5$, the same range than for LBGs and $sBzK$ galaxies. In all cases we have ruled out AGNs by discarding galaxies with X-ray detections.

	\item We have analysed the rest-frame UV-to-near-IR SED of our galaxies with BC03 templates. The templates have been built by assuming a constant SFR and a fixed value of metallicity $Z = 0.2 Z_\odot$. In the comparison between the three kinds of SF galaxies analysed in this work we focus only on the GOODS-S field since it provides the deepest, most homogeneous, best photometric coverage of the fields studied. According to the SED-fitting procedure, LBGs at $z \sim 2$ tend to be brighter and bluer in the rest-frame UV, have a less prominent Balmer break (compatible with their younger ages) than $sBzK$ and UV-selected galaxies at their same redshift. They also have higher dust-corrected total SFR and sSFR. No significant difference in the median stellar mass or dust attenuation is found between these galaxies. The SED-derived physical properties of $sBzK$ and UV-selected galaxies are similar and indicate that they both represent the general population of SF galaxies at $z \sim 2$ better than LBGs.	

	\item In a colour versus stellar mass diagram, LBGs at $z \sim 2$ are mainly located in the blue cloud of galaxies at that redshift. Only those galaxies with older ages, higher dust attenuation and redder UV continuum slope deviate towards the green valley or the red sequence. This is in agreement with the findings of previous studies of LBGs at lower redshifts. For a given stellar mass, LBGs tend to have bluer optical colours than $sBzK$ and UV-selected galaxies.

	\item The LBGs and $sBzK$ galaxies studied follow the MS of galaxies at $z \sim 2$ defined in \cite{Daddi2007} only if their total SFR is recovered with the dust correction factors obtained with the \cite{Meurer1999} IRX-$\beta$ relation defined for local starburst. The definition of the slope and zero-point of the MS are found to be sensitive to the dust correction method employed to derive an estimation of the total SFR.

	\item We report individual clean PACS (100$\mu$m or 160$\mu$m) detections for 48 LBGs, 89 $sBzK$, and 91 UV-selected galaxies, that measure their dust emission directly. PACS-detected LBGs are dustier, redder in the UV continuum, more massive and have higher dust-corrected total SFR than those that are PACS-undetected. The total IR luminosities of the PACS-detected sources are mostly $L_{\rm IR} > 10^{12} L_\odot$ and thus belong to the ULIRG class. They are also massive galaxies typically with $\log{\left( M_*/M_\odot \right)} > 10$ and whose IR+UV-derived total SFR is higher than 100 $M_\odot \, {\rm yr}^{-1}$ for most of the galaxies. PACS-detected galaxies are mostly located in the green valley or the red sequence of galaxies at that redshift, in agreement with their high dust attenuation. Furthermore, we find that most of the PACS-detected galaxies studied  are located above the \cite{Daddi2007} MS at $z \sim 2$; therefore, their star formation mode is dominated by starburst. This is consistent with the shape of their IR SEDs.

	\item We find that the locus of the PACS-detected galaxies studied in an IRX-$\beta$ diagram is above the \cite{Meurer1999}, \cite{Takeuchi2012}, \cite{Buat2012} and \cite{Heinis2013} relations. Therefore, the dust correction factors obtained with those relations tend to underestimate the total SFR of our PACS-detected galaxies. This is similar to what happens at higher redshifts. However, PACS-detected LBGs at $z \sim 1$ are located around the \cite{Meurer1999} relation and the dust correction factors obtained with that relation recovers total SFRs that are in good agreement with those determined with direct UV and IR measurements. Furthermore, we find that for a given UV continuum slope, PACS-detected galaxies at higher redshifts are more attenuated. The dust attenuation at higher redshifts is higher even though the UV continuum slope of the PACS-detected LBGs at $z \sim 1$, $z \sim 2$ and $z \sim 3$ span a similar range. In the $L_{\rm IR}$--mass plane, the evolution can be seen only between $z \sim 1$ and $z \sim 2$, and in the dust-mass plane no evolution can be significantly constrained.

	\item We have built the IR luminosity functions of our PACS-detected LGBs, $sBzK$ and UV-selected galaxies by using the $1/V_{\rm max}$ accessible volume. These tend to follow well the functional form of a Schechter function. For each bin of $L_{\rm IR}$ the number density of PACS-detected LBGS is lower than that for $sBzK$ and UV-selected galaxies owing to the lower number of FIR detections. The IR luminosity function of our PACS-detected LBGs indicates that they contribute with a lower limit of 7\% to the cosmic SFRD of the universe at their redshift. The contribution of PACS-detected $sBzK$ and UV-selected galaxies is higher, with a lower limit of 12\% in both cases.

	\item We find a subpopulation of 17, 26 and 27 LBGs, $sBzK$, and UV-selected galaxies which are detected in PACS and in any of the SPIRE bands (250 $\mu$m, 350 $\mu$m and 500 $\mu$m). We find no significant difference in the SED-derived or FIR-derived properties between the SPIRE-detected and the remaining PACS-detected galaxies. We speculate that SPIRE-detected LBGs are the bridging population between submm galaxies and LBGs.

\end{enumerate}

\section*{Acknowledgments}

The authors thank the anonymous referee for the comments provided, which have improved the paper and the presentation of results. This research has been supported by the Spanish Ministerio de Econom\'ia y Competitividad (MINECO) under the grant AYA2011-29517-C03-01. Some/all of the data presented in this paper were obtained from the Multimission Archive at the Space Telescope Science Institute (MAST). STScI is operated by the Association of Universities for Research in Astronomy, Inc., under NASA contract NAS5-26555. Support for MAST for non-HST data is provided by the NASA Office of Space Science via grant NNX09AF08G and by other grants and contracts. Based on observations made with the European Southern Observatory telescopes obtained from the ESO/ST-ECF Science Archive Facility. Based on zCOSMOS observations carried out using the Very Large Telescope at the ESO Paranal Observatory under Programme ID: LP175.A-0839. Based on observations made with ESO Telescopes at the La Silla or Paranal Observatories under programme ID 171.A-3045.

{\it Herschel} is an ESA space observatory with science instruments provided by European-led Principal Investigator consortia and with important participation from NASA. The Herschel spacecraft was designed, built, tested, and launched under a contract to ESA managed by the Herschel/Planck Project team by an industrial consortium under the overall responsibility of the prime contractor Thales Alenia Space (Cannes), and including Astrium (Friedrichshafen) responsible for the payload module and for system testing at spacecraft level, Thales Alenia Space (Turin) responsible for the service module, and Astrium (Toulouse) responsible for the telescope, with in excess of a hundred subcontractors. PACS has been developed by a consortium of institutes led by MPE (Germany) and including

UVIE (Austria); KUL, CSL, IMEC (Belgium); CEA, OAMP (France); MPIA (Germany); IFSI, OAP/AOT, OAA/CAISMI,

LENS, SISSA (Italy); IAC (Spain). This development has been supported by the funding agencies BMVIT (Austria), ESA-

PRODEX (Belgium), CEA/CNES (France), DLR (Germany), ASI (Italy) and CICYT/MICINN (Spain). The HerMES data was accessed through the HeDaM database (http://hedam.oamp.fr) operated by CeSAM and hosted by the Laboratoire d'Astrophysique de Marseille.

Funding for the SDSS and SDSS-II has been provided by the Alfred P. Sloan Foundation, the Participating Institutions, the National Science Foundation, the U.S. Department of Energy, the National Aeronautics and Space Administration, the Japanese Monbukagakusho, the Max Planck Society, and the Higher Education Funding Council for England. The SDSS Web Site is http://www.sdss.org/.

The SDSS is managed by the Astrophysical Research Consortium for the Participating Institutions. The Participating Institutions are the American Museum of Natural History, Astrophysical Institute Potsdam, University of Basel, University of Cambridge, Case Western Reserve University, University of Chicago, Drexel University, Fermilab, the Institute for Advanced Study, the Japan Participation Group, Johns Hopkins University, the Joint Institute for Nuclear Astrophysics, the Kavli Institute for Particle Astrophysics and Cosmology, the Korean Scientist Group, the Chinese Academy of Sciences (LAMOST), Los Alamos National Laboratory, the Max-Planck-Institute for Astronomy (MPIA), the Max-Planck-Institute for Astrophysics (MPA), New Mexico State University, Ohio State University, University of Pittsburgh, University of Portsmouth, Princeton University, the United States Naval Observatory, and the University of Washington. Financial support from the Spanish grant AYA2010-15169 and from the Junta de Andalucia through TIC-114 and the Excellence Project P08-TIC-03531 is acknowledged.

\bibliographystyle{mn2e}

\bibliography{ioteo_biblio}


%


%











%











\bsp

\label{lastpage}

\end{document}